\documentclass[twocolumn]{aastex631}
\shortauthors{Akaho et al.}
\graphicspath{{./}{figures/}}

\begin{document}

\title{Protoneutron Star Convection Simulated with a New General Relativistic Boltzmann Neutrino Radiation-Hydrodynamics Code}

\author[0000-0002-9234-813X]{Ryuichiro Akaho}
\affiliation{Graduate School of Advanced Science and Engineering,Waseda University,\\
3-4-1 Okubo, Shinjuku, Tokyo 169-8555, Japan}

\author[0000-0003-1409-0695]{Akira Harada}
\affiliation{Interdisciplinary Theoretical and Mathematical Sciences Program (iTHEMS), RIKEN, Wako, Saitama 351-0198, Japan}

\author[0000-0002-7205-6367]{Hiroki Nagakura}
\affiliation{Division of Science, National Astronomical Observatory of Japan, \\ 2-21-1 Osawa, Mitaka, Tokyo 181-8588, Japan}

\author[0000-0003-4959-069X]{Wakana Iwakami}
\affiliation{Advanced Research Institute for Science and Engineering, Waseda University, 3-4-1 Okubo, Shinjuku, Tokyo 169-8555, Japan}

\author{Hirotada Okawa}
\affiliation{Waseda Institute for Advanced Study, Waseda University, 1-6-1 Nishi-Waseda, Shinjuku-ku, Tokyo, 169-8050, Japan}

\author{Shun Furusawa}
\affiliation{College of Science and Engineering, Kanto Gakuin University, 1-50-1 Mutsuurahigashi, Kanazawa-ku, Yokohama, Kanagawa 236-8501, Japan}

\author{Hideo Matsufuru}
\affiliation{High Energy Accelerator Research Organization, 1-1 Oho, Tsukuba, Ibaraki 305-0801, Japan}

\author[0000-0002-9224-9449]{Kohsuke Sumiyoshi}
\affiliation{National Institute of Technology, Numazu College,\\
Ooka 3600, Numazu, Shizuoka 410-8501, Japan}

\author[0000-0002-2166-5605]{Shoichi Yamada}
\affiliation{Advanced Research Institute for Science and Engineering, Waseda University,\\
3-4-1 Okubo, Shinjuku, Tokyo 169-8555, Japan}



\begin{abstract}
We investigate the protoneutron star (PNS) convection using our newly developed general relativistic Boltzmann neutrino radiation-hydrodynamics code.
This is a pilot study for more comprehensive investigations later. As such, we take a snapshot of a PNS at 2.3 seconds after bounce from a 1D PNS cooling calculation and run our simulation for $\sim160\,\mathrm{ms}$ in 2D under axisymmetry. The original PNS cooling calculation neglected convection entirely and the initial condition is linearly unstable to convection. We find in our 2D simulation that convection is instigated there indeed and expands inward after being full-fledged. The convection is then settled to a quasi-steady state in $\sim100\,\mathrm{ms}$, being sustained by the negative $Y_e$ gradient, which is in turn maintained by neutrino emissions. It enhances the luminosities and mean energies of all species of neutrinos compared to 1D. Taking advantage of the Boltzmann solver, we analyze the possible occurrence of the neutrino fast flavor conversion (FFC). We found that FFC is likely to occur in the regions, where $Y_e$ is lower, and that the growth rate can be as high as $\sim 10^{-1}\,{\mathrm{cm}^{-1}}$.
\end{abstract}



\section{Introduction} \label{sec:intro}
Massive stars heavier than $\sim8M_\odot$ will undergo core-collapse supernovae (CCSNe) and eventually form neutron stars (NSs) or black holes (BHs).
Sophisticated CCSN simulations have been performed in the last decades, and successful explosions have been indeed obtained (\citet{Burrows2021} for a recent review).
However, the explosion energy and the Ni yield suggested by the observation remain to be reproduced systematically by 3D simulations (\citet{Bollig2021}). 

Recently, the subsequent phase, i.e., the protoneutron star (PNS) cooling, is also getting attention. For Galactic CCSNe, a large amount of neutrinos will be observed for $\sim100$ seconds by the current and future detectors \citep{Li2021}.
This phase will be hence useful for tightly constraining the nuclear matter equation of state (EOS) \citep{Pons2001b,Pons2001a,Roberts2012a,Nakazato2020,Nagakura2022,Nakazato2022}. Putting aside the initial condition, the evolution of PNS in its cooling phase may be (and has been indeed) studied independently of the supernova explosion preceding it.
The theoretical study of PNS cooling phase has its own difficulties, though. Its long duration of $\sim10\sim100$ seconds makes it difficult to simulate numerically with detailed microphysics taken into account.

In fact, the long-term ($\gtrsim10\,\mathrm{s}$) PNS cooling calculations have been performed in 1D under the assumption of spherical symmetry (see \citet{Roberts2017} for a summary of recent studies.) \deleted{Multi-dimensional behavior of the PNS has been investigated only in the early phase ($\sim$several seconds) \citep{Dessart2006,Nagakura2020,Nagakura2021c}.}
The largest drawback of 1D calculations, however, is that convection cannot be treated self-consistently. In the previous 1D simulations, it was either simply ignored \citep{Nakazato2013, Nakazato2019}, or the mixing-length treatment was employed to model the matter mixing by convection \citep{Roberts2012a}. \deleted{It has been pointed out that luminosity and the energy is enhanced by the convection \citep{Keil1996,Roberts2012a,Pascal2022}, hence its inclusion is crucial for better understanding of PNS.
In addition, there are also other multidimensional features such as rotation, NS kick, and asymmetric accretion, whose effects may systematically change the result.}
Other multidimensional features such as rotation, if any, cannot be treated, either (but see also \citet{Margalit2022}).

The multi-dimensional simulations of PNS cooling are very much scarce mainly because of the difficulty mentioned above. For example, \citet{Keil1996} studied the PNS convection up to $\sim1\mathrm{s}$ after core bounce in 2D under axisymmetry with the radial, gray neutrino transport taken into account. The initial condition is constructed by extracting a central portion ($1.1M_\odot$) of the supernova core at $25\,\mathrm{ms}$ post bounce in their 2D simulation for a $15M_\odot$ progenitor. They showed that a lepton-driven convection occurs inside PNS, and the convective zone is extended inward with time as the positive entropy gradient gets weaker. They also found that the neutrino luminosity and the mean energy are both enhanced by the convection. \citet{Mezzacappa1998}, employing the 1D MGFLD transport coupled with 2D hydrodynamics, reported that the PNS convection is suppressed by the neutrino transport in their simulations up to $\sim100\,\mathrm{ms}$ after bounce. This may be an artifact of the angle-averaged 1D neutrino transport, though, in which the lateral transfer will be overestimated. In fact, the simulations with the 2D neutrino transport performed by \citet{Buras2006b} and \citet{Dessart2006} observed a convection, which enhances the neutrino luminosities. Note that these studies are all limited to the very early phase up to a few $100\,\mathrm{ms}$ after bounce, since their main focus is the explosion mechanism.
To our knowledge, the paper by \citet{Nagakura2020} is the only one that is devoted to the study of the PNS convection in 3D. It is based on their 3D supernova simulations, which obtained shock revival. They demonstrated that the convection enhances the luminosity of $\nu_x$, but not of $\nu_e$ and $\bar{\nu}_e$. There are even longer simulations in 2D \citep{Burrows2021} and 3D \citep{Bollig2021} up to several seconds post bounce with the two-moment approximation employed for neutrino transfer. Unfortunately, the PNS convection was not discussed there. Hence there has been no  detailed multi-dimensional numerical investigation on the PNS convection after $\sim1\,\mathrm{s}$. That is what we want to do in this and later papers.

In so doing, neutrino transport plays an important role, since the PNS is mostly optically thick but the notable exception is near the surface, where  neutrinos decouple from matter and their luminosity and spectrum are formed. The latter region is a part of the convectively unstable zone as shown later and the multi-dimensional neutrino transfer coupled with hydrodynamics is required. Since neutrinos are neither in chemical nor in thermal equilibrium with matter there, the neutrino distribution in phase space should be calculated in general. There are broadly two categories of methods to solve the transfer equation without artificial approximations (see, e.g., \citet{Richers2017} for their comparison): (1) the deterministic method, which directly solves the Boltzmann equation using the $S_N$ method \citep{Mezzacappa1993,Sumiyoshi2005,Sumiyoshi2012} or the spectral method \citep{Peres2014}, (2) the probabilistic method, of which the Monte Carlo method is a representative \citep{Abdikamalov2012,Kato2020}.

Since these rigorous transport schemes are computationally demanding, various approximate transport schemes have been also employed in supernova simulations: the two-moment transport method \citep{Thorne1981,Shibata2011,Cardall2013}, the flux-limited diffusion (FLD) approximation \citep{Arnett1977}, the isotropic diffusion source approximation (IDSA) method \citep{Liebendorfer2009}, and the ray-by-ray plus approximation \citep{Buras2006b}. Comparison of these methods were also conducted and some differences were observed \citep{Skinner2016,Cabezon2018,Just2018}. 

\deleted{In addition, rigorous transport schemes are necessary in terms of the neutrino fast flavor conversion (FFC), which is recently suggested to occur inside CCSNe \citep{Abbar2019,Nagakura2019b,Azari2020,Harada2022}. Occurrence of the FFC is determined by the neutrino distribution in the momentum space. Hence the approximate transport schemes, which only have reduced information about it, is inappropriate for its judgement or analysis. (but also see \citet{Nagakura2021a,Nagakura2021b,Nagakura2021d}, where they could reasonably judge the occurrence of FFC in the results by the two-moment scheme if there is only one crossing).}

\deleted{The use of approximate method or simplifications is unavoidable, given the computational cost of the multi-dimensional simulation with the sophisticated neutrino transport. However, the approximate method should be well calibrated by the rigorous method. The feasible strategy is to perform short-term simulations with rigorous setup, and use the results for calibrating the approximate method.}

{
In this paper we employ the $S_N$ method, which we have developed over the years for supernova simulations \citep{Nagakura2018,Nagakura2019a,Harada2019,Harada2020,Iwakami2020}. Unlike the computations cited, we use a general relativistic hydrodynamics code coupled to the Boltzmann solver, which is already fully general relativistic \citep{Akaho2021}. It is needless to say that general relativity is indispensable in CCSNe \citep{Bruenn2001,Lentz2012a,OConnor2018a}, particularly at late times, when the PNS becomes more compact. Note that many of recent state-of-the-art supernova simulations are still not fully general relativistic but use effective potentials \citep{Glas2019,Burrows2020,Mezzacappa2022,Vartanyan2022}. In this paper, the time evolution of spacetime is still not considered and its metric is fixed. The use of the Boltzmann solver is also motivated by our interest in the collective neutrino oscillation via the fast flavor conversion (FFC). Since FFC is driven by the crossing of the neutrino flavor lepton numbers in the angular distributions in momentum space \citep{Morinaga2021}, the Boltzmann solver, which keeps track of the neutrino angular distributions in momentum space faithfully, is best suited for the study of FFC.}

{As we saw above, the long-term multi-dimensional study of the PNS cooling at $t\gtrsim 1\,\mathrm{s}$ is absent. We have started a project to tackle this issue. We have conceived a new Lagrangian formulation in full general relativity to numerically construct equilibrium configurations possibly with rotation in 2D under axisymmetry \citep{Okawa2022} to be used for their long-term evolutions. In this paper, we study the convection in PNS, particularly its influences on the neutrino emission, by rather short-term ($\sim100\,\mathrm{ms}$) 2D simulations conducted for a snapshot at $t=2.3\,\mathrm{s}$ provided by a conventional 1D simulation of PNS cooling \citep{Nakazato2019}. Note that this is a first ever attempt of such kinds and meant to be a pilot study for more systematic explorations later. It is known that the convection and turbulent motions it produces in 2D are qualitatively different from those in 3D. Hence we need to be cautious in considering the implications of this study for the realistic 3D PNS cooling. However, the difference between 1D and 2D is much larger than that between 2D and 3D \citep{Vartanyan2018} and we think that the 2D study is still useful.}

This paper is organized as follows. Section \ref{sec_method} describes the numerical method, and the simulation setups are summarized in section \ref{sec_setup}. The simulation results are presented in section \ref{sec_result}. We summarize our findings and discuss future prospects in section \ref{sec_summary}. The details for the newly developed hydrodynamics code are given in appendix \ref{sec_hydroformulation}, and the code verification tests are performed in appendix \ref{sec_codetests}. Throughout the paper, we use the metric signature $-+++$, and the Greek ($\alpha$, $\beta$, $\mu$, $\phi$) and Latin ($i$, $j$, $k$) indices run over 0-3 and 1–3, respectively.

\section{Numerical Method}
\label{sec_method}
We use the the Boltzmann radiation hydrodynamics code. It simultaneously solves the Boltzmann neutrino transport and hydrodynamics equations in general relativity. The details are summarized in the following subsections.

\subsection{Boltzmann Solver}
For the neutrino transport, we solve the Boltzmann equation with respect to the distribution function $f$ in phase space \citep{Lindquist1966,Ehlers1971}. The equation is written in the conservative form \citep{Shibata2014}:
\begin{eqnarray}
\label{eq_conservBoltz}
&& \left.\frac{1}{\sqrt{-g}}\frac{\partial}{\partial x^\mu}\right|_{q_i}\left[\left(e_{(0)}^{\mu} + \sum_{i=1}^{3}l_i {e_{(i)}^\mu}\right)\sqrt{-g}f\right] \nonumber \\
&& -\frac{1}{\epsilon^2}\frac{\partial}{\partial\epsilon}\left(\epsilon^3f\omega_{(0)}\right) + \frac{1}{\mathrm{\mathrm{sin}\theta_\nu}}\frac{\partial}{\partial\theta_\nu}\left(\mathrm{sin}\theta_\nu f\omega_{(\theta_\nu)}\right) \nonumber \\
&& -\frac{1}{\mathrm{sin}^2\theta_\nu}\frac{\partial}{\partial{\phi_\nu}}\left(f\omega_{({\phi_\nu})}\right) = S_{\rm rad},
\end{eqnarray}
where $\epsilon\equiv-p_\mu e_{(0)}^\mu$ is the neutrino energy; $\theta_\nu$ and $\phi_\nu$ are the zenith and azimuth angles in momentum space, respectively. The symbol $q_i$ represents the momentum space coordinates as $q_1=\epsilon$, $q_2=\theta_\nu$, $q_3=\phi_\nu$. The symbol $g$ denotes the determinant of the metric tensor. The directional cosines in momentum space $l_{(i)}$ are given as
\begin{eqnarray}
&& l_{(1)} = \mathrm{cos}\theta_\nu, \nonumber \\
&& l_{(2)} = \mathrm{sin}\theta_\nu\mathrm{cos}{\phi_\nu}, \nonumber \\
&& l_{(3)} = \mathrm{sin}\theta_\nu\mathrm{sin}{\phi_\nu}. \nonumber
\end{eqnarray}
The factors $\omega_{(0)}$, $\omega_{(\theta_\nu)}$, and $\omega_{(\phi_\nu)}$ are defined as
\begin{eqnarray}
&& \omega_{(0)} \equiv \epsilon^{-2} p^\mu p_\nu \nabla_\mu e_{(0)}^\mu, \nonumber \\
&& \omega_{(\theta_\nu)} \equiv \sum_{i=1}^{3}\omega_{(i)}\frac{\partial l_{(i)}}{\partial \theta_\nu}, \nonumber \\
&& \omega_{(\phi_\nu)} \equiv \sum_{i=2}^{3}\omega_{(i)}\frac{\partial l_{(i)}}{\partial \phi_\nu},  \nonumber \\
&& \omega_i \equiv \epsilon^{-2} p^\mu p_\nu \nabla_\mu e_{(i)}^\nu. \nonumber
\end{eqnarray}
In the configuration space, the spherical polar coordinates are deployed, where $r$, $\theta$, $\phi$ corresponds to the radius, the zenith and azimuth angles, respectively. The tetrad bases are given as
\begin{eqnarray}
e^\mu_{(0)} && = (1/\alpha, -\beta^i/\alpha),
\nonumber \\
e^\mu_{(1)} && = 
\gamma_{rr}^{-1/2}\left(\frac{\partial}{\partial r}\right)^\mu,
\nonumber \\
e^\mu_{(2)} && = - \frac{\gamma_{r\theta}}{\sqrt{\gamma_{rr}(\gamma_{rr}\gamma_{\theta\theta} - \gamma_{r\theta}^2)}}
\left(\frac{\partial}{\partial r}\right)^\mu
\nonumber \\
&&
+ \sqrt{\frac{\gamma_{rr}}{\gamma_{rr}\gamma_{\theta\theta} - \gamma_{r\theta}^2}}
\left(\frac{\partial}{\partial\theta}\right)^\mu,
\nonumber \\
e^\mu_{(3)} && = 
\frac{\gamma^{r\phi}}{\sqrt{\gamma^{\phi\phi}}}
\left(\frac{\partial}{\partial r}\right)^\mu
+
\frac{\gamma^{\theta\phi}}{\sqrt{\gamma^{\phi\phi}}}
\left(\frac{\partial}{\partial \theta}\right)^\mu
\nonumber \\
&& +
\sqrt{\gamma^{\phi\phi}}
\left(\frac{\partial}{\partial \phi}\right)^\mu, \label{eq_tetrad}
\end{eqnarray}
where the symbols $\alpha$, $\beta^i$ and $\gamma_{ij}$ stand for the lapse function, the shift vector, and the spatial metric on the time-constant hypersurface, respectively.
Further details and the code tests can be found in \citet{Akaho2021}.

\subsection{Hydrodynamics Solver}
\label{sec_method_hydro}
In this paper, we newly developed a general relativistic hydrodynamics solver coupled with the Boltzmann neutrino transport solver. It is a general relativistic extension of our previous Newtonian hydrodynamics code. Code verification tests are performed in appendix \ref{sec_codetests}.

We solve the general relativistic hydrodynamics equations written in the 3+1 decomposition (see \S4 in \citet{Shibata2015}):
\begin{equation}\label{eq_massconserv}
\partial_t \rho_* + \partial_j(\rho_* v^j) = 0,
\end{equation}
\begin{eqnarray}\label{eq_Euler}
&& \partial_t S_i + \partial_j (S_i v^j + \alpha\sqrt{\gamma}Pc^2\delta_i{}^j) 
\nonumber \\
&=& -S_0 c^2\partial_i \alpha + S_j c\partial_i \beta^j - \frac{1}{2}\alpha c^2 \sqrt{\gamma} S_{jk}\partial_i \gamma^{jk} \nonumber \\
&-& \alpha \sqrt{\gamma}\gamma_i{}^\mu G_\mu,
\end{eqnarray}
\begin{eqnarray}\label{eq_energyconserv}
&& \partial_t (S_0 - \rho_*c^2) + \partial_k ((S_0 - \rho_*c^2)v^k + \sqrt{\gamma}P (v^k + c\beta^k)) 
\nonumber \\
&=& \alpha\sqrt{\gamma}S^{ij}K_{ij} - S_i D^i \alpha + \alpha \sqrt{\gamma}n^\mu G_\mu,
\end{eqnarray}
where
\begin{eqnarray}
v^j &\equiv& \frac{u^j}{u^t}, \\
\rho_* &\equiv& \alpha \sqrt{\gamma}\rho_0 u^t = \sqrt{\gamma}\rho_0\omega, \\
S_j &\equiv& \rho_* h u_j c,\\
S_0 &\equiv& \sqrt{\gamma}(\rho h w^2 - P), \\
w &\equiv& \alpha u^t, \\
S_{ij} &\equiv& \rho h u_i u_j + P\gamma_{ij},
\end{eqnarray}
and $\rho$, $P$, $u^\mu$, $h$, represent the density, the pressure, the four velocity, and the specific enthalpy, respectively. The symbol $G^\mu$ stands for the feedback from neutrinos, which are calculated in the same way as in our previous Newtonian hydrodynamics code \citep{Nagakura2017}.

We need a special care when we discretize the equations. In a near-equilibrium situation, there is a severe cancellation between the source terms and the derivatives of fluxes. However, unlike analytical formulae, it is not necessarily guaranteed in the finite difference. The mismatch between the terms to be cancelled will cause unphysical matter motions. {We indeed found in the stability test of the relativistic star in appendix \ref{sec_TOV} that ensuring this cancellation as much as possible is crucially important. We give in appendix \ref{sec_hydroformulation} the explicit formulation that satisfies this cancellation exactly in the flat spacetime. Although the cancellation is not perfect in curved spacetimes in general, it turns out that the formulation still works well also in such cases as demonstrated in the tests presented in sections \ref{sec_TOV} and \ref{sec_BHacc}.} Further details of the hydrodynamics code are described in appendix \ref{sec_hydroformulation}.

\section{Setup}
\label{sec_setup}
{We use a snapshot taken from a 1D PNS model \citep{Nakazato2019} as our initial condition. In their paper, the collapse of a $15\,M_\odot$ progenitor \citep{Woosley1995} was followed with the Lagrangian radiation hydrodynamics code by \citet{Sumiyoshi2005} up to $300\,\mathrm{ms}$. Although this simulation did not lead to shock revival, they extracted the central portion of the supernova core that includes a PNS and its subsequent evolution was calculated for $\sim80\,\mathrm{s}$ by another code for the quasi-static PNS cooling developed by the same authors. This cooling computation was done under the assumption of spherical symmetry, and Togashi EOS \citep{Togashi2017} was employed. We pick up the snapshot at $2.3\,{\rm s}$ after bounce and use it as the initial data for our own calculation. 
Note that the original PNS cooling by \citet{Nakazato2019} employed the flux-limited diffusion approximation. The use of the Boltzmann transport in this paper is hence inconsistent, resulting in some differences in the luminosities and spectra of neutrinos obtained in the two calculations. Since the neutrino reactions incorporated and their treatments are also different between the two simulations, the detailed comparison is not very useful and not needed indeed for the current purpose of the paper, i.e., the study of the dynamics of PNS convection and the difference it could make in the neutrino luminosity and energy. We will hence not consider this issue further.}

In order to mitigate the possible unphysical transient at the beginning of the 2D simulation, we first perform a 1D calculation with our code until the initial wobbling is settled. Then we start the 2D calculation, using the resultant configuration as initial condition. We impose the following perturbations to the four-velocity of matter by hand to instigate convection:
\begin{eqnarray}\label{eq_perturb}
\delta u_r &=& V \mathrm{cos}\left(m\pi\frac{r-r_\mathrm{min}}{r_\mathrm{max}-r_\mathrm{min}}\right)\mathrm{cos}(n\theta), 
\nonumber \\
\delta u_{\theta} &=& Vr\mathrm{sin}\left(m\pi\frac{r-r_\mathrm{min}}{r_\mathrm{max}-r_\mathrm{min}}\right)\mathrm{sin}(n\theta),
\end{eqnarray}
where $V\equiv1\times10^{-4}\times c$, $r_\mathrm{min}\equiv3\mathrm{km}$, $r_\mathrm{max}\equiv13\mathrm{km}$, $m\equiv10$ and $n\equiv10$.

The mesh setup is the same for the 1D and 2D calculations except that $\theta$ and $\phi_\nu$ degrees of freedom are excluded in 1D. The spatial coordinates are divided into $N_r=384$ and $N_\theta=128$ cells in the $r$ and $\theta$ directions, respectively. The radial grid covers the range $r\in[0:50]\mathrm{km}$, where it has a finer resolution around the PNS surface. As for the momentum space, the energy and $\theta_\nu$ and $\phi_\nu$ grids have $N_\epsilon=20$, $N_{\theta_\nu}=10$, $N_{\phi_\nu}=6$, respectively. The energy grid covers the range $\epsilon\in[0:300]\mathrm{MeV}$, where it is logarithmically spaced. {In order to see the resolution dependence of numerical results, we also run a simulation with a higher spatial resolution: $N_r=512$ , $N_\theta=192$ (referred to as model \texttt{HR-S}) and another with a higher angular resolution in momentum space: $N_{\theta_\nu}=14$, $N_{\phi_\nu}=10$ (model \texttt{HR-M}). The boundary conditions are imposed as follows: the reflective condition at the center $r=0$ and on the poles $\theta=0,\,\pi$, and the free-streaming condition at the outer boundary located at $r=50\,\mathrm{km}$, and the periodic condition for $\phi_\nu=0,\,2\pi$.}

In this study, we employ the Furusawa-Togashi EOS \citep{Furusawa2017}. Although it is another inconsistency with the initial data prepared with the Togashi EOS, this difference is minor. We employ the neutrino reactions based on \citet{Bruenn1985} with some extensions: energy-changing electron scattering, and the nucleon–nucleon bremsstrahlung are incorporated.

\begin{figure}[t]
    \begin{tabular}{c}
    \begin{minipage}[t]{0.8\hsize}
        \centering
        \includegraphics[scale=0.7]{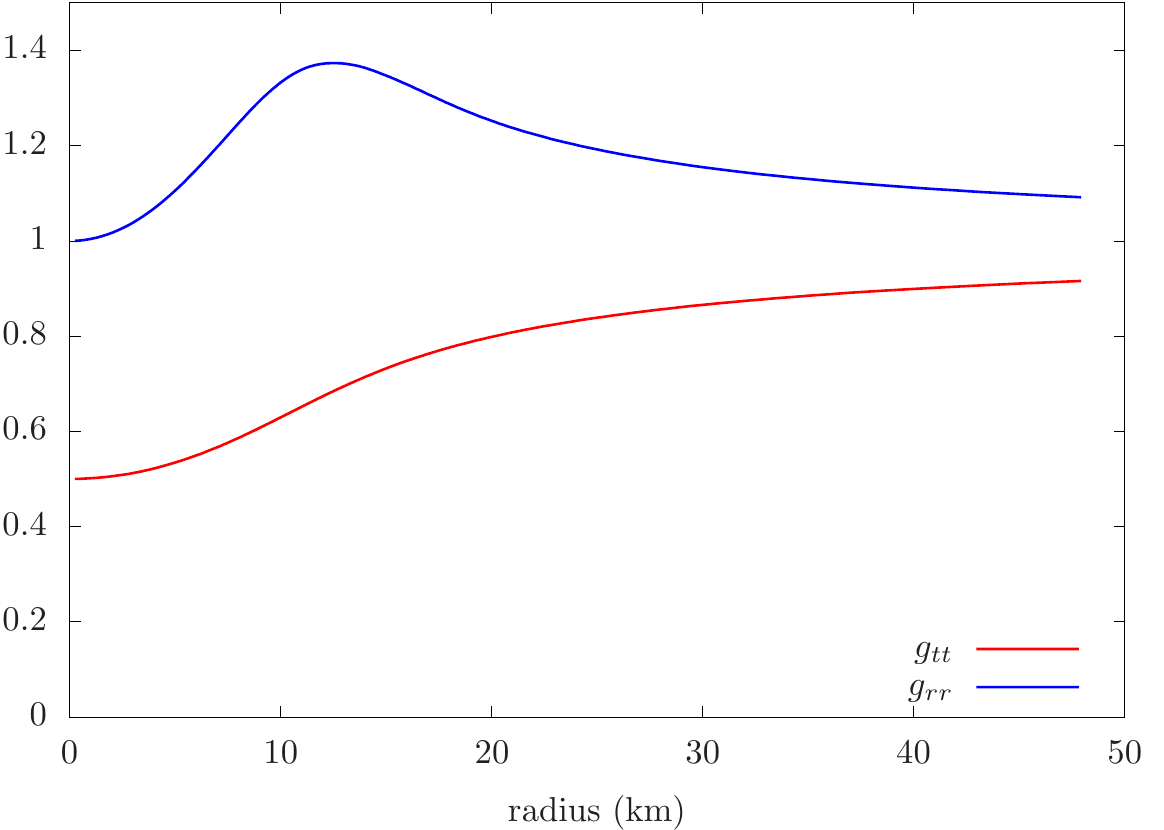}
      \end{minipage} 
    \end{tabular}
  \caption{Radial profiles of metric components $g_{tt}$ (red) and $g_{rr}$ (blue).}
  \label{fig_metric}
\end{figure}

\deleted{The spacetime metric is derived by solving the Tolman-Oppenheimer-Volkoff (TOV) equations.} 
{The spacetime metric is determined in the following manner. We adopt the metric ansatz:}
\begin{equation}
ds^2 = -e^{2\Phi(r)}c^2dt^2 + (X(r))^2dr^2 + r^2d\theta^2 + r^2\mathrm{sin}^2\theta d\phi^2.
\end{equation}
{Then the Einstein equation gives the function $X(r)$ as \citep{Oppenheimer1939}}
\begin{equation}
X(r)=\left(1-\frac{2GM(r)}{r}\right)^{-\frac{1}{2}},
\end{equation}
{in which $M(r)$ is the enclosed mass and is given by the following equation with the density $\rho$:}
\begin{eqnarray}
\frac{dM(r)}{dr} = 4\pi r^2\rho(r).
\end{eqnarray}
{The function $\Phi(r)$, on the other hand, is derived as \citep{Oppenheimer1939}}
\begin{eqnarray}
\frac{d\Phi(r)}{dr} = \frac{G}{r^2c^2}\left(M(r)+\frac{4\pi r^3P(r)}{c^2}\right)\left(1-\frac{2GM(r)}{rc^2}\right)^{-1},
\end{eqnarray}
{where $P(r)$ denotes the pressure. The subsequent evolution of the metric is not considered and the spacetime geometry is fixed in this paper.} Since the density distribution is not globally asymmetric and the fluid velocities are orders of magnitude smaller than the speed of light, this assumption is reasonable. {The nontrivial metric components, $g_{tt}$ and $g_{rr}$, are shown in figure \ref{fig_metric}.}

\section{Results}
\label{sec_result}
\subsection{Basic Characteristics}
\begin{figure}[t]
    \begin{tabular}{c}
    \begin{minipage}[t]{1\hsize}
        \centering
        \includegraphics[scale=0.7]{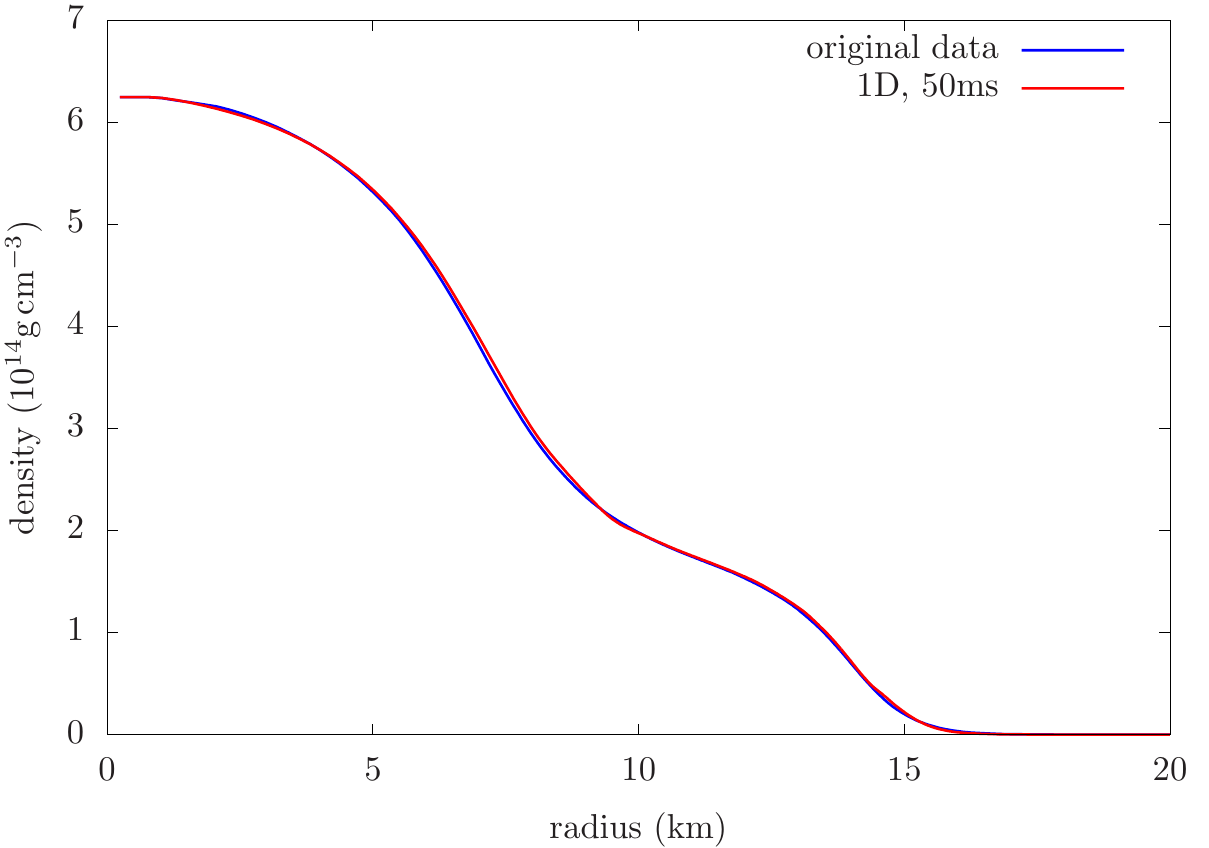}
      \end{minipage} \\
    \begin{minipage}[t]{1\hsize}
        \centering
        \includegraphics[scale=0.7]{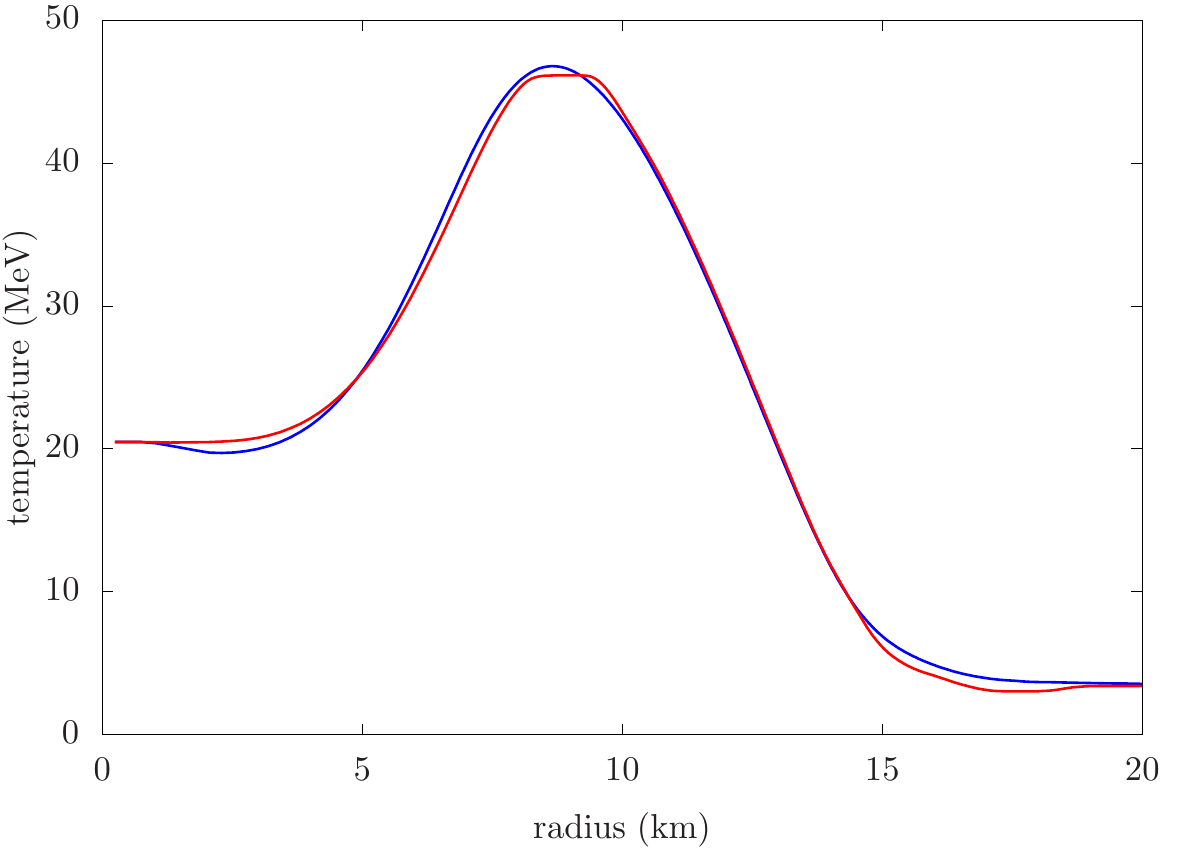}
      \end{minipage} \\
    \begin{minipage}[t]{1\hsize}
        \centering
        \includegraphics[scale=0.7]{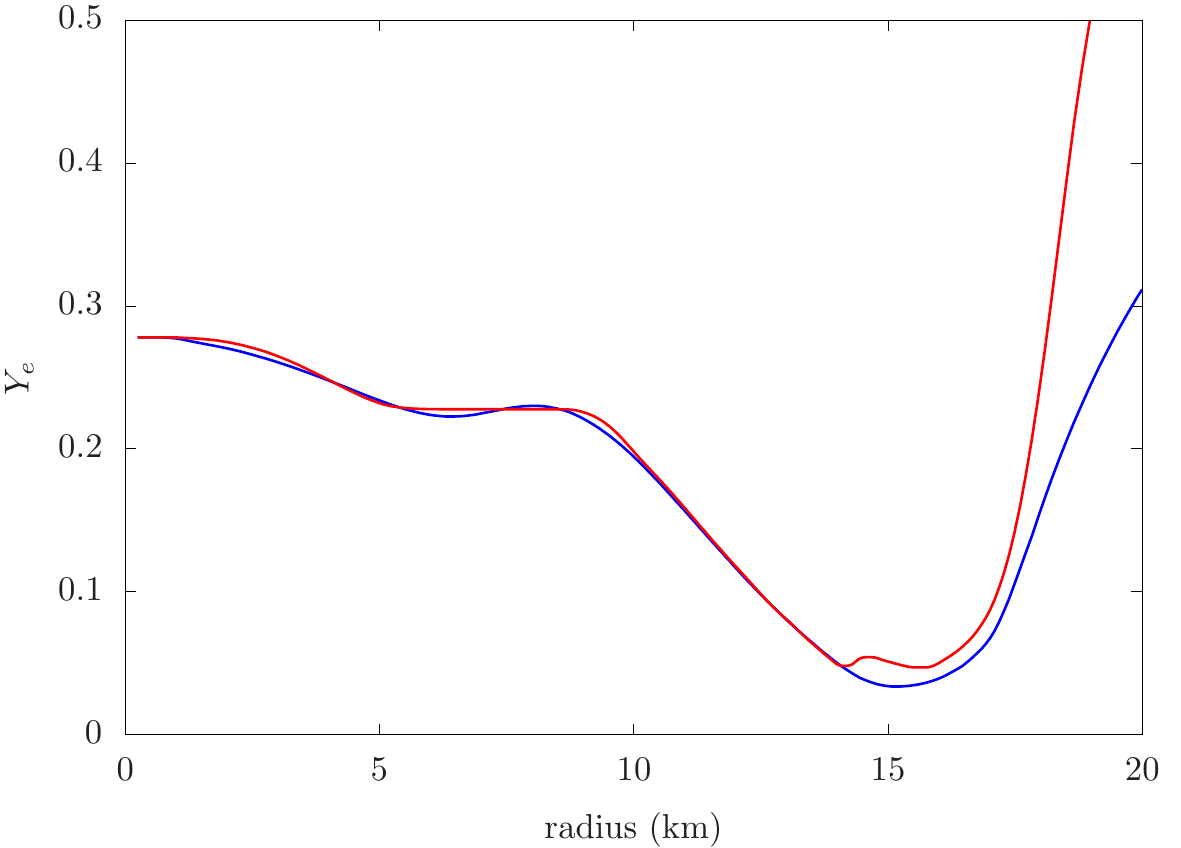}
      \end{minipage} \\
    \end{tabular}
    \caption{Radial profiles of the density (top), the temperature (middle), and the electron fraction (bottom) are shown with red lines. The original data (provided by another code) is shown with blue lines.}
    \label{fig_1Dresult}
\end{figure}

We first perform the 1D calculation with our own code until initial transients subside as stated above. Figure-\ref{fig_1Dresult} compares the radial profiles of some hydrodynamical variables at $50\,\mathrm{ms}$ with those at the initial time. {Note that the computational range in this calculation is extended from the original one: $r\lesssim17\,\mathrm{km}$ to $r<50\,\mathrm{km}$ for numerical convenience by adding a low-density matter smoothly. In so doing, all hydrodynamical variables are extrapolated as shown in the figure.} 

The density and the temperature distributions are almost the same as the original ones with the deviations being at most several \%. As for the electron fraction, the dip gets a bit shallower from the initial data at $r\sim15\,\mathrm{km}$. Further out we observe a continuous rise of $Y_e$ toward the outer boundary. This is because the absorption of $\nu_e$ is dominant over that of $\bar\nu_e$ while the emissions of these neutrinos are much smaller in this optically thin region. As explained above, the initial $Y_e$ distribution at $r\gtrsim17\,\mathrm{km}$ is set by the extrapolation and the large discrepancy found there is just as expected. Since the density in this region is low and the region with negative lepton gradients inside are little affected by the initial transients, we think it is rather unlikely that the convection of our interest is severely modified by the presence of the extended layer. The overall agreement between the data before and after the relaxation with no anomalous behavior is an indirect indication that our code is working properly and can treat PNS correctly in general relativity.
From the snapshot at $50\mathrm{ms}$, which we refer to as \texttt{2Dinitial} hereafter, we start the 2D calculation. For comparison we continue the 1D calculation also. 

\begin{figure}[htbp]
    \begin{tabular}{cc}
    \begin{minipage}[t]{1\hsize}
        \centering
        \includegraphics[scale=0.7]{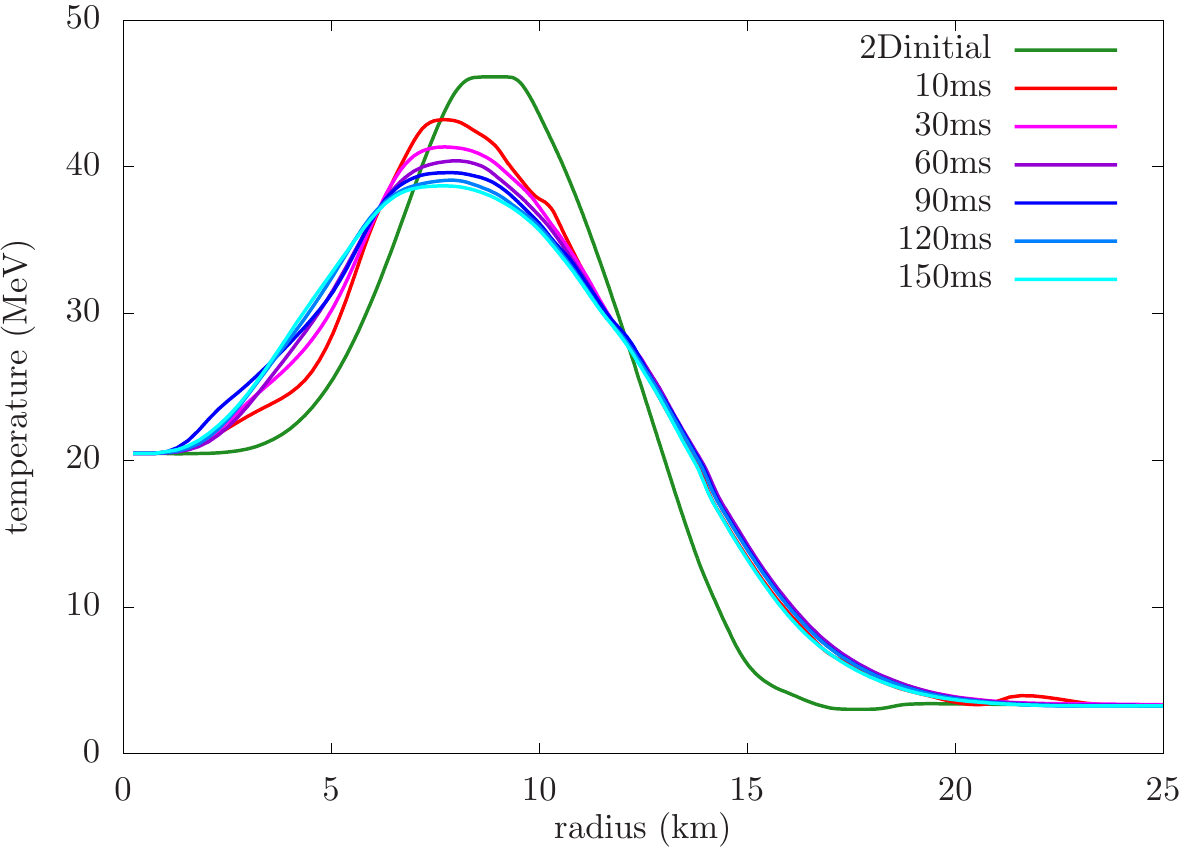}
      \end{minipage} \\
    \begin{minipage}[t]{1\hsize}
        \centering
        \includegraphics[scale=0.7]{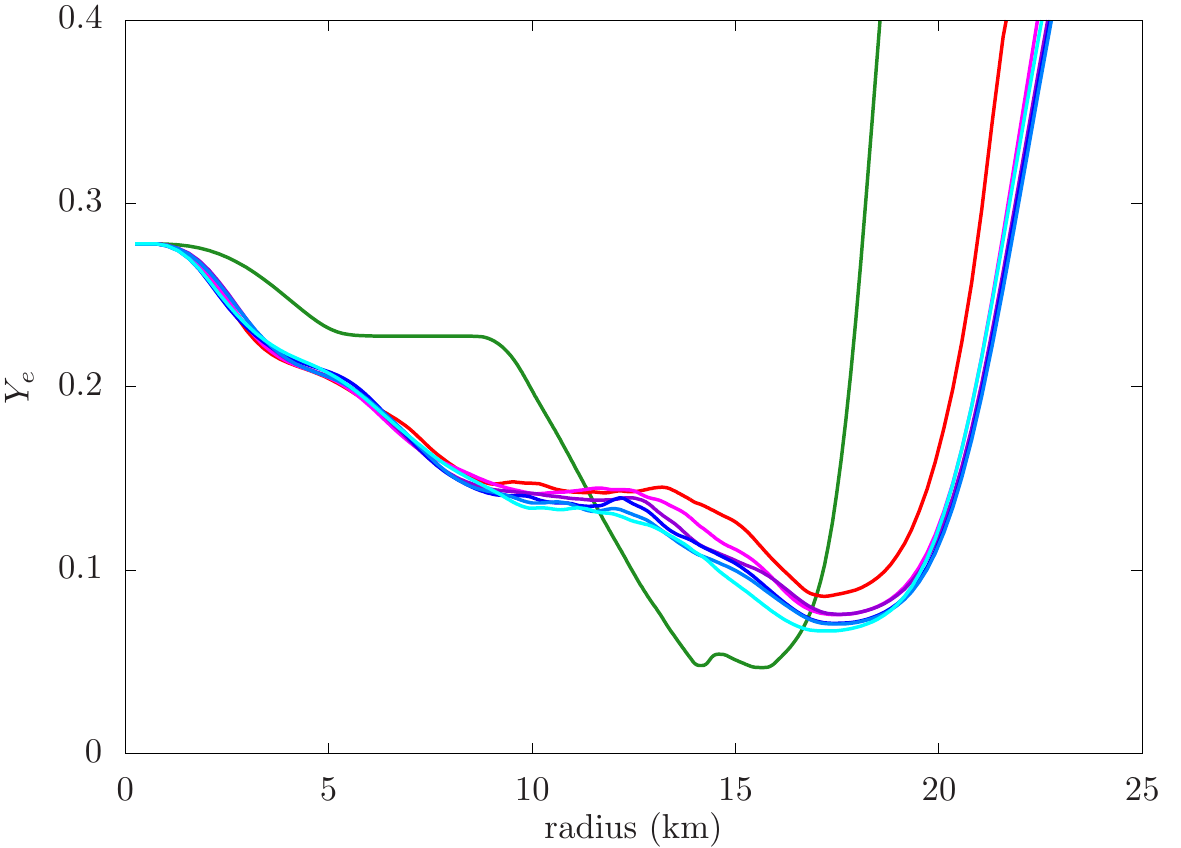}
      \end{minipage} \\
    \begin{minipage}[t]{1\hsize}
        \centering
        \includegraphics[scale=0.7]{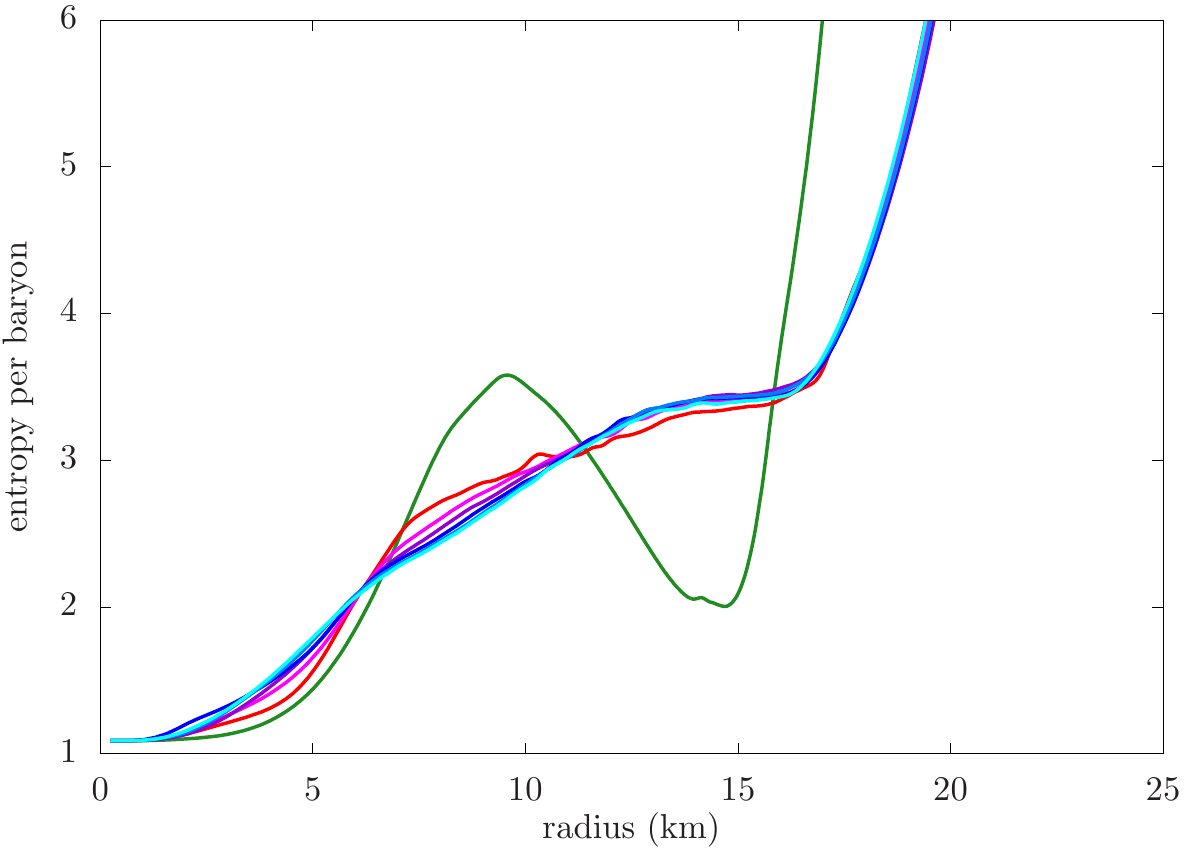}
      \end{minipage} \\
    \end{tabular}
    \caption{Radial profiles of temperature (top), the electron fraction (middle), and the entropy per baryon (bottom). Green line denote \texttt{2Dinitial}, and the rest denote 2D data for time snapshots 10, 30, 60, 90, 120, and 150ms, respectively. Note that $\theta$-averaged quantities are shown for 2D data.}
    \label{fig_tempye_evo}
\end{figure}

Figure \ref{fig_tempye_evo} compares the angle-averaged radial profiles of the temperature, the electron fraction, and the entropy per baryon between \texttt{2Dinitial} and some snapshots from the 2D simulation at later times. For all quantities, the radial distributions get smoothed out with time due to the matter mixing driven by the convection. The most notable difference appears in the entropy per baryon: the peak and dip feature is gone and the gradient is positive at all radii in the 2D case. 

\begin{figure}[t]
    \begin{tabular}{c}
    \begin{minipage}[t]{0.8\hsize}
        \centering
        \includegraphics[scale=0.7]{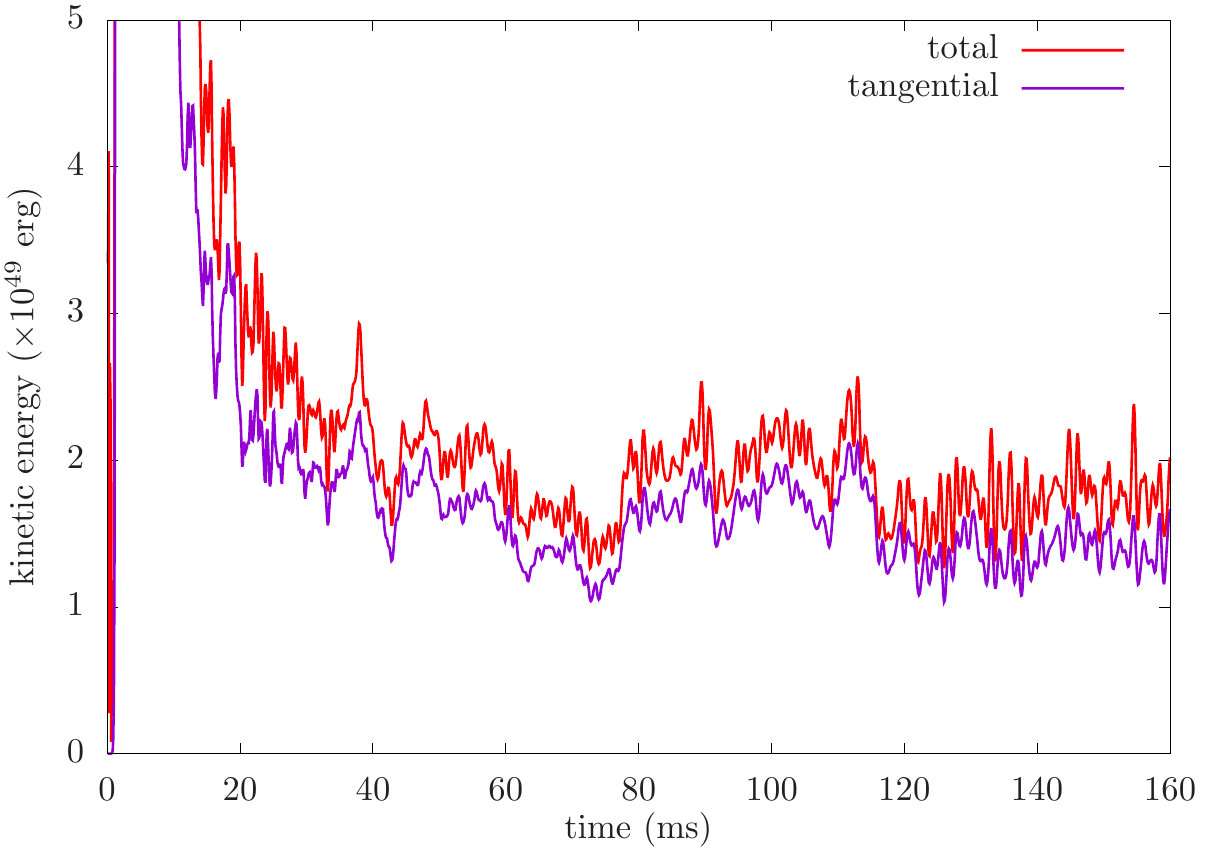}
      \end{minipage} 
    \end{tabular}
  \caption{Time evolution of the kinetic energy. The red and purple lines denote total and tangential ones, respectively.}
  \label{fig_kineticene}
\end{figure}

In order to see the vigor of convection, we show the time evolution of the kinetic energy in figure \ref{fig_kineticene}. The perturbation given in equation \ref{eq_perturb} instigates violent convective motions in several milliseconds. \deleted{The tangential kinetic energy accounts for the large portion of the total kinetic energy.} The kinetic energy decreases until $30\,\mathrm{ms}$ and settles gradually down to a fluctuation around a constant value thereafter. The initial violent turbulence is an artifact produced by the transition from 1D to 2D. After $\sim100\,\mathrm{ms}$, however, the transient is subsided and the convection has reached a (quasi) steady state, in which the angle-averaged matter profiles change much more slowly on the secular time scale. Regarding this state as representative of the PNS convection around this time of the cooling phase, we analyze it further in the following.

\begin{figure*}[t]
 \begin{center}
  \includegraphics[width=17cm]{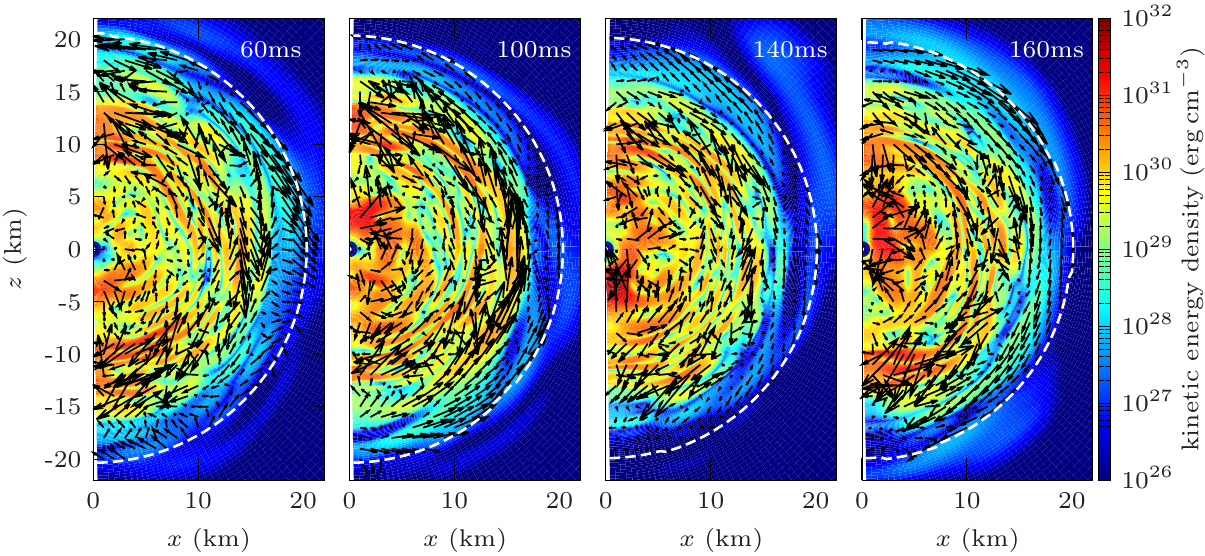}
  \caption{Kinetic energy density (total) on the meridian slice for some time snapshots; from left to right, 60ms, 100ms, 140ms and 160ms. Matter velocity field inside the PNS is also plotted as black arrows. The white broken line denote the PNS surface.}
  \label{fig_2Dmap_kineticene}
 \end{center}
\end{figure*}

We first look at where the PNS convection occurs.
Figure \ref{fig_2Dmap_kineticene} is the 2D color map of the kinetic energy in the meridian section at different times. The velocity field inside the PNS is superimposed. Note that the PNS is defined as the region where the density is higher than $10^{11}\,\mathrm{g\,cm^{-3}}$. 
The kinetic energy and velocity distributions are non-spherical from the center up to $r\sim17\,\mathrm{km}$, with the lateral motions dominant over the radial ones (see also figure \ref{fig_kineticene}). This seems to be consistent with the convection that \citet{Nagakura2020} investigated. \deleted{It indicates that convection commonly occurs in the wide range of PNS. As we saw in the figure \ref{fig_tempye_evo}, the entropy gradient is positive for all region and the $Y_e$ gradient is negative up to $\sim17\,\mathrm{km}$. As can be seen in figure \ref{fig_2Dmap_kineticene}, the kinetic energy and the velocity are large for inside ($\lesssim17\,\mathrm{km}$), and small for outer convectively stable region.} This is in contrast to other earlier works, though, in which either almost no convection was found \citep{Lattimer1981,Mezzacappa1993} or it was observed in a rather limited region \citep{Dessart2006,Nagakura2021a}. This discrepancy may be attributed to the differences in the treatment of neutrino transfer as well as to the difference in the phase we focus to. As we saw in figure \ref{fig_tempye_evo}, the entropy gradient is positive in the entire region while the $Y_e$-gradient remains negative in the convective region at this rather late time of $t\sim\,2\mathrm{s}$. Although our simulation is not fully self-consistent as the initial matter profile is taken from the 1D PNS cooling calculation \citep{Nakazato2019}, in which convection was ignored, the entropy and $Y_e$ distributions are self-adjusted and sustained once the (quasi) steady state is established, and hence the results obtained in this paper will have some generality.

In order to see the convectively unstable region more quantitatively, we evaluate the relativistic Brunt-V\"ais\"al\"a frequency 
\begin{equation}
N^2=\frac{\partial\alpha}{\partial r}\frac{\alpha}{\rho h \gamma_{rr}}\left(\frac{1}{c_s^2}\frac{\partial P}{\partial r}-\frac{\partial \rho(1+\epsilon)}{\partial r}\right),
\end{equation}
where $N^2<0$ indicates the linear convective instability. We adopted this formula from \citet{Muller2013}, in which the conformal flatness is assumed, by simply replacing the conformal factor with $\gamma_{rr}$. 
It is conventional \citep{Gossan2020} to define the following quantity:
\begin{equation}
f_\mathrm{BV}=\mathrm{sign}(N^2)\sqrt{|N^2|}.
\end{equation}

\begin{figure*}[t]
 \begin{center}
  \includegraphics[width=17cm]{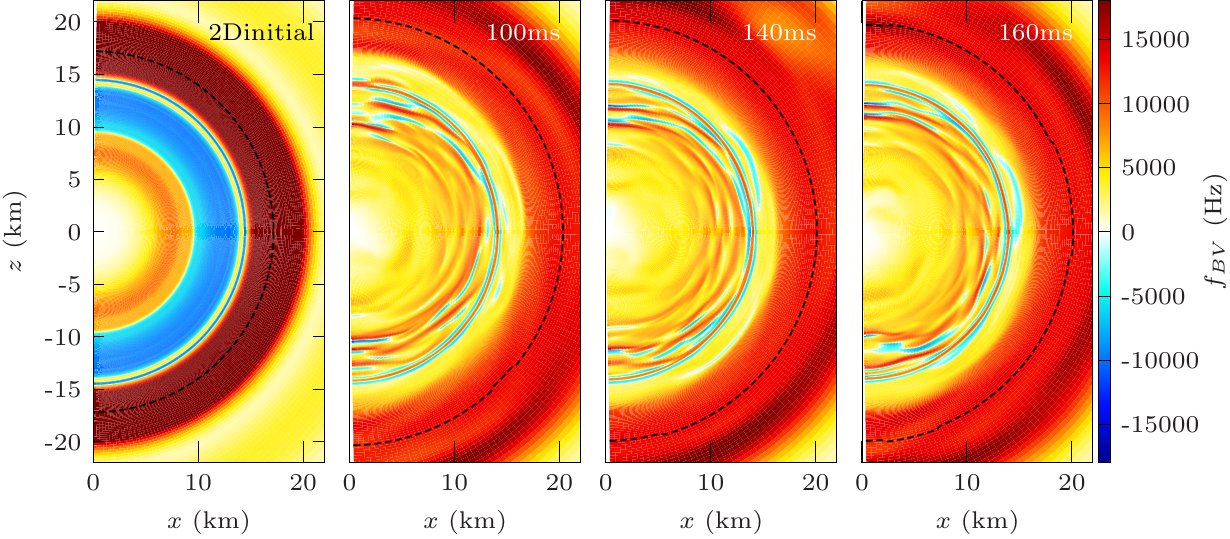}
  \caption{Brunt-V\"ais\"al\"a frequency on the meridian slice for \texttt{2Dinitial} and rest for 2D (100, 140, and 160ms). The black broken lines denote the PNS surface.}
  \label{fig_BVfreq}
 \end{center}
\end{figure*}

{Figure \ref{fig_BVfreq} shows the spatial distributions of $f_\mathrm{BV}$ in the meridian section for some time snapshots. The leftmost panel presents the result for the initial time. We can see clear boundaries between the (linearly) unstable region with bluish colors and the stable regions with reddish colors. It is found that the inner region $r<9\,\mathrm{km}$ is linearly stable although it has negative $Y_e$-gradients (see figure \ref{fig_tempye_evo}). The convection occurs in this linearly unstable region initially indeed. As it grows with time and becomes nonlinear, the convective region is extended inward by overshooting and the inner stable region is extinct eventually. At the same time, the value of $f_\mathrm{BV}$ gets smaller inside the convective region as can be observed in other panels to the right. This happens because the convection mixes the matter up until its profiles become marginally stable, since the neutrino emissions still work in the opposite way to produce unstable $Y_e$-profiles. As a result, the convection is sustained even if the Brunt-V\"ais\"al\"a frequency is marginally positive.}

\begin{figure}[htbp]
    \begin{tabular}{c}
      \begin{minipage}[t]{1.0\hsize}
        \centering
        \includegraphics[scale=0.7]{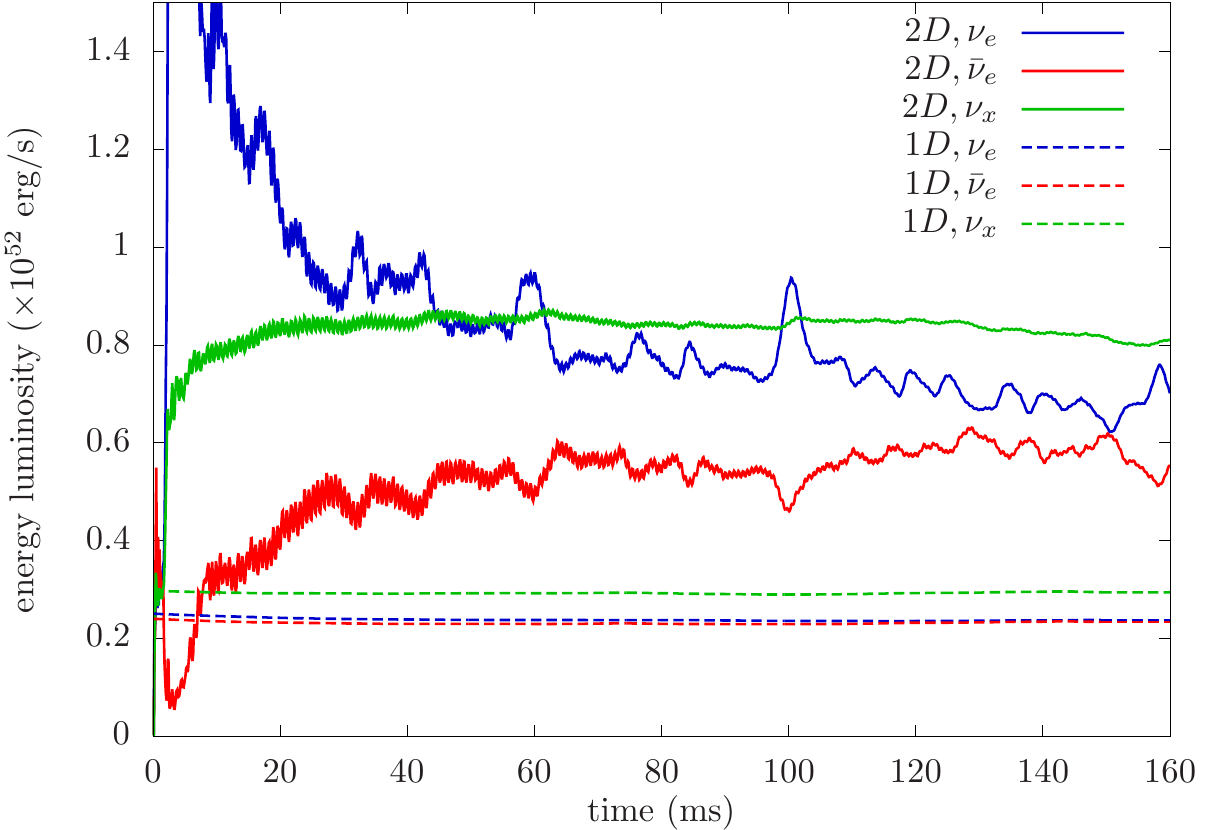}
      \end{minipage} \\
      \begin{minipage}[t]{1.0\hsize}
        \centering
        \includegraphics[scale=0.7]{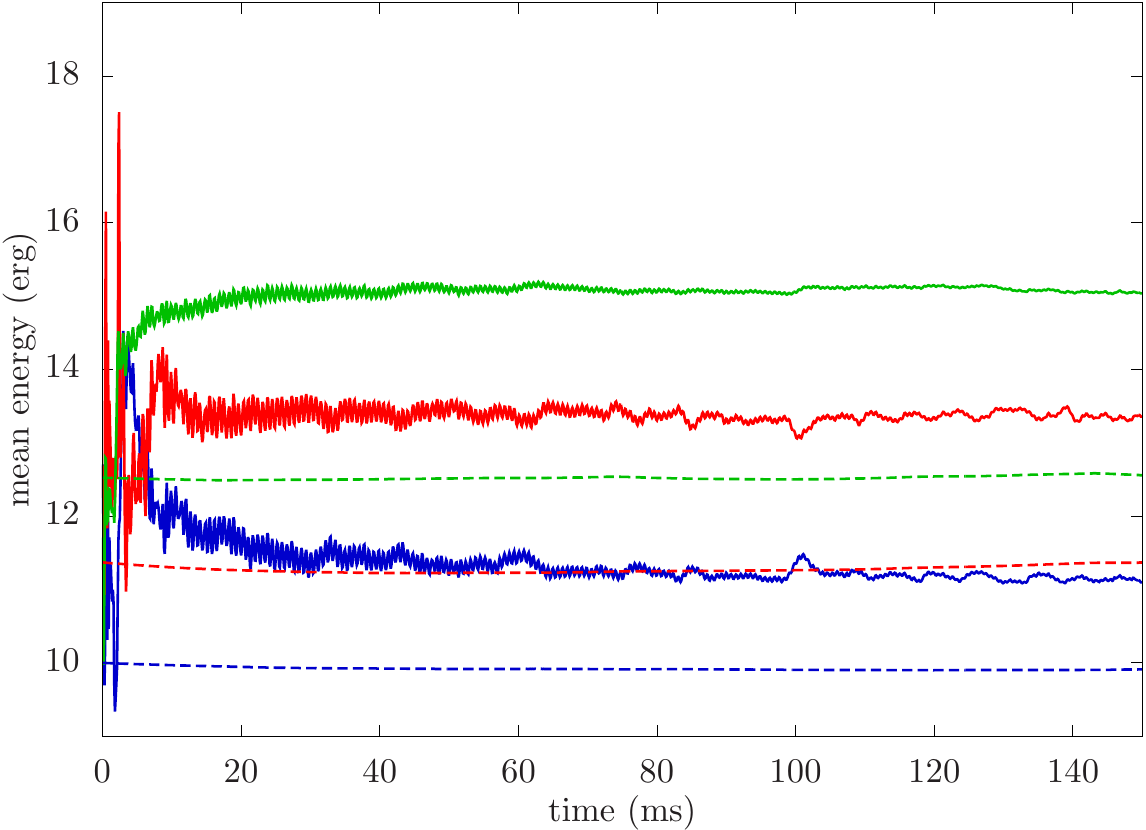}
      \end{minipage}
    \end{tabular}
    \caption{Time evolution of the energy luminosity (left) and mean neutrino energy (right). The solid lines correspond to 2D, and the broken line correspond to 1D. Blue, red, and green lines denote $\nu_e$, $\bar\nu_e$, and $\nu_x$, respectively.}
    \label{fig_lumi}
\end{figure}

Next, we discuss the properties of neutrino emissions from the PNS. Figure \ref{fig_lumi} shows the time evolution of the luminosity and mean energy of neutrinos. 
Similarly to the kinetic energy, we found a sudden increase of the $\nu_e$ luminosity right after the start of the simulation, which is followed by a gradual decline as the initial transient subsides, and is settled to a roughly constant value in $\sim70\,\mathrm{ms}$. 
{The rise of the $\bar{\nu}_e$ luminosity is turned to a decline very quickly, followed by a gradual increase to a constant value over several tens milliseconds. It is interesting that the time variations are anti-correlated with those of $\nu_e$. The $\nu_x$ luminosity, on the other hand, increases monotonically and approaches an asymptotic value a bit more quickly than other neutrinos without producing a pronounced peak. These different behaviors reflect the difference in their decoupling with matter.} 
This early time evolution is an artifact, though, caused by the switch from 1D to 2D. {We hence focus on the asymptotic phase after the (quasi) steady convection is established.
The luminosity and the mean energy in 2D are both higher than those in 1D (compare the solid lines with the dashed lines). Such an enhancement was also observed in the previous studies on the earlier phase \citep{Keil1996,Buras2006b,Dessart2006,Roberts2012a,Pascal2022}.}

\deleted{Overall, the luminosity is the order of $o(10^{51})\sim o(10^{52})\,\mathrm{erg\,s^{-1}}$ and the energy $10\sim20\,\mathrm{MeV}$, which are in reasonable agreement with the results obtained with previous works with different code \citep{Nagakura2020,Nagakura2021c}.}

\begin{figure}[t]
    \begin{tabular}{c}
    \begin{minipage}[t]{0.8\hsize}
        \centering
        \includegraphics[scale=0.7]{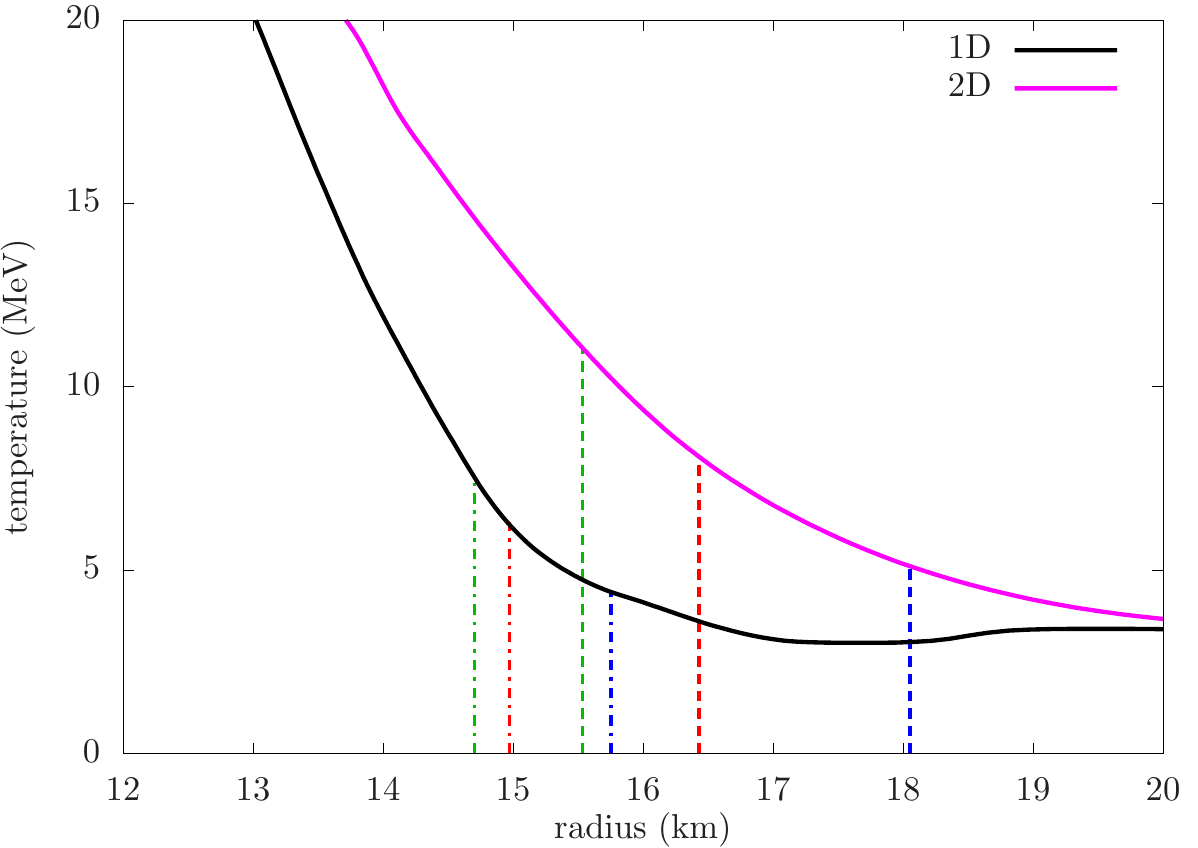}
      \end{minipage} 
    \end{tabular}
  \caption{Temperature profiles near the neutrino sphere of the neutrinos with the energy of 12.8 MeV. The black and magenta lines denote the result for 1D and 2D, respectively. We pick up the snapshot $t=150\,\mathrm{ms}$. The vertical lines denote the positions of the neutrino sphere for each flavor; $\nu_e$ (blue), $\bar\nu_e$ (red), and $\nu_x$ (green), and the chain lines denote 1D, and the broken lines for 2D. Note that angle-averaged quantities are shown for the 2D result.}
  \label{fig_neutrinosph}
\end{figure}

\begin{figure}[t]
    \begin{tabular}{c}
    \begin{minipage}[t]{0.8\hsize}
        \centering
        \includegraphics[scale=0.7]{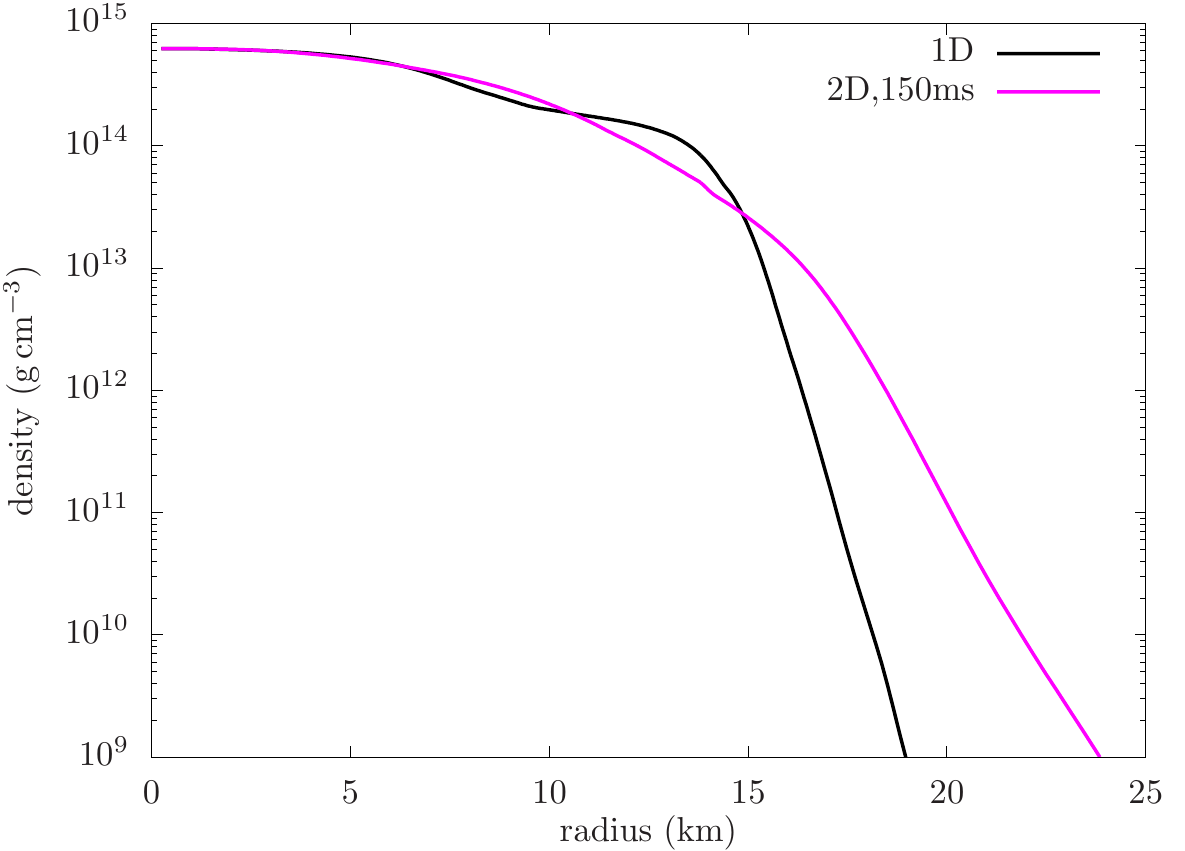}
      \end{minipage} 
    \end{tabular}
  \caption{Comparison of density between 1D and 2D.}
  \label{fig_rhoevo}
\end{figure}

The enhancement of the mean energy in 2D can be understood from the angle-averaged radial profiles of temperature near the neutrino sphere shown in figure~\ref{fig_neutrinosph}. The vertical lines indicate the positions of the neutrino sphere, which is defined to be the surface with the optical depth of 2/3, for individual neutrino species at the energy of $12.8\,\mathrm{MeV}$. {The dash-dotted and dashed lines indicate the positions of the neutrino sphere for 1D and 2D, respectively.}
\deleted{Since we focus on the neutrino emission, the absorption reactions are only taken into account to calculate the mean free path here.} 
It is evident that the temperature is higher in 2D than in 1D. This is due to the dredge-up by convection of the hotter matter located originally deeper inside. If the location of the neutrino sphere were unchanged, the mean energy would be even higher. As should be also apparent in figure \ref{fig_neutrinosph}, the neutrino spheres are all shifted outwards to lower temperatures. This is due to the density rise at $r\gtrsim15\,\mathrm{km}$ (see figure \ref{fig_rhoevo}) associated with the expansion of PNS, which is in turn driven by the convection. For all flavors, the temperatures at the neutrino spheres are higher, which naturally leads to their greater mean energies as well as luminosities as observed in figure \ref{fig_lumi}. In addition, the smaller neutrino sphere in 1D means that it is located deeper in the gravitational well, and the neutrinos emitted from it experience a greater gravitational redshift. This effect further lowers the luminosity and the mean energy for all flavors of neutrinos in 1D.

\subsection{Asymmetric Properties}
\label{sec_asymmetric}
In this subsection, we discuss the asymmetry caused by the convection and its effect on the neutrino emission.
Since we assume axisymmetry in this study, the results are inevitably affected by the artifacts unique to 2D. Nevertheless, the results will be useful as the reference for the 3D study in the future.

\begin{figure}[htbp]
    \begin{tabular}{c}
      \begin{minipage}[t]{1.0\hsize}
        \centering
        \includegraphics[scale=0.7]{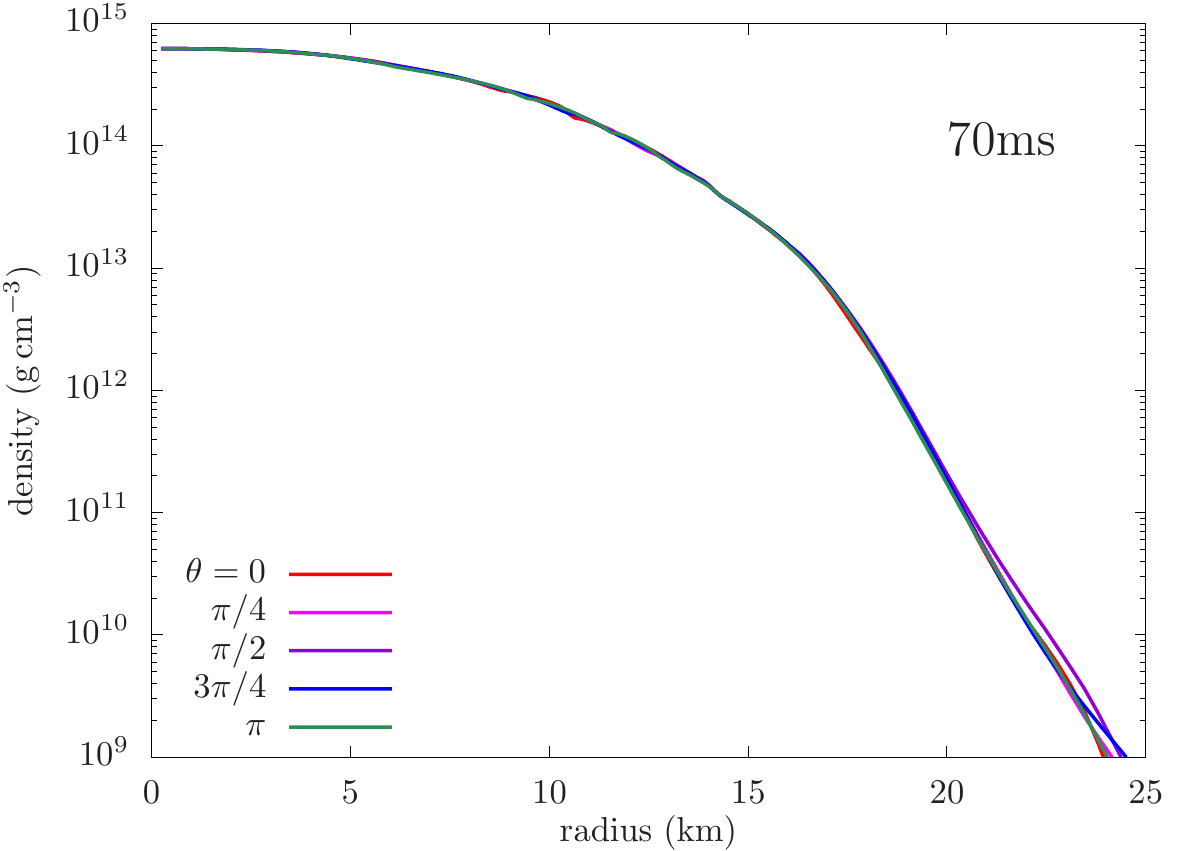}
      \end{minipage} \\
      \begin{minipage}[t]{1.0\hsize}
        \centering
        \includegraphics[scale=0.7]{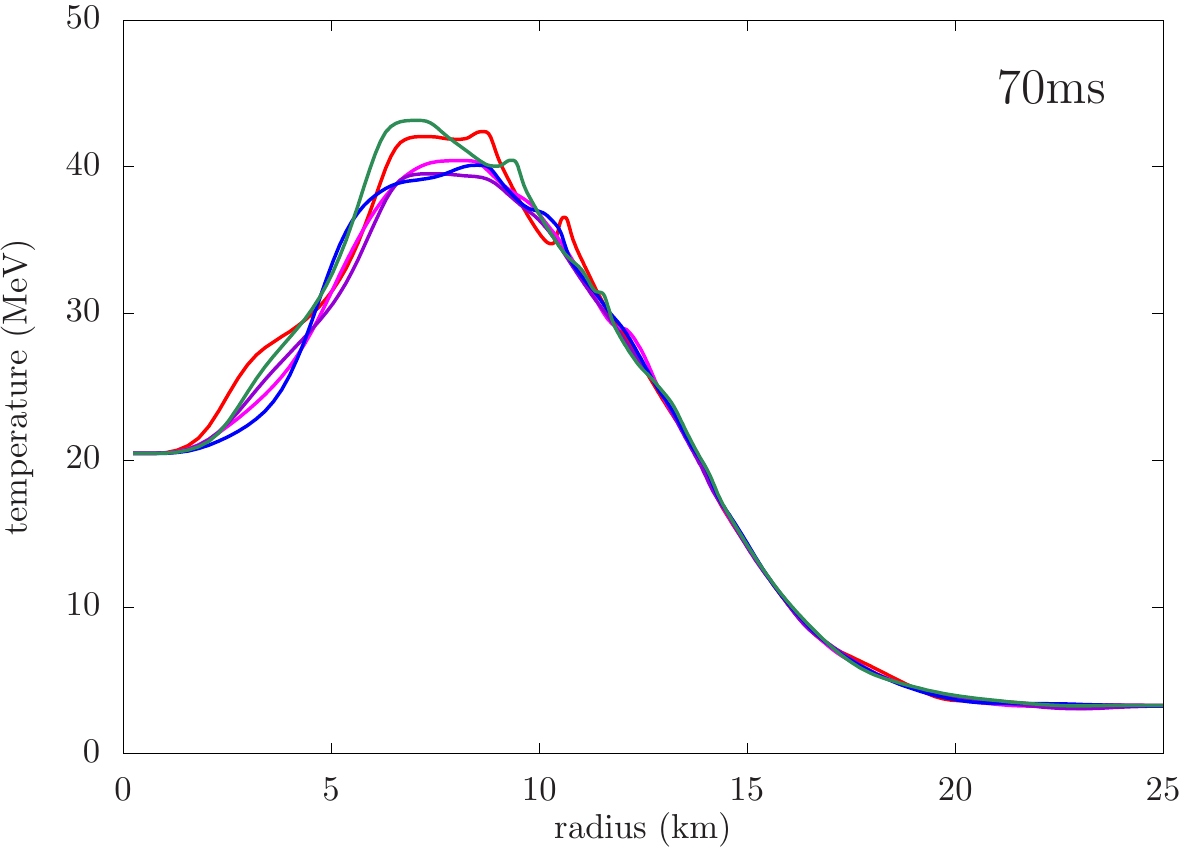}
      \end{minipage} \\
      \begin{minipage}[t]{1.0\hsize}
        \centering
        \includegraphics[scale=0.7]{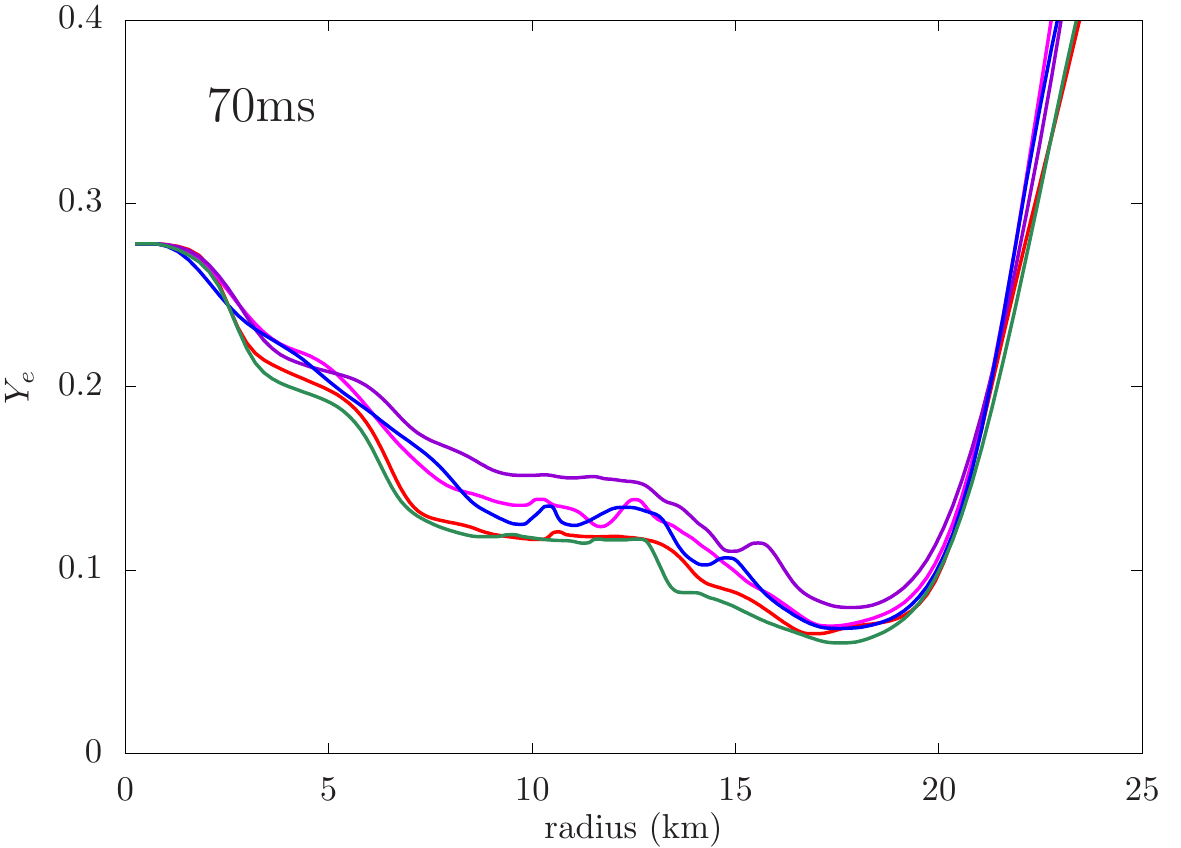}
      \end{minipage} 
    \end{tabular}
    \caption{Comparison of the density (top), and the electron fraction (middle), and the temperature (bottom) for different angles at a time snapshot 70ms. The red, magenta, purple, blue, and green lines correspond to $\theta=0$, $\pi/4$, $\pi/2$, $3\pi/4$, and $\pi$, respectively.}
    \label{fig_thetadistri_70}
\end{figure}

\begin{figure}[htbp]
    \begin{tabular}{c}
      \begin{minipage}[t]{1.0\hsize}
        \centering
        \includegraphics[scale=0.7]{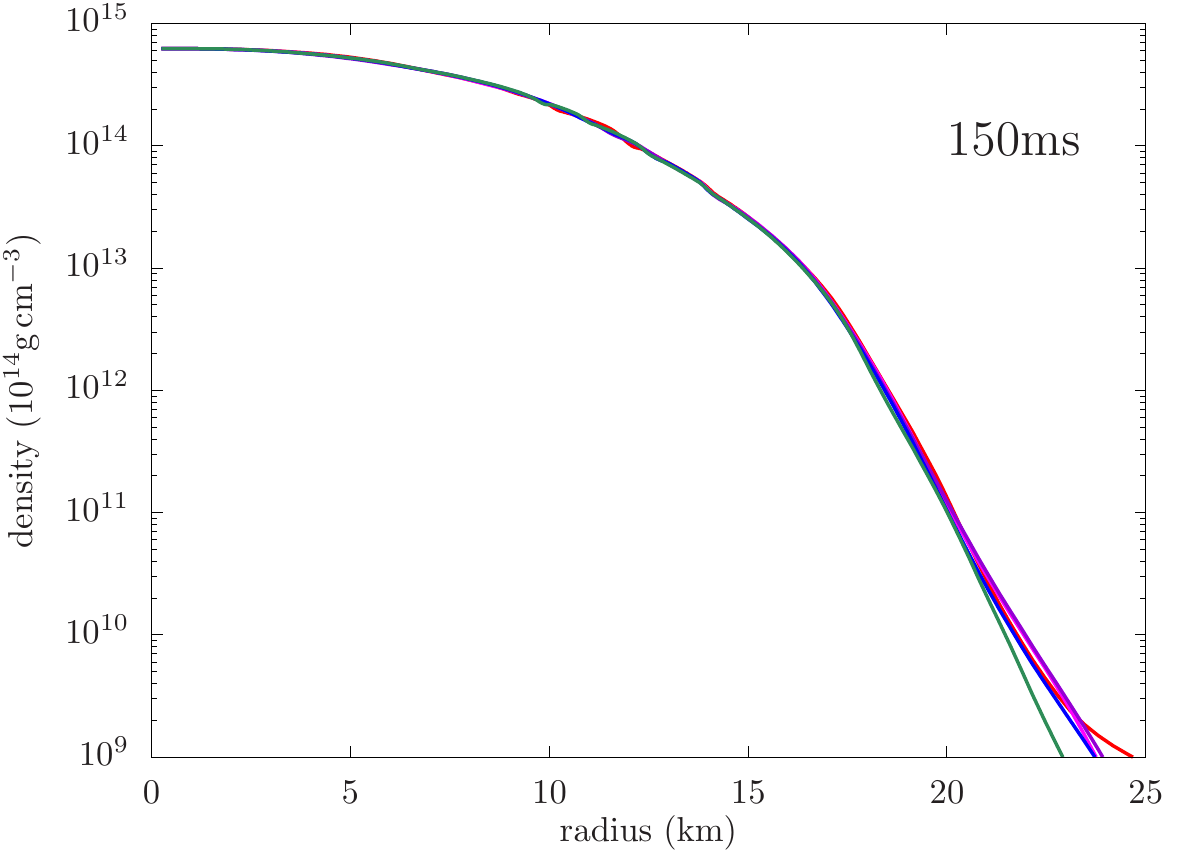}
      \end{minipage} \\
      \begin{minipage}[t]{1.0\hsize}
        \centering
        \includegraphics[scale=0.7]{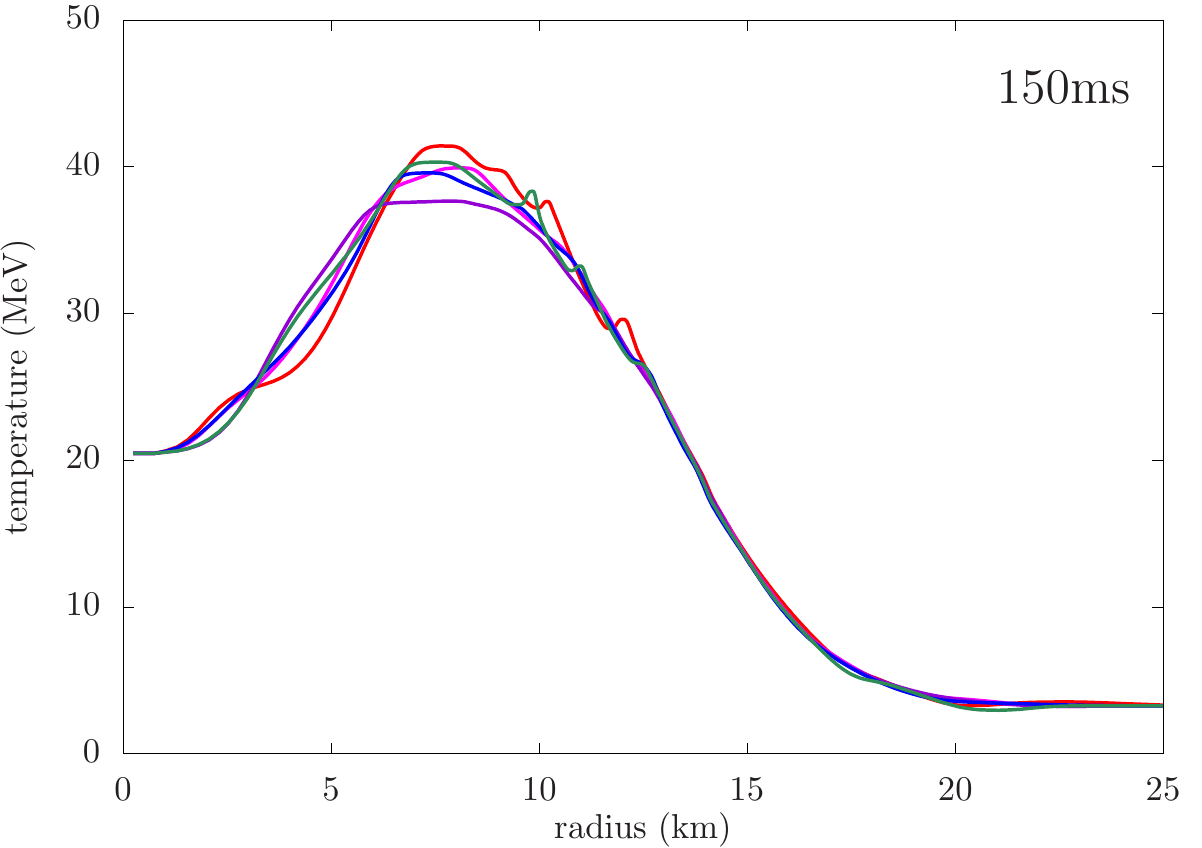}
      \end{minipage} \\
      \begin{minipage}[t]{1.0\hsize}
        \centering
        \includegraphics[scale=0.7]{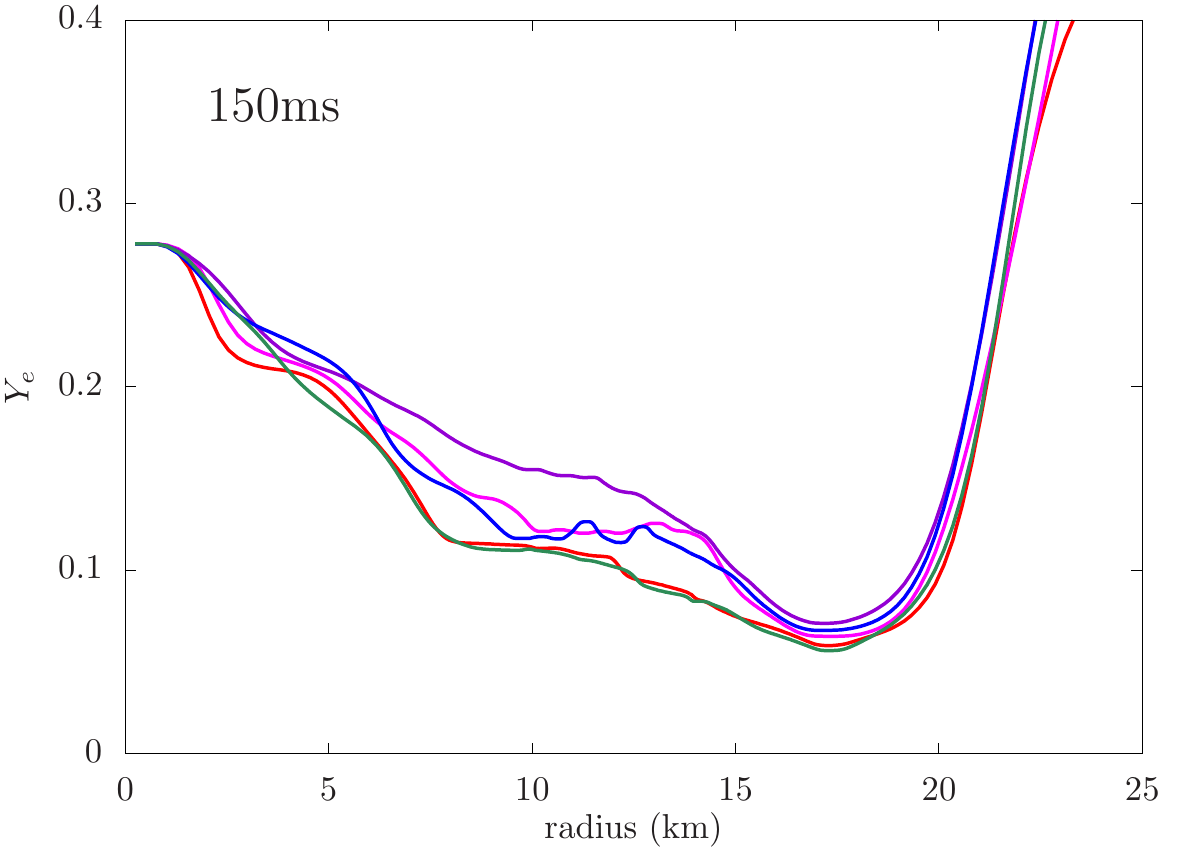}
      \end{minipage} 
    \end{tabular}
    \caption{Same as figure \ref{fig_thetadistri_70}, but for the time snapshot 150ms.}
    \label{fig_thetadistri_150}
\end{figure}

Figures \ref{fig_thetadistri_70} and \ref{fig_thetadistri_150} show the angle-dependence of density, temperature, and electron fraction at different times. The density variation is rather small except for the outer low-density layer at $r\gtrsim20\,\mathrm{km}$, which we do not consider.
The temperature distribution shows remarkable variations at $r\lesssim13\,\mathrm{km}$ due to the convection, whereas it is almost symmetric at larger radii. The temperature profiles are especially bumpy at the north ($\theta=0$) and the south ($\theta=\pi$) poles, suggesting that the convective motion is more violent there.
The $Y_e$ profile is also highly asymmetric in a more extended region, in which the maximum deviation from the average, $\Delta Y_e\gtrsim0.05$, occurs around $r=10\,\mathrm{km}$. In addition, $Y_e$ tends to be lower on the poles than on the equator. This again indicates that the convection is stronger near the poles than around the equator. 

\begin{figure}[t]
    \begin{tabular}{c}
    \begin{minipage}[t]{1\hsize}
        \centering
        \includegraphics[scale=1]{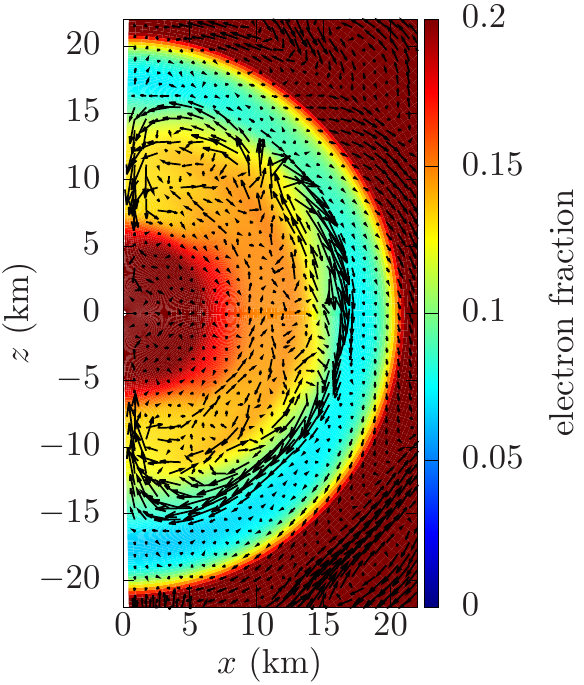}
      \end{minipage} 
    \end{tabular}
  \caption{Time-averaged electron fraction and the velocity field on the meridian slice. The average is taken for the time period of $t\in[20,40]\,\mathrm{ms}.$}
  \label{fig_2Dmap_yevelo}
\end{figure}

{
In order to see the convective pattern more clearly, we plot the time-averaged electron fraction and velocity in figure \ref{fig_2Dmap_yevelo}. 
One can see three large vortices: one near the north pole, another centered at an intermediate latitude in the northern hemisphere and the other covering most of the southern hemisphere. Note that we do not impose the equatorial symmetry. They are extended radially from $\sim7\,\mathrm{km}$ to $\sim16\,\mathrm{km}$. Near the both poles matter is moving downwards, causing lower values of $Y_e$ to prevail at $r\lesssim15\,\mathrm{km}$.}
At the same time, the central region with higher values ($\gtrsim0.15$) of $Y_e$ becomes a bit oblate. This is why we observed $Y_e$ tends to be lower near the poles in figures \ref{fig_thetadistri_70} and \ref{fig_thetadistri_150}. {The converging flows observed at the poles are mostly due to the artifact well-known in the axisymmetric simulation. However, a similar $Y_e$ anisotropy was observed by \citet{Keil1996} (Fig. 3), in their 2D simulation, in which the poles were avoided by choosing a $45^\circ$-wedge region centered at the equator as their computational zone. They found that the zone at $15\lesssim r\lesssim20\,\mathrm{km}$ is divided into two, large vortices with lepton-rich matter rising and deleptonized matter sinking.}{We hence think that such configurations are rather generic and expect that the downdraft of low-$Y_e$ matter and the updraft of high-$Y_e$ matter will occur also in 3D at several points.} 
{Moreover, if the PNS is rotating rapidly, there may occur converging flows at the poles indeed. We have to wait for 3D studies, however.}

\begin{figure}[htbp]
    \begin{tabular}{c}
      \begin{minipage}[t]{1.0\hsize}
        \centering
        \includegraphics[scale=0.7]{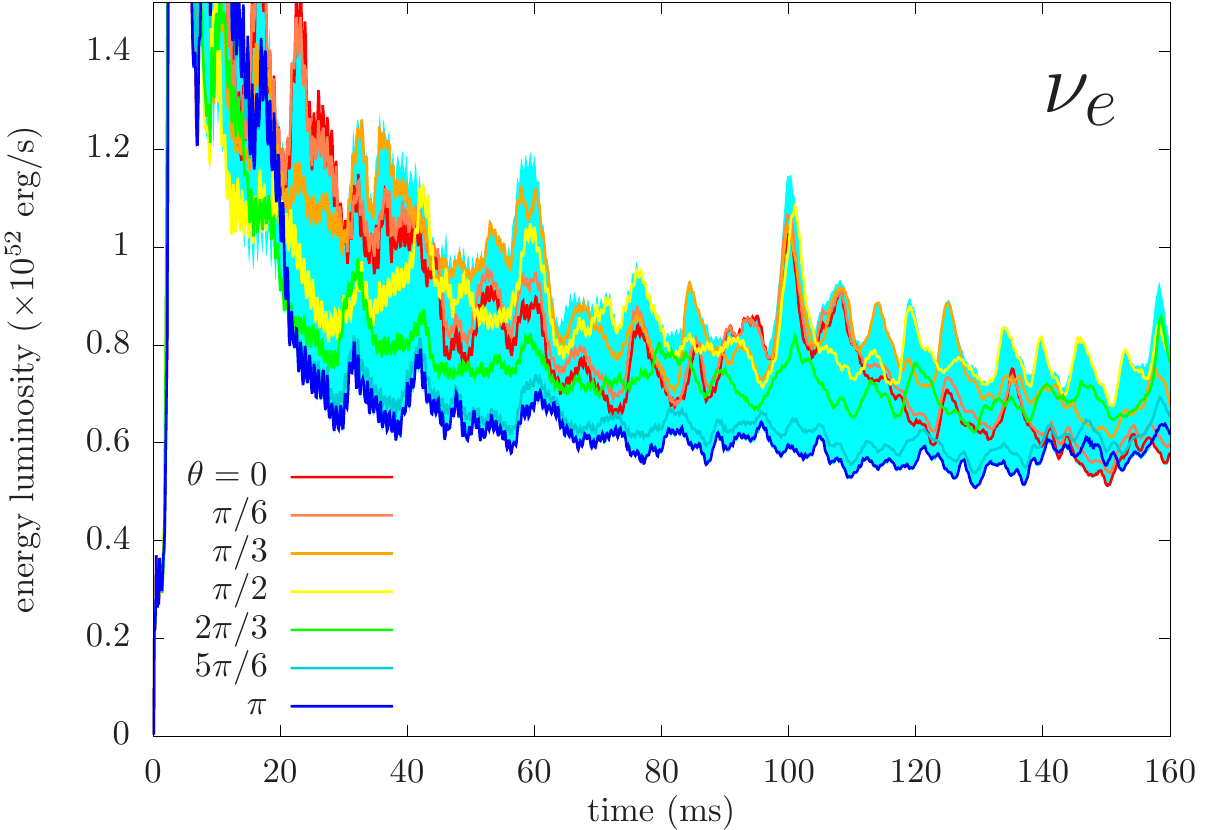}
      \end{minipage} \\
      \begin{minipage}[t]{1.0\hsize}
        \centering
        \includegraphics[scale=0.7]{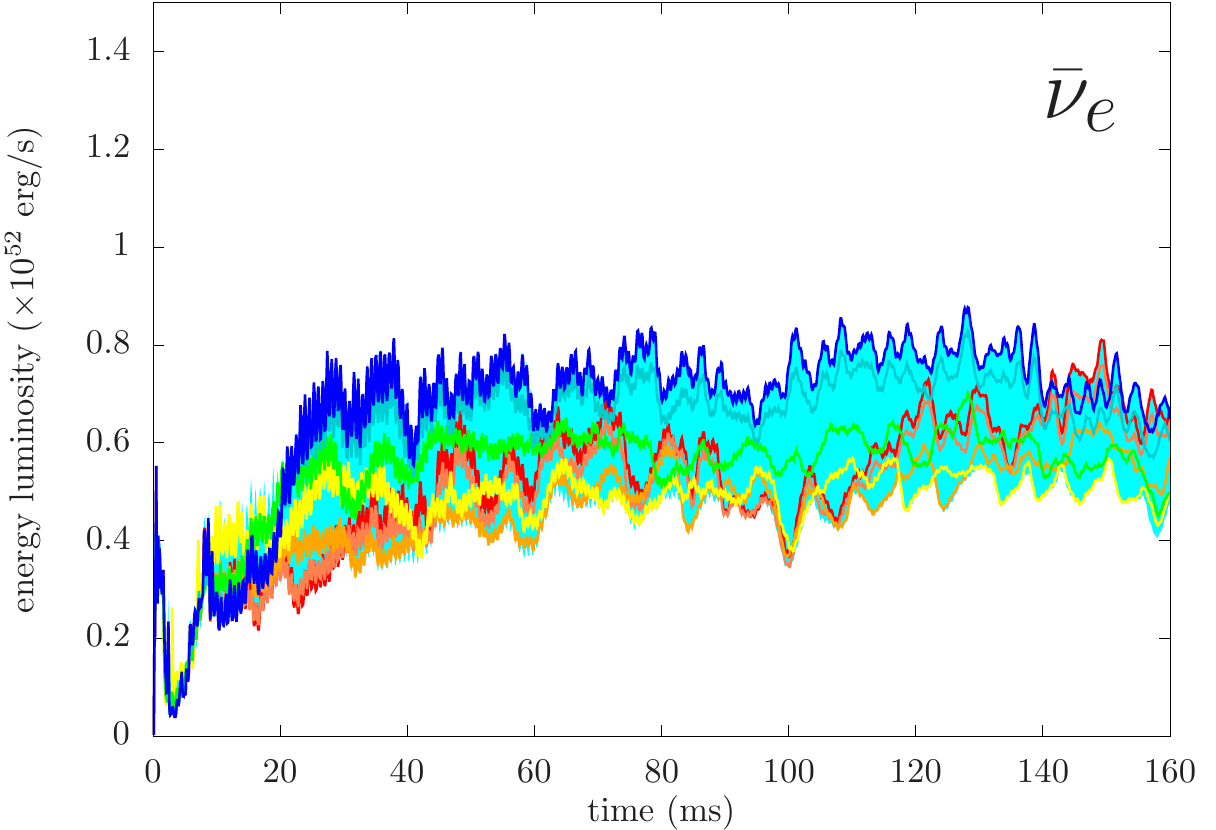}
      \end{minipage} \\
        \begin{minipage}[t]{1.0\hsize}
        \centering
        \includegraphics[scale=0.7]{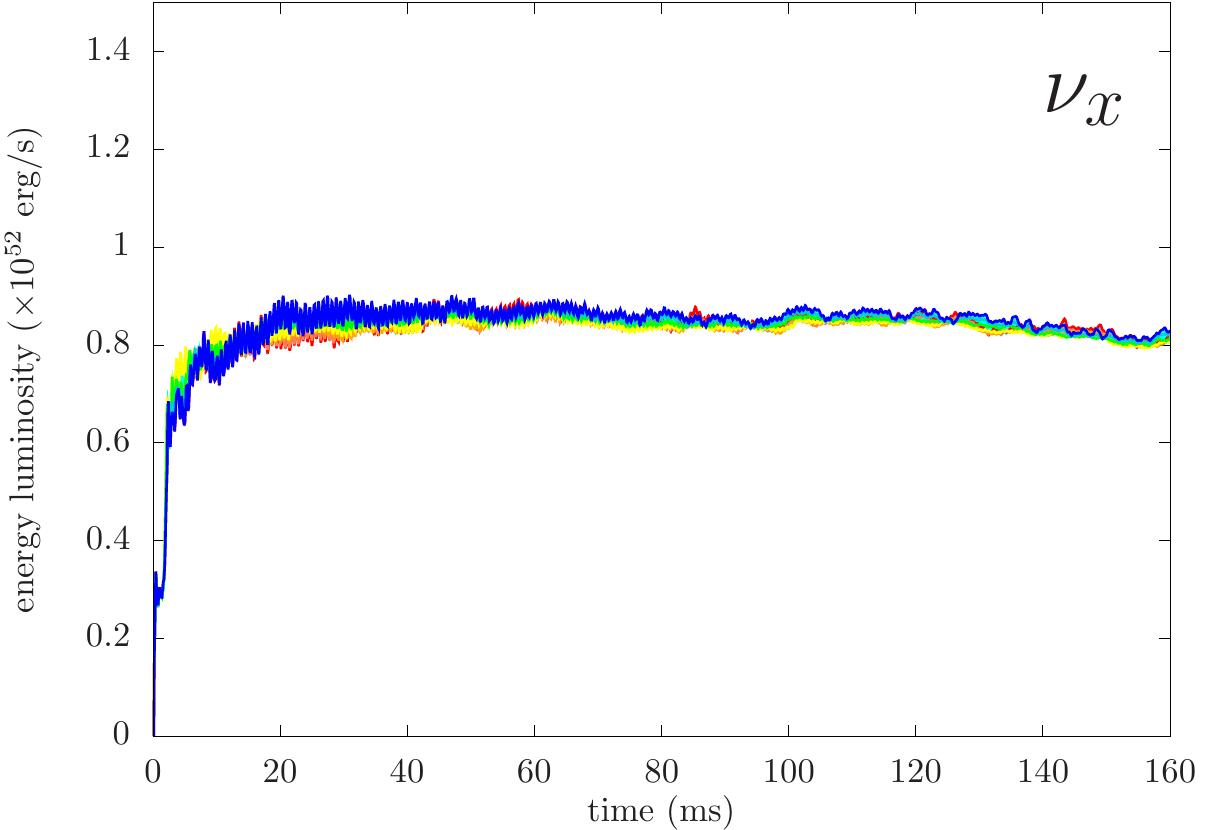}
      \end{minipage}
    \end{tabular}
    \caption{Comparison of the energy luminosity for different directions. Three panels correspond to different flavors; $\nu_e$ (top), $\bar\nu_e$ (middle), and $\nu_x$ (bottom). The red, orange, and the blue lines denote the luminosity for the north pole, the equator, and the south pole, respectively. The region shaded with light-blue represent the value between the maximum and the minimum luminosity.}
    \label{fig_lumi_aniso}
\end{figure}

{
Bearing this possible caveat in mind, we discuss the directional dependence of the neutrino emission caused by the asymmetric matter distribution derived above.
Figure \ref{fig_lumi_aniso} shows a comparison of the luminosities at several angles in space. Both $\nu_e$ and $\bar\nu_e$ show large angular variations, where the maximum value becomes twice as large as the minimum value. The temporal changes at different angles for $\bar\nu_e$ are inversely correlated with that of $\nu_e$. The $\nu_e$ luminosities near the south pole ($\theta=5\pi/6$, $\pi$) tends to be lower than other angles. This is because the low-$Y_e$ environment there (figures \ref{fig_thetadistri_70}, \ref{fig_thetadistri_150}) is preferable for the absorption of $\nu_e$ and emission of $\bar\nu_e$. Although $\nu_e$ luminosities near the north pole ($\theta=0$, $\pi/6$) are higher than those near the south pole at early times, they become similar later as the transient subsides and the convection becomes quasi-steady. It is important that there is $\sim30$\% of anisotropy existent even if these pole regions are excluded. The $\nu_x$ luminosity, on the other hand, shows much smaller angular variations ($\lesssim10$\%) compared to those of $\nu_e$ and $\bar\nu_e$ $\lesssim10\%$ in the late phase. This is because the $\nu_x$ emission is barely affected by the anisotropy of $Y_e$ and the temperature variations are smaller.}

\subsection{Resolution Dependence}
{Here we investigate the numerical resolution in our 2D simulation. 
As mentioned earlier in section \ref{sec_setup}, we run the additional simulations either with a higher spatial resolution (model \texttt{HR-S}) or with a higher angular resolution in momentum space (model \texttt{HR-M}). Since we are interested in the (quasi) steady convection, we start these runs from the normal-resolution result at $110\mathrm{ms}$.}

\begin{figure}[t]
    \begin{tabular}{c}
    \begin{minipage}[t]{0.8\hsize}
        \centering
        \includegraphics[scale=0.7]{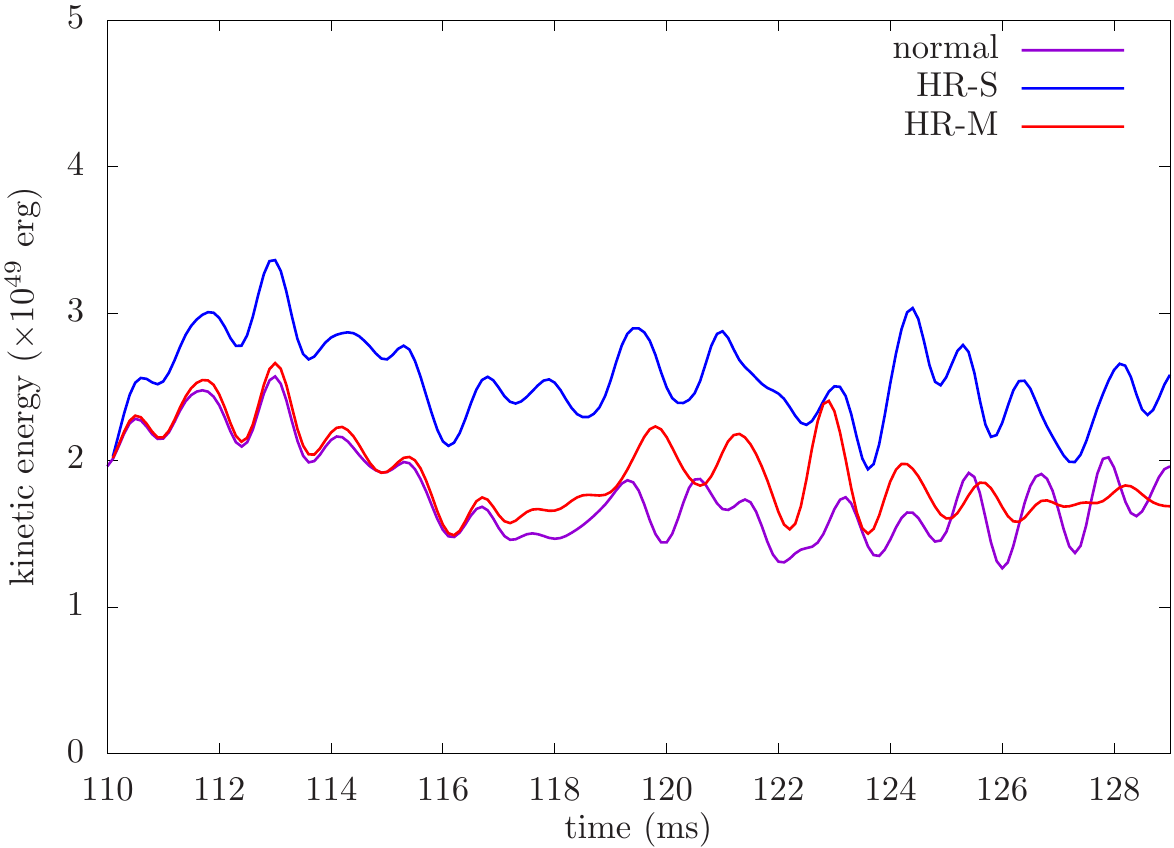}
      \end{minipage} 
    \end{tabular}
  \caption{Time evolution of the matter kinetic energy for the normal resolution (purple), \texttt{HR-S} (blue), and \texttt{HR-M} (red).}
  \label{fig_kineticene_reso}
\end{figure}

{Figure \ref{fig_kineticene_reso} shows the time evolutions of the kinetic energy for the three runs. Since the convection is stochastic, the time variations are a bit different from model to model but they show the same trend. It is apparent that model \texttt{HR-S} has larger values of kinetic energy than the other two by $\sim20$\% for most of the time. The enhancement of turbulence due to higher spatial resolution is already reported in the context of CCSN simulation \citep{Nagakura2019c}. Model \texttt{HR-M} also tends to have higher values than the normal-resolution model but the difference is much smaller and it may be a temporary.
The period of the oscillation $\sim1\,\mathrm{ms}$, is the same for different resolution models.}

\begin{figure}[htbp]
    \begin{tabular}{c}
      \begin{minipage}[t]{1.0\hsize}
        \centering
        \includegraphics[scale=0.7]{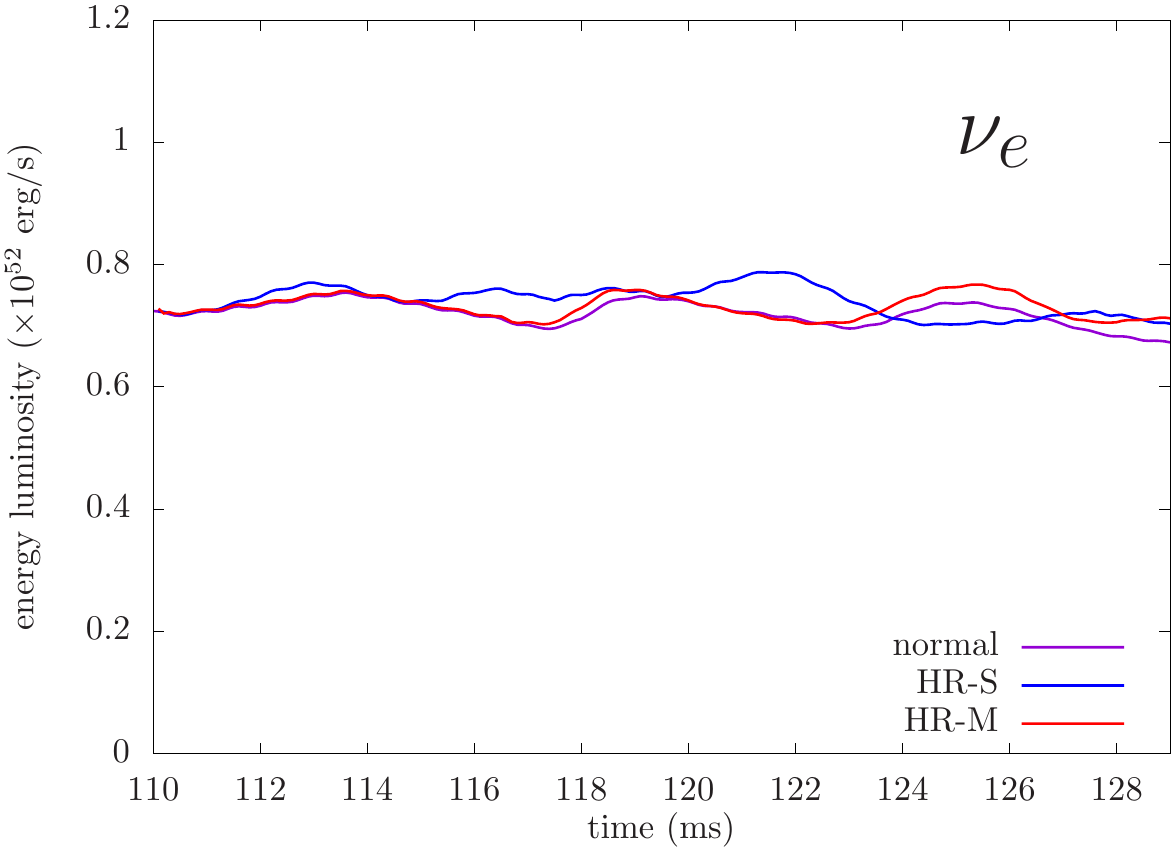}
      \end{minipage} \\
      \begin{minipage}[t]{1.0\hsize}
        \centering
        \includegraphics[scale=0.7]{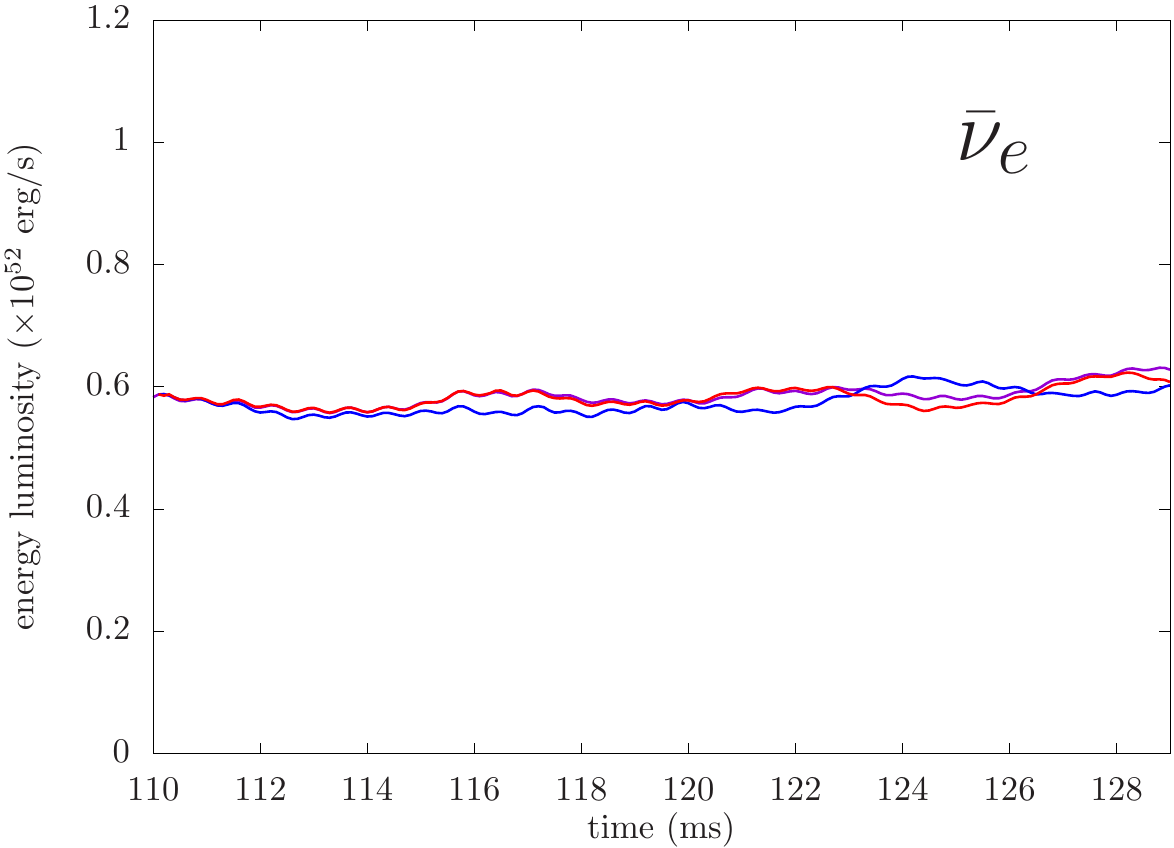}
      \end{minipage} \\
      \begin{minipage}[t]{1.0\hsize}
        \centering
        \includegraphics[scale=0.7]{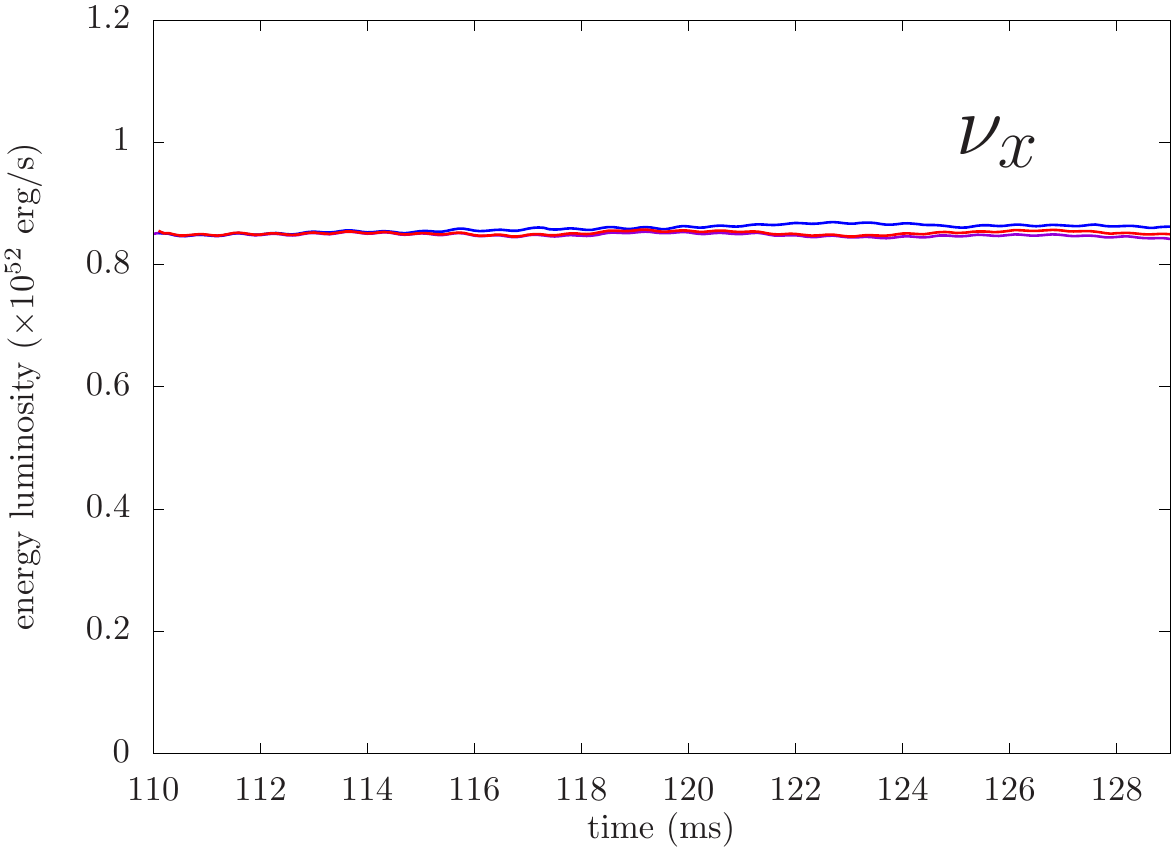}
      \end{minipage} 
    \end{tabular}
    \caption{Time evolution of the energy luminosity for three different flavors; $\nu_e$ (top), $\bar\nu_e$ (middle), and $\nu_x$ (bottom). Different colors denote the different resolution; normal resolution (purple), \texttt{HR-S} (blue), and \texttt{HR-M} (red).}
    \label{fig_enelumi_reso}
\end{figure}

\begin{figure}[t]
    \begin{tabular}{c}
    \begin{minipage}[t]{0.8\hsize}
        \centering
        \includegraphics[scale=0.7]{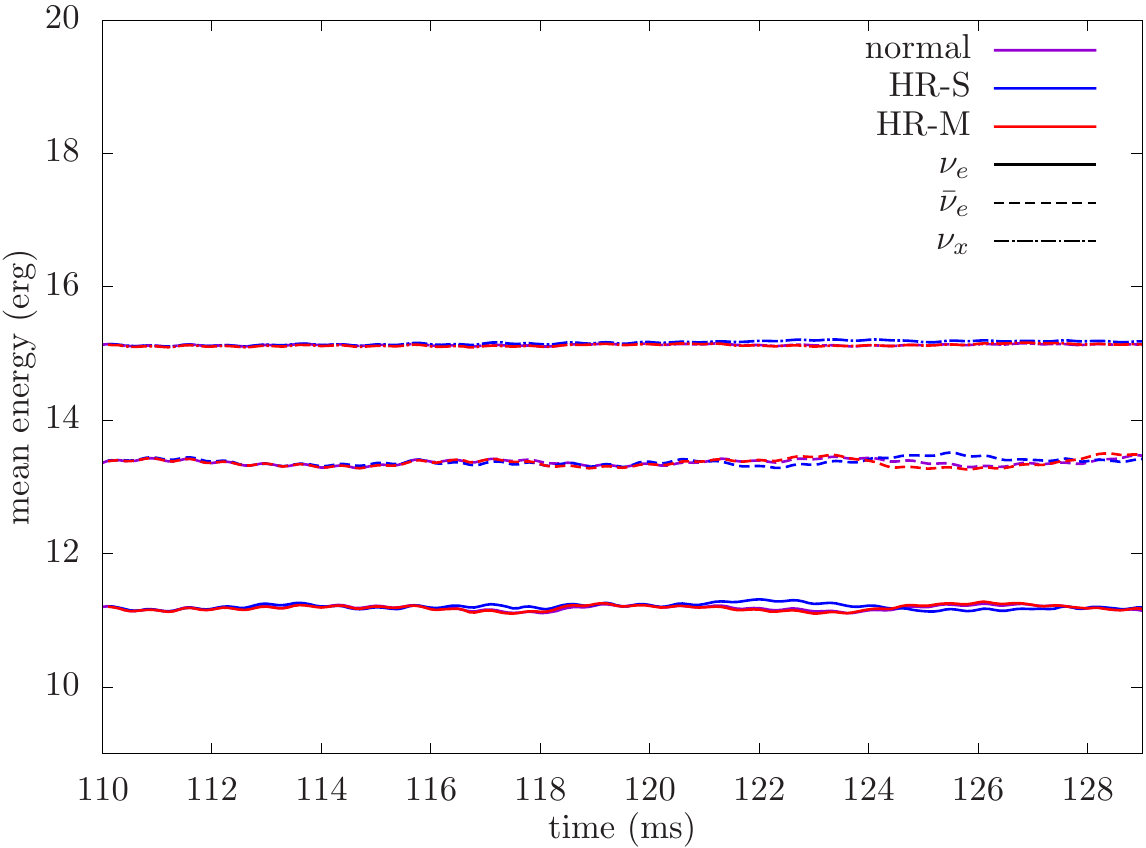}
      \end{minipage} 
    \end{tabular}
  \caption{Time evolution of the mean neutrino energy for the normal resolution (purple), \texttt{HR-S} (blue), and \texttt{HR-M} (red).}
  \label{fig_aveneutrinoene_reso}
\end{figure}

{Figure \ref{fig_enelumi_reso} shows the resolution dependence of the neutrino luminosity. The $\nu_e$ luminosity for model \texttt{HR-S} is slightly higher than that for the normal resolution whereas the $\bar\nu_e$ luminosity shows opposite behavior. The $\nu_x$ luminosity gets greater for the higher spatial resolution but the difference is even smaller. The luminosities in model \texttt{HR-M} are almost the same as those in the normal-resolution model.  
Figure \ref{fig_aveneutrinoene_reso} shows the comparison of the (angle-averaged) mean neutrino energy among the three runs. For all flavors, the resolution dependence is very minor.
We can say that although the turbulence induced by the convection is under-resolved spatially by a few tens percent in our 2D simulation, the enhancement in the neutrino luminosities and mean energies is much less affected by the resolution.}

\subsection{Occurrence of the Neutrino Fast Flavor Conversion}
Since we directly solve the Boltzmann equation, we have an advantage that the full information on the neutrino distribution in phase space can be obtained. This allows us to analyze, albeit linearly, the occurrence of the neutrino fast flavor conversion (FFC). In fact, our group has already reported the possibility of FFC \citep{Nagakura2019b,Azari2020,Harada2022} for the results of their CCSN simulations performed with the Boltzmann radiation hydrodynamics code. {We extend them to to a later time in the PNS cooling phase here.}

\begin{figure*}[t]
 \begin{center}
  \includegraphics[width=17cm]{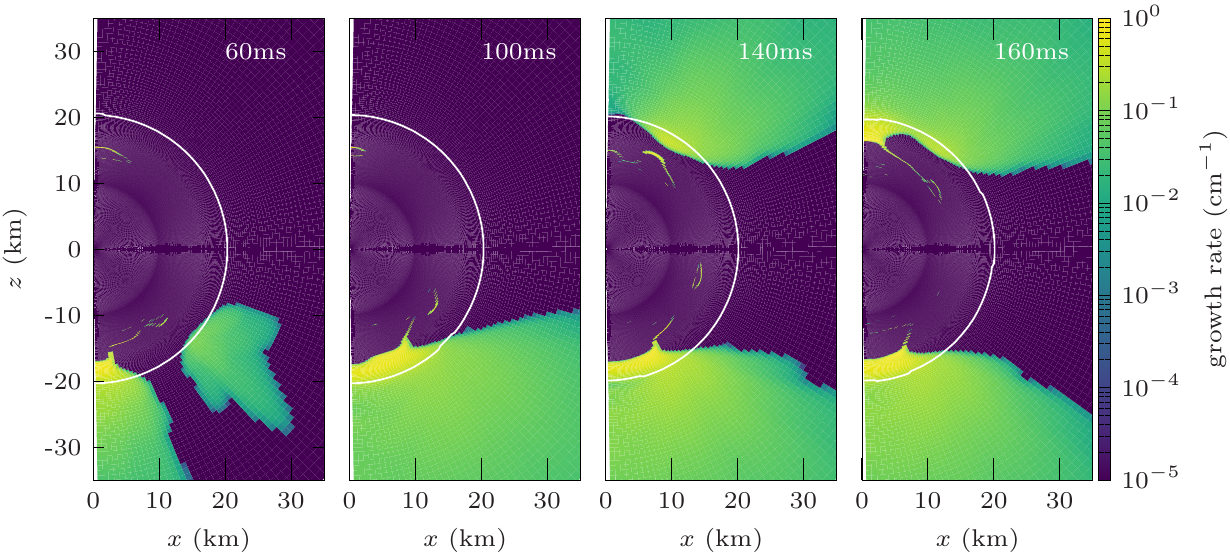}
  \caption{Growth rate of the fast flavor conversion for some time snapshots; from left to right, 60ms, 100ms, 140ms and 160ms. The white lines denote the PNS surface for each time.}
  \label{fig_fastconvgrowth}
 \end{center}
\end{figure*}

The occurrence of FFC can be judged by the existence of the ELN(electron-lepton number) crossing, as proven by \citet{Morinaga2021}.
We estimate the growth rate of FFC using the following approximate formula, motivated by the two-beam model \citep{Morinaga2020}:
\begin{equation}
\label{eq_growth}
\sigma\sim\sqrt{-\left(\int_{\Delta G>0}\frac{d\Omega}{4\pi}\Delta G\right)
\left(\int_{\Delta G<0}\frac{d\Omega}{4\pi}\Delta G\right)},
\end{equation}
where
\begin{equation}
\Delta G = \frac{\sqrt{2}G_F}{2\pi^2}\int\left(f_{\nu_e}-f_{\bar\nu_e}\right)\nu^2d\nu.
\end{equation}
We show the results in Figure \ref{fig_fastconvgrowth} at four different times. In the greenish regions $\sigma$ is real and FFC is expected to occur, whereas in the dark regions $\sigma$ is actually imaginary and FFC is not expected. At early times ($\lesssim 100\,\mathrm{ms}$), the FFC regions are confined in the  southern hemisphere. At later times, when the initial transient is subsided and the convection becomes quasi-stationary, it appears also in the northern hemisphere and is extended horizontally. The growth rates are as high as $\gtrsim10^{-1}\,\mathrm{cm}^{-1}$.

\begin{figure}[htbp]
    \begin{tabular}{c}
      \begin{minipage}[t]{1.0\hsize}
        \centering
        \includegraphics[scale=0.7]{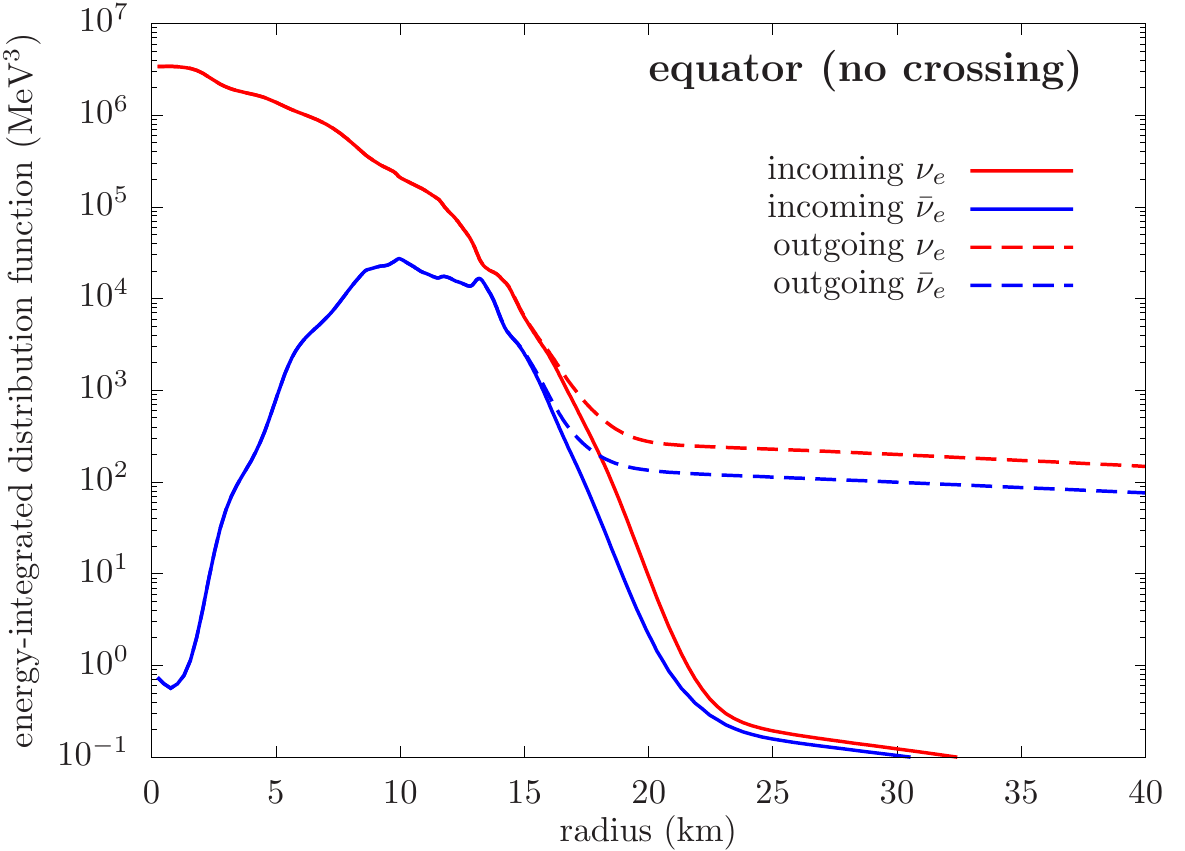}
      \end{minipage} \\
      \begin{minipage}[t]{1.0\hsize}
        \centering
        \includegraphics[scale=0.7]{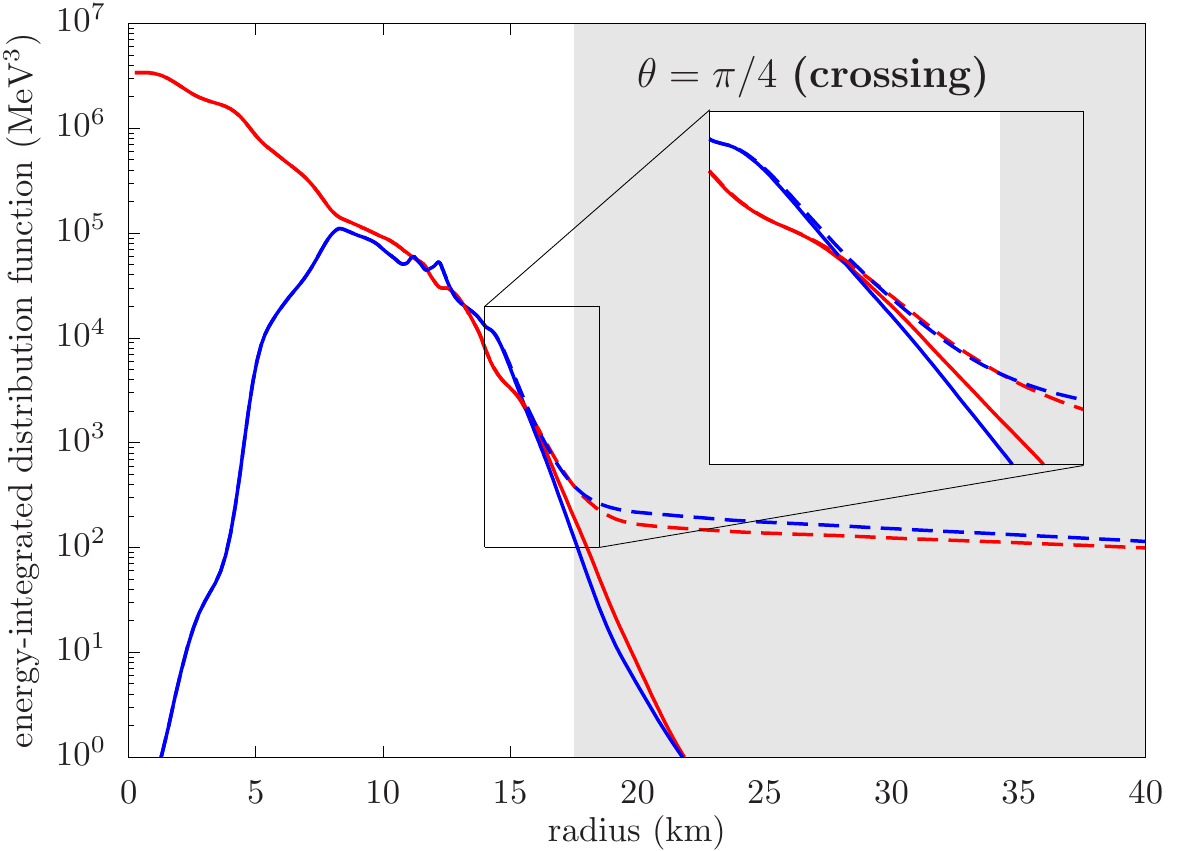}
      \end{minipage}
    \end{tabular}
    \caption{Radial profile of the energy-integrated distribution function on the equator (top) and the angular slice $\theta=\pi/4$ (bottom) for the time $t=160\,\mathrm{ms}$. The red, and blue lines denote the $\nu_e$ and $\bar\nu_e$ distribution, respectively. The solid and the dashed lines denote incoming and outgoing ones, respectively. The gray-shaded region denote where the crossing occurs.}
    \label{fig_G_in_out}
\end{figure}

Figure \ref{fig_G_in_out} shows a comparison of the energy-integrated distribution functions for the inward- and outward-going $\nu_e$ and $\bar{\nu}_e$ between the equator and the intermediate latitude ($\theta=\pi/4$). for the snapshot at $t=160\,\mathrm{ms}$. On the equator, where no crossing is observed, the abundance of $\bar{\nu}_e$ is clearly smaller than that of $\nu_e$ both for the outgoing and incoming directions, which makes it impossible to realize a crossing. At $\theta = \pi/4$, on the other hand, the abundance of $\bar\nu_e$ is comparable or sometimes larger than that of $\nu_e$. As discussed in section \ref{sec_asymmetric}, this is caused by the low-$Y_e$ environment produced by the downdraft in the convective motion.
At $r\gtrsim18.5\,\mathrm{km}$, $\bar\nu_e$ dominates over $\nu_e$ for the outward direction, whereas the opposite holds for the inward direction (the so-called type-II crossing in the language of \citet{Nagakura2021d}). This reflects the well-known difference in their interactions with matter between the two; $\bar\nu_e$ is decoupled from matter deeper inside compared to $\nu_e$, which produces the region where $\bar\nu_e$ is more forward-peaked than $\nu_e$ is (see the magnified figure).

{The downdrafts observed near the poles and the resultant low-$Y_e$ regions are likely to be exaggerated by the 2D artifact, such structures are probably generic as we argued earlier. We hence think that the FFC is likely to occur also in the 3D PNS cooling via the mechanism mentioned above. 
It should be also added that the ELN crossing was also observed in the post-shock region by \citet{Harada2022} in their 2D simulations of a rapidly rotating CCSN although the time and the location are different from ours and the centrifugal forces play an important role in the prevalence of the low-$Y_e$ region.} 

\section{Summary and Conclusions}
\label{sec_summary}

We investigated the PNS convection in 2D under axisymmetry, using our newly developed general relativistic Boltzmann neutrino radiation-hydrodynamics code. 
{This is meant to be a pilot study for more comprehensive explorations of the PNS cooling in multi-dimensions. To our knowledge, it has not been conducted yet in the literature for a couple of reasons. In fact, most of the previous works on the PNS were done either in 1D \citep{Roberts2012a,Roberts2012b,Nakazato2019,Nakazato2020,Li2021,Nakazato2022,Pascal2022} or in the early phase ($\lesssim1\,\mathrm{s}$) of core-collapse supernova explosion \citep{Mezzacappa1998,Dessart2006,Buras2006b,Nagakura2020} (but see \citet{Bollig2021}). For the purpose of this paper, we extracted from a conventional 1D PNS-cooling calculation in spherical symmetry by \citet{Nakazato2019} a snapshot of PNS at $2.3\,\mathrm{s}$ post bounce, mapped it onto a 2D grid and ran our code, adding some perturbations initially, to see the subsequent convective activity.}

{The Brunt-V\"ais\"al\"a frequency calculated at each grid point shows that this model has indeed a radially extended zone inside the PNS, that is linearly unstable against convection.}
{We observed that the PNS convection is actually instigated in that region. The convective motion is particularly violent in the first $\sim100\,\mathrm{ms}$, extending itself inward by overshooting and rendering the entropy gradient positive there. We also saw a rapid rise of the neutrino luminosities and mean energies.
These are all transients, though, which are induced by the switch from 1D to 2D and the subsequent growth of the convective motions. They subside gradually in $\sim100\,\mathrm{ms}$. Then the convection enters a quasi-steady phase sustained up to the end of the simulation at $\sim160\,\mathrm{ms}$ by the negative radial gradient of $Ye$, which remains thanks to the lasting neutrino emissions from the PNS surface. The PNS, on the other hand, is settled to a new expanded configuration, emitting neutrinos at higher luminosities and mean energies. The density, temperature and $Y_e$ as well as the neutrino luminosities and spectra change much more slowly on the secular time scale thereafter. The sustained convection with the extension of the convective zone inward is consistent with \citet{Keil1996} but is at odds with \citet{Mezzacappa1998}. The latter is due probably to their approximation in the neutrino transport as well as to the difference in the phase focused. The higher neutrino luminosities and mean energies in 2D than in 1D were also observed by other earlier works done for the core-collapse supernova simulations \citep{Dessart2006}.}

{Supposing that the self-sustained state above is representative of the PNS around this time in its cooling, we investigated it further. Note that the difference between the original matter profile obtained in the 1D PNS cooling calculation, and the self-sustained state in our 2D simulation indicates clearly that the multi-dimensional PNS calculation should be done from much earlier on, possibly from right after a successful launch of the shock wave from the supernova core.}

\deleted{
The obtained results are in reasonable agreement with previous studies by different groups. Since this is our first result by our newly developed code, this is a important demonstration to confirm the credibility of our code.}

We analyzed angular variations in the matter distributions that are produced by the convective motions.
We found that the temperature and $Y_e$ showed larger deviations from spherical symmetry than the density. In particular, $Y_e$ tends to be lower near the poles and higher around the equator. This asymmetry in the $Y_e$ distribution in turn gives rise to anisotropic emissions of $\nu_e$ and $\bar\nu_e$. {The time-averaged convective pattern revealed that this $Y_e$ distribution is generated by the subduction of low-$Y_e$ matter and the dredge-up of high-$Y_e$ matter by the convective motion. In our 2D simulation with axisymmetry imposed, the former occurs predominantly in the polar region and is most likely to be exaggerated. On the other hand, the downdraft of low-$Y_e$ matter and the updraft of high-$Y_e$ matter were also observed in other's simulations \citep{Keil1996} that avoided the polar artifact. Hence we think it is a rather generic feature and expect that a similar asymmetry will occur also in 3D.}

{This paper is also meant for the check of our newly developed code. In fact, before starting the 2D run, we ran our code in the 1D mode to see a possible inconsistency between the 1D PNS-cooling code \citep{Nakazato2019} and our code. It turns out that the resultant matter profiles were not much different from the original ones, showing that our code is working as expected. We also investigated the resolution dependence of the findings on the convection and its consequences. We increased either the spatial resolution or the angular resolution in momentum space. We found that the latter have little effect on the results. On the other hand, roughly doubling the number of grid points in space increased the turbulence kinetic energy by $\sim20$\%.}

Finally, taking advantage of the Boltzmann solver, which can provide the full information on the angular distribution of neutrino in momentum space, we analyzed the possible occurrence of FFC in our model by searching for the ELN crossing. We found that FFC is unlikely to occur in the high-$Y_e$ region, i.e., around the equator in our model, where $\nu_e$ dominates $\bar\nu_e$ in all directions in momentum space. In the low-$Y_e$ region, on the other hand, we found the type-II ELN crossing, since the $\bar\nu_e$ distribution become more forward-peaked than the $\nu_e$ distribution because $\bar\nu_e$ decouples from matter deeper inside.

As repeatedly mentioned, convection in 2D under axisymmetry is different from that in 3D qualitatively \citep{Lentz2015} and the  converging flows in the polar regions are most likely due to the 2D artifact. The results on the FFC are hence also affected by them. As argued above, however, the convective pattern with low-$Y_e$ matter sinking and high-$Y_e$ matter rising will be rather generic and likely to occur also in 3D. Then the possibility of FFC in the low-$Y_e$ region may also survive in 3D. It is encouraging that the FFC region is much wider than the converging flows. In the rapidly rotating PNS, the converging flows may be realized indeed.
FFC is important as it will change the neutrino signals observed at the terrestrial detectors and may also affect the PNS cooling. With the Boltzmann solver, we will be able to study these issues in detail.

As we mentioned at the beginning, we are planning to conduct a comprehensive study of the long-term PNS cooling in multi-dimensions. We are particularly interested in the rapidly rotating PNS, which may be also produced by a neutron star merger. For that purpose we have developed a code to build a general relativistic rotational equilibrium configuration on the Lagrangian coordinates \citep{Okawa2022}. We will run the Boltzmann-$\nu$-radiation-hydrodynamics code on top of those configurations and, based on these models, the convections, neutrino emissions and oscillations as well as the gravitational waves from neutrinos \citep{Fu2022} will be investigated quantitatively. The results will be reported elsewhere in the near future.



\begin{acknowledgments}
We thank Ken'ichi Sugiura, Ken'ichiro Nakazato and Hideyuki Suzuki for providing the PNS model. We thank Kyohei Kawaguchi for helpful advice on the hydrodynamics solver. This research used the K and Fugaku supercomputers provided by RIKEN, the FX10 provided by Tokyo University, the FX100 provided by Nagoya University, the Grand Chariot provided by Hokkaido University, and Oakforest-PACS provided by JCAHPC through the HPCI System Research Project (Project ID: hp130025, 140211, 150225, 150262, 160071, 160211, 170031, 170230, 170304, 180111, 180179, 180239, 190100, 190160, 200102, 200124, 210050, 210051, 210164, 220047, 220173), and the Cray XC50 at Center for Computational Astrophysics, National Astronomical Observatory of Japan (NAOJ). This work is supported by 
Grant-in-Aid for Scientific Research
(19K03837, 20H01905)
and 
Grant-in-Aid for Scientific Research on Innovative areas 
"Gravitational wave physics and astronomy:Genesis" 
(17H06357, 17H06365) and ”Unraveling the History of the Universe and Matter Evolution with Underground Physics” (19H05802 and 19H05811)
from the Ministry of Education, Culture, Sports, Science and Technology (MEXT), Japan.
R. A. is supported by JSPS Grant-in-Aid for JSPS Fellows (Grant No. 22J10298) from MEXT. 
S. Y. is supported by the Institute for Advanced
Theoretical and Experimental Physics, and Waseda University and the Waseda University Grant for Special Research Projects (project numbesr: 2021C-197, 2022C-140).
For providing high performance computing resources, Computing Research Center, KEK, JLDG on SINET of NII, Research Center for Nuclear Physics, Osaka University, Yukawa Institute of Theoretical Physics, Kyoto University, Nagoya University, and Information Technology Center, University of Tokyo are acknowledged. This work was supported by 
MEXT as "Program for Promoting Researches on the Supercomputer Fugaku" 
(Toward a unified view of the universe: from large scale structures to planets, JPMXP1020200109) and the Particle, Nuclear and Astro Physics Simulation Program (Nos. 2020-004, 2021-004, 2022-003) of Institute of Particle and Nuclear Studies, High Energy Accelerator Research Organization (KEK).
\end{acknowledgments}

%

\vspace{5mm}





\appendix
\restartappendixnumbering
\section{Formulation of the Hydrodynamics Solver}
\label{sec_hydroformulation}
In this appendix, we describe the formulation of the hydrodynamics equations solver. It is a general relativistic extension of the Newtonian counterpart employed in our previous papers \citep{Nagakura2014, Nagakura2017, Nagakura2019a}. By following \cite{Kawaguchi2021}, we first decompose the variables as follows:
\begin{eqnarray}
\Lambda^r_{(r)} \equiv 1,\;
\Lambda^\theta_{(\theta)} \equiv \frac{1}{r},\;
\Lambda^\phi_{(\phi)} &\equiv& \frac{1}{r\sin\theta},\;
\Lambda^j_{(i)} \equiv0\,(i\neq j), \\
\Lambda_r^{(r)} \equiv 1,\;
\Lambda_\theta^{(\theta)} \equiv r,\;
\Lambda_\phi^{(\phi)} &\equiv& r\sin\theta,\;
\Lambda_j^{(i)} \equiv 0\,(i\neq j), \\
\tilde{\gamma}_{(i)(j)} &\equiv& \Lambda^k_{(i)} \Lambda^\ell_{(j)} \gamma_{k\ell}, \\
\tilde{K}_{(i)(j)} &\equiv& \Lambda^k_{(i)} \Lambda^\ell_{(j)} K_{k\ell}, \\
\sqrt{\tilde{\gamma}} &\equiv& \frac{1}{r^2\sin\theta}\sqrt{\gamma}, \\
\tilde{\rho_*} &\equiv& \frac{1}{r^2\sin\theta}\rho_*, \\
v^{(i)} &\equiv& \Lambda^{(i)}_j v^j, \\
\beta^{(i)} &\equiv& \Lambda^{(i)}_j \beta^j, \\
\tilde{S}_{(i)} &\equiv& \frac{1}{r^2 \sin\theta} \Lambda^j_{(i)}S_j, \\
\tilde{S}_{(i)(j)} &\equiv& \Lambda^k_{(i)} \Lambda^\ell_{(j)} S_{k\ell}.
\end{eqnarray}
Note that if the spacetime is flat, the parenthesis indices ($(r)$, $(\theta)$, and $(\phi)$) means the orthonormal components of the flat spacetime. Therefore, similarly to the philosophy of \cite{Baumgarte2013}, we factor out the trivial coordinate dependence from the coordinate components of the tensors. This decomposition gives better accuracy of the interface value reconstruction.

With the variables defined above, we discretized the equations (\ref{eq_massconserv}--\ref{eq_energyconserv}) into the following form:
\begin{eqnarray}
\partial_t(\rho_{*i}) &=& - \frac{r_i^2}{r_I^3/3 - r_{I-1}^3/3} \Delta_r (r_I^2 \sin\theta_j \tilde{\rho_*}_I v^{(r)}_I) \\
&+&\frac{\sin \theta_j}{\mu_J - \mu_{J-1}} \Delta_\theta (r_i \sin\theta_J \tilde{\rho_*}_J v^{(\theta)}_J) \\
&-& \frac{1}{\phi_K - \phi_{K-1}}\Delta_\phi (r_i \tilde{\rho_*}_K v^{(\phi)}_K) = 0,
\end{eqnarray}
\begin{eqnarray}
\partial_t(S_{r,i}) &=& - \frac{r_i^2}{r_I^3/3 - r_{I-1}^3/3} \Delta_r (r_I^2 \sin\theta_j (\tilde{S}_{(r),I}v^{(r)}_I + P_I \alpha_I \sqrt{\tilde{\gamma}}_I)) \\
&+&\frac{\sin \theta_j}{\mu_J - \mu_{J-1}} \Delta_\theta (r_i\sin\theta_J \tilde{S}_{(r),J}v^{(\theta)}_J) \\
&-& \frac{1}{\phi_K - \phi_{K-1}}\Delta_\phi (r_i \tilde{S}_{(r),K}v^{(\phi)}_K) \\
&-& S_{0,i} (\partial_r \alpha)_i + \Lambda^{\ell}_{(m)}S_{\ell,i} (\partial_r \beta^{(m)})_i -\frac{1}{2}\alpha_i \sqrt{\gamma}_i \Lambda^{j}_{(\ell)} \Lambda^{k}_{(m)}S_{jk,i}(\partial_r \tilde{\gamma}^{(\ell)(m)})_i \\
&+&\frac{r_I^2 - r_{I-1}^2}{r_I^3/3 - r_{I-1}^3/3} ( \alpha_i \sqrt{\gamma}_i P_i + \frac{1}{2}( S_{\theta,i}v^{\theta}_i + S_{\phi,i}v^{\phi}_i)) \\
&-& \alpha_i \sqrt{\gamma}_i G_{r,i},
\end{eqnarray}
\begin{eqnarray}
\partial_t(S_{\theta,i}) &=& - \frac{r_i^2}{r_I^3/3 - r_{I-1}^3/3} \Delta_r (r_I^3 \sin\theta_j \tilde{S}_{(\theta),I}v^{(r)}_I) \\
&+&\frac{\sin \theta_j}{\mu_J - \mu_{J-1}} \Delta_\theta (r_i^2\sin\theta_J (\tilde{S}_{(\theta),J}v^{(\theta)}_J + P_J \alpha_J \sqrt{\tilde{\gamma}}_J)) \\
&-& \frac{1}{\phi_K - \phi_{K-1}}\Delta_\phi (r_i^2 \tilde{S}_{(\theta),K}v^{(\phi)}_K) \\
&-& S_{0,j} (\partial_\theta \alpha)_j + \Lambda^\ell_{(m)}S_{\ell,j} (\partial_\theta \beta^{(m)})_j -\frac{1}{2}\alpha_j \sqrt{\gamma}_j \Lambda^{j}_{(\ell)} \Lambda^{k}_{(m)}S_{jk,j}(\partial_\theta \tilde{\gamma}^{(\ell)(m)})_j \\
&+& \frac{\sin\theta_{J-1} - \sin\theta_{J}}{\mu_J - \mu_{J-1}} (\alpha_j \sqrt{\gamma}_j P_j + S_{\phi,j}v^{\phi}_j) \\
&-& \alpha_j \sqrt{\gamma}_j G_{\theta,j},
\end{eqnarray}
\begin{eqnarray}
\partial_t(S_{\phi,i}) &=& - \frac{r_i^2}{r_I^3/3 - r_{I-1}^3/3} \Delta_r (r_I^3 \sin^2\theta_j \tilde{S}_{(\phi),I}v^{(r)}_I) \\
&+&\frac{\sin \theta_j}{\mu_J - \mu_{J-1}} \Delta_\theta (r_i^2\sin^2\theta_J \tilde{S}_{(\phi),J}v^{(\theta)}_J) \\
&-& \frac{1}{\phi_K - \phi_{K-1}}\Delta_\phi (r_i^2\sin\theta_j (\tilde{S}_{(\phi),K}v^{(\phi)}_K + P_K \alpha_K \sqrt{\tilde{\gamma}}_K)) \\
&-&S_{0,k} (\partial_\phi \alpha)_k +\Lambda^{\ell}_{(m)}S_{\ell,k} (\partial_\phi \beta^{(m)})_k -\frac{1}{2}\alpha_k \sqrt{\gamma}_k \Lambda^{j}_{(\ell)} \Lambda^{k}_{(m)}S_{jk,k}(\partial_\phi \tilde{\gamma}^{(\ell)(m)})_k \\
&-& \alpha_k \sqrt{\gamma}_k G_{\phi,k},
\end{eqnarray}
\begin{eqnarray}
\partial_t (S_{0,i} - \rho_{*i}) &=&  - \frac{r_i^2}{r_I^3/3 - r_{I-1}^3/3} \Delta_r (r_I^2 \sin\theta_j ((\tilde{S}_{0,I} - \tilde{\rho_*})_I) v^{(r)}_I + \sqrt{\tilde{\gamma}}_IP_I (v^{(r)}_I + \beta^{(r)}_I)) \\
&+&\frac{\sin \theta_j}{\mu_J - \mu_{J-1}} \Delta_\theta (r_i \sin\theta_J ((\tilde{S}_{0,J} - \tilde{\rho_*})_J) v^{(\theta)}_J +\sqrt{\tilde{\gamma}}_JP_J (v^{(\theta)}_J+\beta^{(\theta)}_J)) \\
&-&\frac{1}{\phi_K - \phi_{K-1}}\Delta_\phi (r_i ((\tilde{S}_{0,K} - \tilde{\rho_*})_K)v^{(\phi)}_K + \sqrt{\tilde{\gamma}}_KP_K (v^{(\phi)}_K + \beta^{(\phi)}_K)) \\
&+& \alpha_i\sqrt{\gamma}_i S^{\ell m}_iK_{\ell m,i} - \gamma^{\ell m}_i S_{\ell,i}(\partial_{m}\alpha)_i \\
&+& \alpha_i \sqrt{\gamma}_i (n^\mu G_\mu)_i,
\end{eqnarray}
where $\mu\equiv\cos \theta$; lower (e.g., $i$) and upper (e.g., $I$) case subscripts denote the cell center and interface values, respectively; $i$, $j$, and $k$ indicates the grid ID numbers of the radial ($r$), zenith ($\mu$), and azimuthal ($\phi$) coordinates, respectively; the symbols $\Delta_r$, $\Delta_\theta$, and $\Delta_\phi$ means the difference between the interface values along the indicated coordinates, e.g., $\Delta_r (f_j g_I) \equiv f_j g_I - f_j g_{I-1}$ for any functions $f$ and $g$. The interface values of the decomposed hydrodynamic variables (rest mass density $\rho_0$, temperature $T$ or pressure $P$, electron fraction $Y_{\rm e}$, and the spatial components of four-velocity $u^{(i)}\equiv u^t v^{(i)}$) are evaluated by the piecewise parabolic method \citep[PPM,][]{Colella1984174} with minmod flux limiter; the interface values of the decomposed metric variables (lapse $\alpha$, shift $\beta^{(i)}$, and spatial metric $\tilde{\gamma}_{(i)(j)}$) are evaluated by the third order Lagrange interpolation; the coordinate values $r_I$, $\theta_J$, and $\phi_K$ is exactly evaluated. The numerical flux is evaluated by the Harten-Lax-van Leer (HLL) method \citep{Harten1983} and the time evolution is solved by the fourth order Runge-Kutta scheme. By evaluating the curvature terms such as $\alpha \sqrt{\gamma}P/r$ and so on in these ways, the steady state of uniform matter in flat spacetime is guaranteed.

\section{Code Verification Tests of the Hydrodynamics Solver}
\label{sec_codetests}
In this section, we performed three kinds of code tests to verify the ability of our newly developed general relativistic hydrodynamics code.
Special relativistic shock tube tests are performed in section \ref{sec_shock}. In the sections \ref{sec_TOV} and \ref{sec_BHacc}, we perform stability tests of the relativistic star, and accreting matter onto BH, respectively.

\subsection{Special Relativistic Shock Tube Test}
\label{sec_shock}
Since our previous code was for Newtonian hydrodynamics, the matter with relativistic motion could not be treated properly. Here, we perform the special relativistic shock tube tests.
As the initial state, we employ the first case in \citet{Nagakura2011}. The initial left state is $(\rho,v,p)=(10,0,13.3)$, and the right state is $(\rho,v,p)=(1,0,10^{-6})$. Since those values are for the geometrical unit, the input hydrodynamics quantities should be re-scaled for our code written in cgs unit. If the length conversion factor is defined as $L$, the conversion are like: $\rho_\mathrm{cgs}=\rho\times c^2/G\times L^2$, $p_\mathrm{cgs}=p\times c^4/G\times L^2$, $v_\mathrm{cgs}=v\times c$ and $t=L/c$.
We choose the value $L=1\times10^{6}\,\mathrm{cm}$. We employ the gamma-law EOS with the adiabatic index $\Gamma=5/3$. 
Analytical solution can be exactly calculated by \citep{Pons2000}.

Since our code employs the polar coordinates, in principle, we cannot perform this test meant for the Cartesian coordinates. Similarly to \citet{Yamada1999}, we simulate in a very thin shell where the curvature can be ignored. The initial discontinuity is placed at the radius $r=10^5\,\mathrm{km}$, and the width of the computational domain is $10\,\mathrm{km}$, which is four-orders of magnitude smaller than the distance from the origin. In order to check the resolution dependence, we test three mesh; 100, 200, and 400 grid points. The computational region is equally divided.

\begin{figure}[htbp]
    \label{fig_shock}
    \begin{tabular}{ccc}
      \begin{minipage}[t]{0.31\hsize}
        \centering
        \includegraphics[scale=0.53]{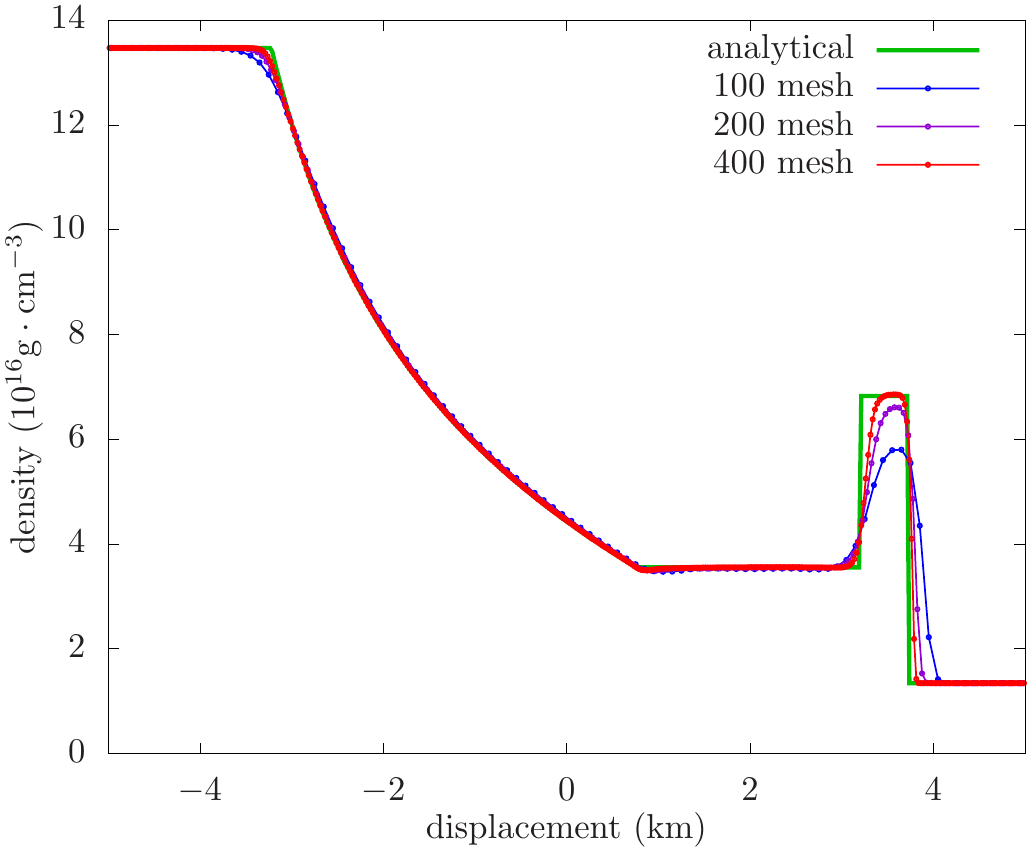}
      \end{minipage} &
      \begin{minipage}[t]{0.31\hsize}
        \centering
        \includegraphics[scale=0.53]{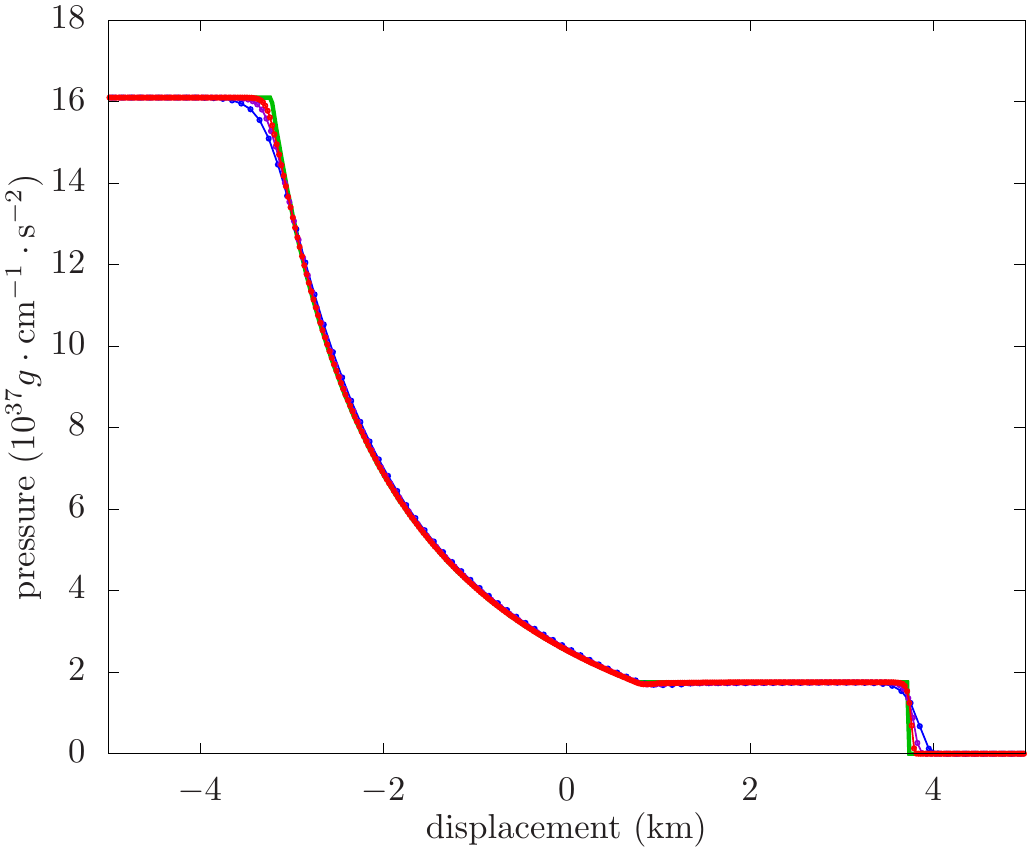}
      \end{minipage} &
         \begin{minipage}[t]{0.31\hsize}
        \centering
        \includegraphics[scale=0.53]{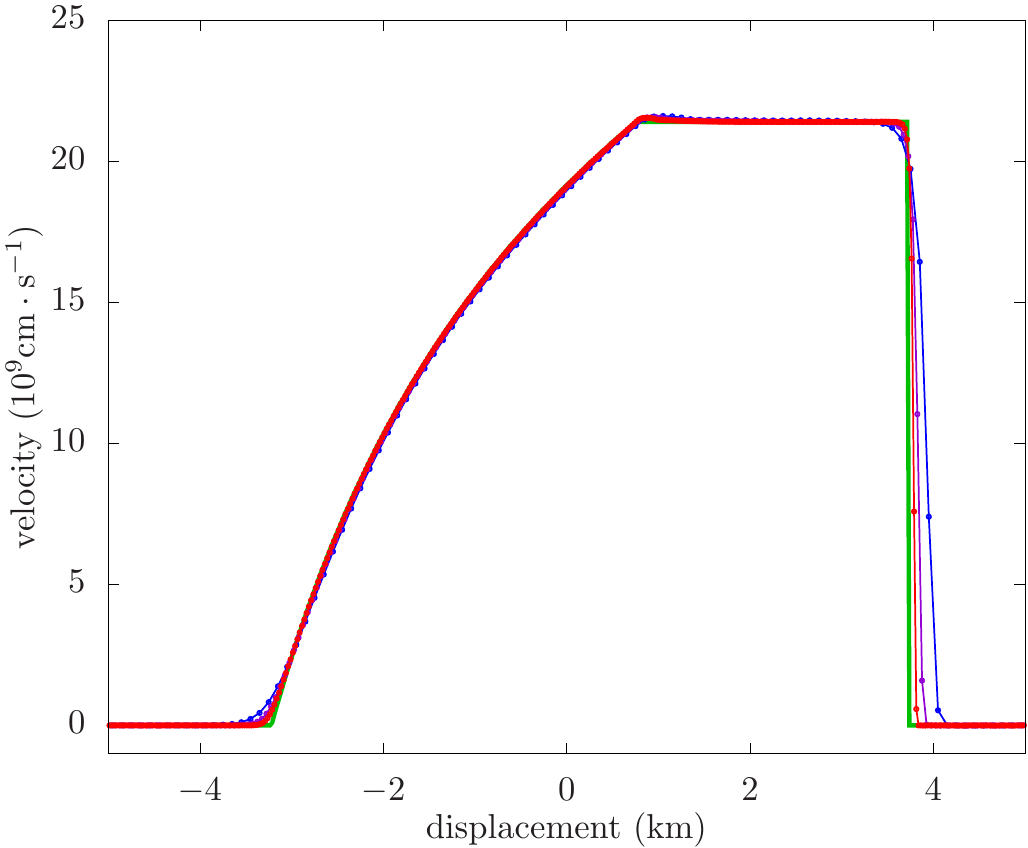}
      \end{minipage} 
    \end{tabular}
    \caption{Radial distributions of the density (left), the pressure (middle), and the velocity (right). The red, purple, and the blue lines denote the results for 400 mesh, 200 mesh, and 100 mesh, respectively.}
\end{figure}

Figure \ref{fig_shock} shows the hydrodynamic quantities at the time snapshot $t=1.5\times10^{-5}\,\mathrm{s}$. The analytical solution is well reproduced for all resolutions. In addition, raising the resolution improves the result, which indicates the resolution convergence of our code.

\subsection{Stability Test of a Relativistic Star}
\label{sec_TOV}
We check the code's ability to test matter under a strong gravitational field. By starting from the stable neutron star model as the initial condition, we check that the initial state is maintained by time evolution.

\begin{figure}[t]
    \label{fig_TOVini}
    \begin{tabular}{cc}
      \begin{minipage}[t]{0.45\hsize}
        \centering
        \includegraphics[scale=0.7]{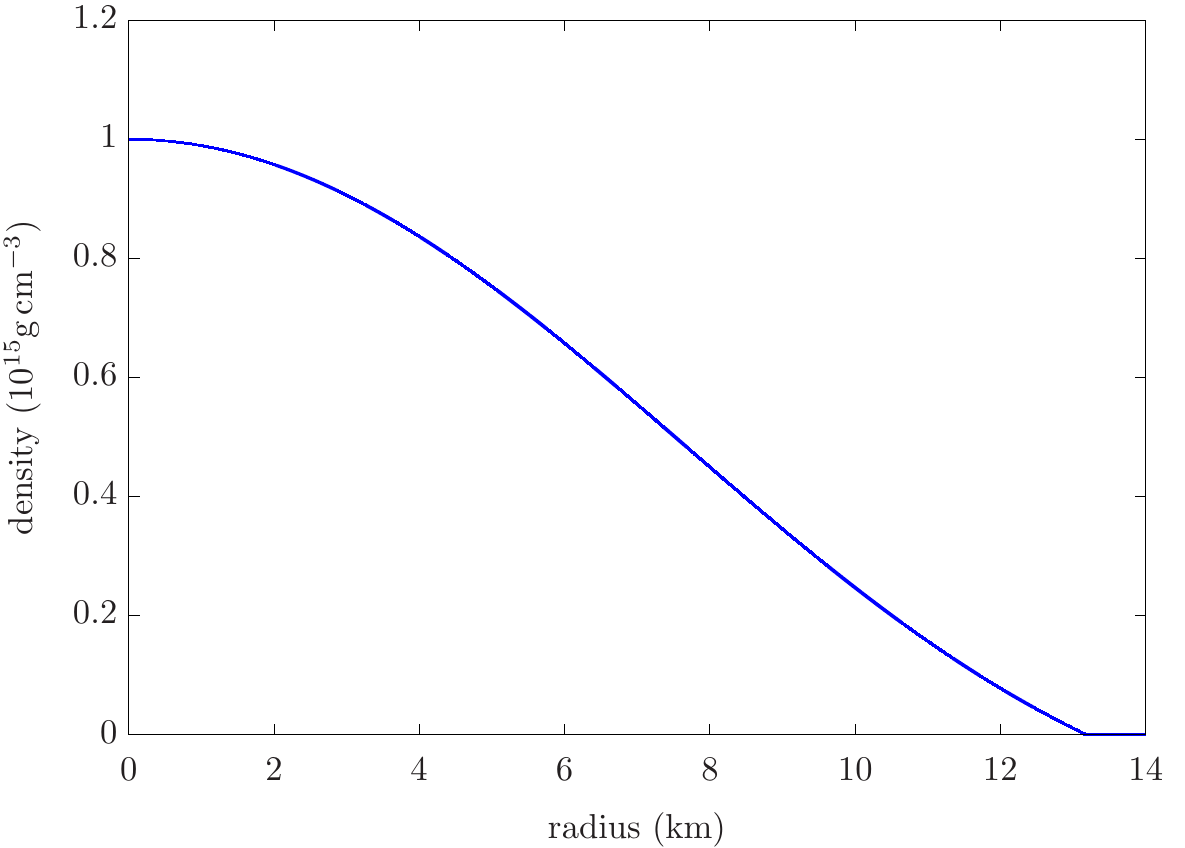}
      \end{minipage} &
      \begin{minipage}[t]{0.45\hsize}
        \centering
        \includegraphics[scale=0.7]{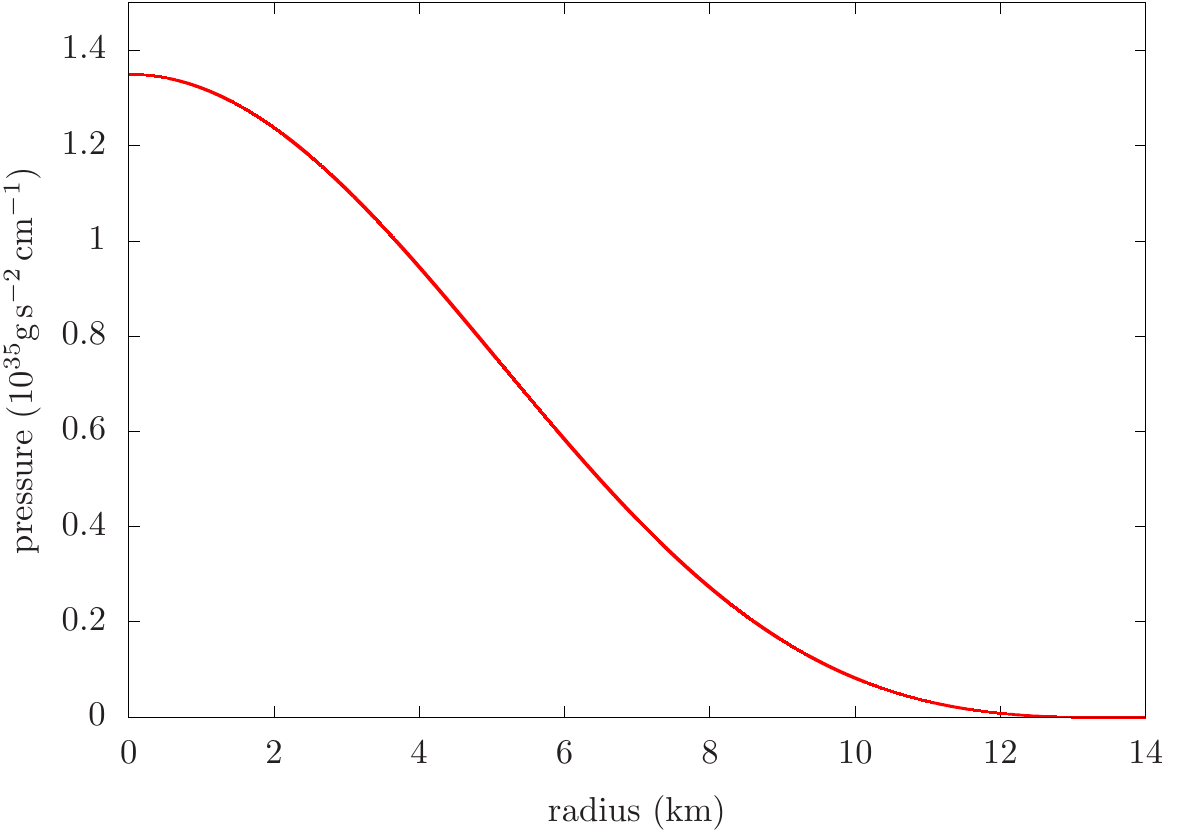}
      \end{minipage} 
    \end{tabular}
    \caption{Radial profile of density (left) and the pressure (right) for the initial model.}
\end{figure}

We constructed a neutron star model with a central density of $1\times10^{15}\,\mathrm{\, g\,cm^{-3}}$ and the central pressure of $p=1.35\times10^{35}\,\mathrm{\, g\,cm^{-1}\,s^{-2}}$. We employ the gamma-law EOS with the adiabatic index of $\Gamma=2$. We solve the TOV equations by using the fourth-order explicit Runge-Kutta method. This results in the neutron star with mass $1.42 M_\odot$ and the radius $13.2\,\mathrm{km}$. The density and the pressure distributions are shown in figure \ref{fig_TOVini}. In order to check the resolution dependence, we test three mesh cases; $N_r=128$, 256, and 384 grid points. The computational region is $r\in[0:15]\,\mathrm{km}$, and it is equally divided.

Figure B.3 shows the time evolution of the relative error of density for different radii.
The error for $N_r=128$ increases with time. 
On the other hand, the error for $N_r=256$ and $384$ shows the oscillation around the initial data.
The amplitude of the oscillation is smaller for the highest resolution case. Hence we can conclude that resolution convergence is obtained for this test.

\begin{figure}[htbp]
    \label{fig_TOVerr}
    \begin{tabular}{ccc}
      \begin{minipage}[t]{0.31\hsize}
        \centering
        \includegraphics[scale=0.49]{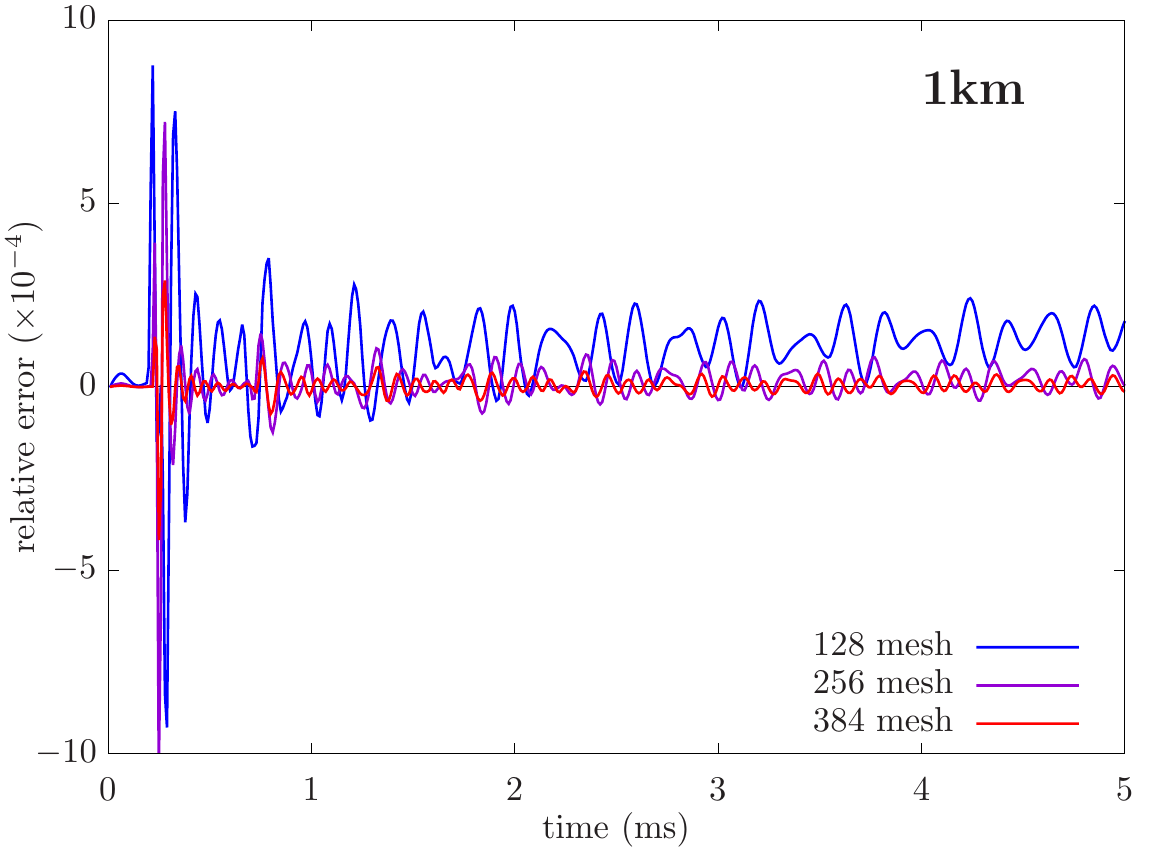}
      \end{minipage} &
      \begin{minipage}[t]{0.31\hsize}
        \centering
        \includegraphics[scale=0.49]{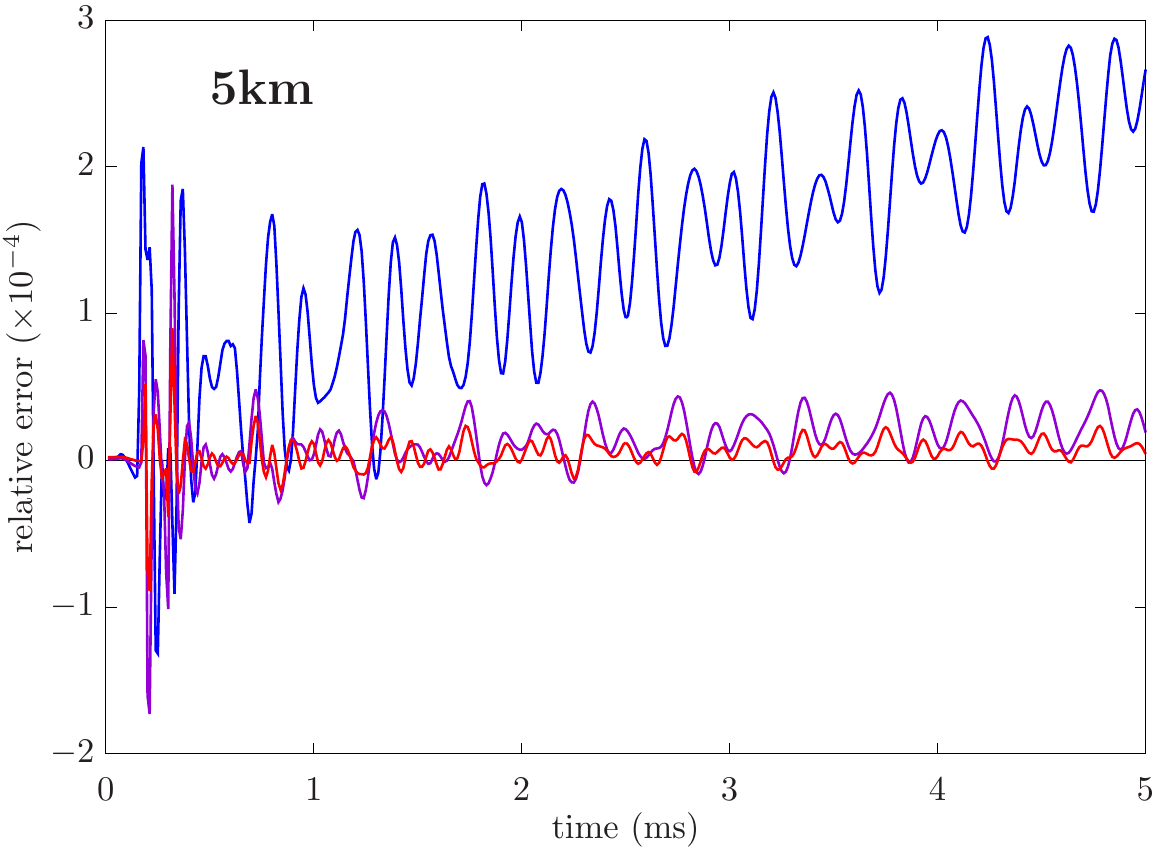}
      \end{minipage} &
         \begin{minipage}[t]{0.31\hsize}
        \centering
        \includegraphics[scale=0.49]{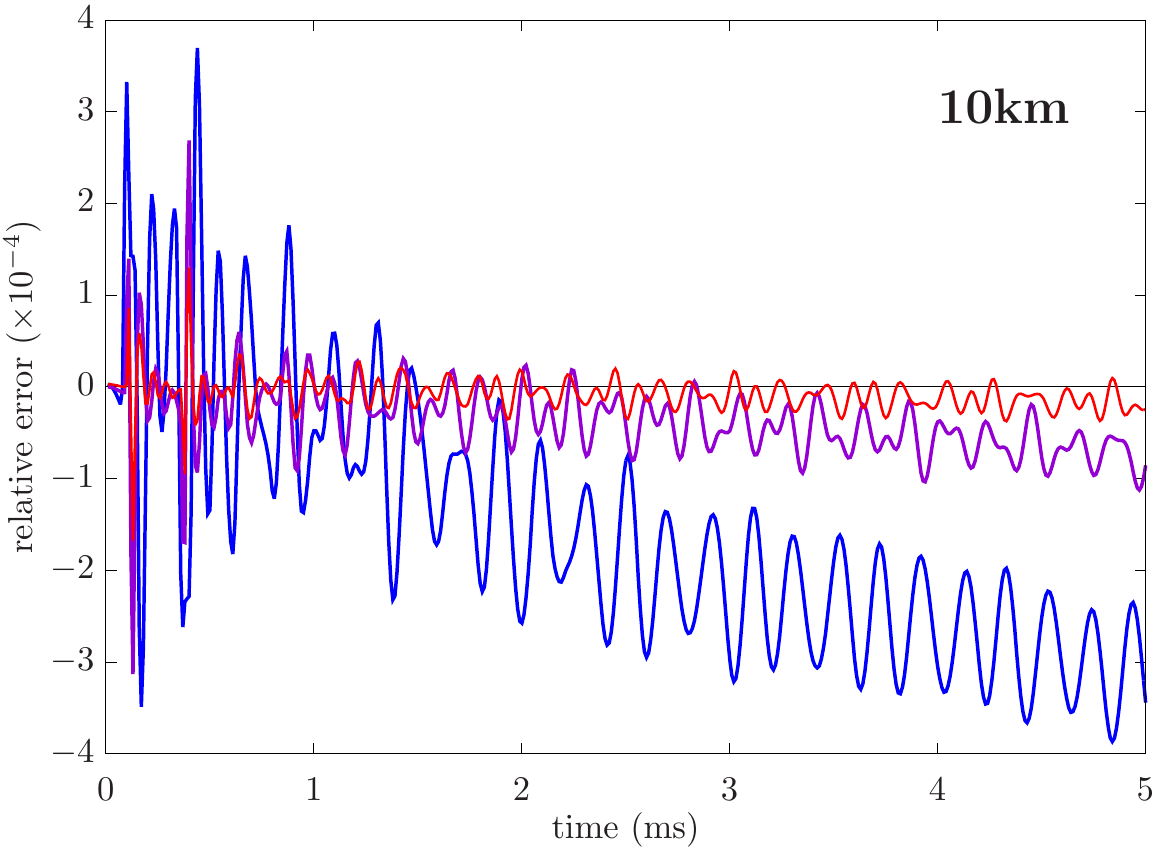}
      \end{minipage}     
    \end{tabular}
    \caption{Relative error of density for different spatial positions; 1km (left), 5km (middle), and 10km (right). Red, purple and blue lines denote the results for 384 mesh, 256 mesh, and 128 mesh, respectively.}
\end{figure}
  
\subsection{Black Hole Accretion Test}
\label{sec_BHacc}
In this test, we test code's ability to treat matter with relativistic speed under a strong gravitational field. We will perform tests for the matter accreting onto a Kerr BH. Similarly to the previous test in \ref{sec_TOV}, we start from a stable state and check that the initial state is maintained with time evolution.

We constructed the reference model in the following way. If we assume steady state and limit the motion only on the equatorial plane ($\theta=\pi/2$), the matter equations around a Kerr BH reduce to the ordinary differential equations with respect to radius \citep{Nagakura2009}:
\begin{equation}\label{eq_S_r}
\partial_r\left(r^2\rho u^r\right) = 0,
\end{equation}
\begin{equation}\label{eq_BHevo}
\partial_r p + \rho_0u^r \partial_r(hu_r) = \frac{1}{2}\rho_0 h\left\{\partial_r g_{rr}(u^r)^2 + \partial_r g_{\phi\phi}(u^\phi)^2 + \partial_r g_{tt}(u^t)^2 + 2\partial_r g_{t\phi}u^t u^\phi\right\},
\end{equation}
\begin{equation}\label{eq_hu_t}
\partial_r(hu_t) = 0,
\end{equation}
\begin{equation}\label{eq_hu_phi}
\partial_r(hu_\phi) = 0.
\end{equation}
Note that Boyer–Lindquist coordinates are used for the Kerr metric. 

By explicitly expanding the specific enthalpy by the pressure, the equation \ref{eq_BHevo} can be rewritten as the equation with a single pressure derivative term as following:
\begin{equation}\label{eq_dpdr}
\frac{dp}{dr} = \frac{\rho hu^ru_r/2 - \rho h (u^r)^2 g_{rr,r} + 2u^r h g_{rr} S_r / r^3}{1 + u^r u_r - u^r h g_rr S_r/(r^2\Gamma p)},
\end{equation}
where $S_r\equiv r^2\rho u^r$.
We solve the equation \ref{eq_dpdr} using the fourth-order explicit Runge-Kutta method. The other equations \ref{eq_S_r}, \ref{eq_hu_t} and \ref{eq_hu_phi} are used to calculate the four-velocity from the pressure.

In this test, we employ a BH with mass $M_\mathrm{BH}=5M_\odot$ with dimensionless spin parameter $0.5$. With these parameters, horizon is located at $13.8\,\mathrm{km}$. We do not solve hydrodynamic equations inside the horizon, where the metric is singular. We employ the gamma-law EOS with $\Gamma=4/3$ in this test.

The parameters for the hydrodynamics variables are as follows. We assume that the matter with the density $\rho=1\times10^7\,\mathrm{g\,cm^{-3}}$ is constantly injected from the radius $r=300\,\mathrm{km}$ with supersonic velocity $u^r=-1.2c_s$ where $c_s$ is the speed of sound $c_s \equiv \sqrt{\Gamma p/\rho}$. As for the specific angular momentum $\lambda=u_\phi/u_t$, we test three cases; $\lambda=0$ (no rotation), $GM_\mathrm{BH}/c^2$ (prograde) and $GM_\mathrm{BH}/c^2$ (retrograde). Figure B.4 shows the initial hydrodynamics profiles for three cases.
\begin{figure}[htbp]
    \label{fig_BHacc_ini}
    \begin{tabular}{ccc}
      \begin{minipage}[t]{0.31\hsize}
        \centering
        \includegraphics[scale=0.49]{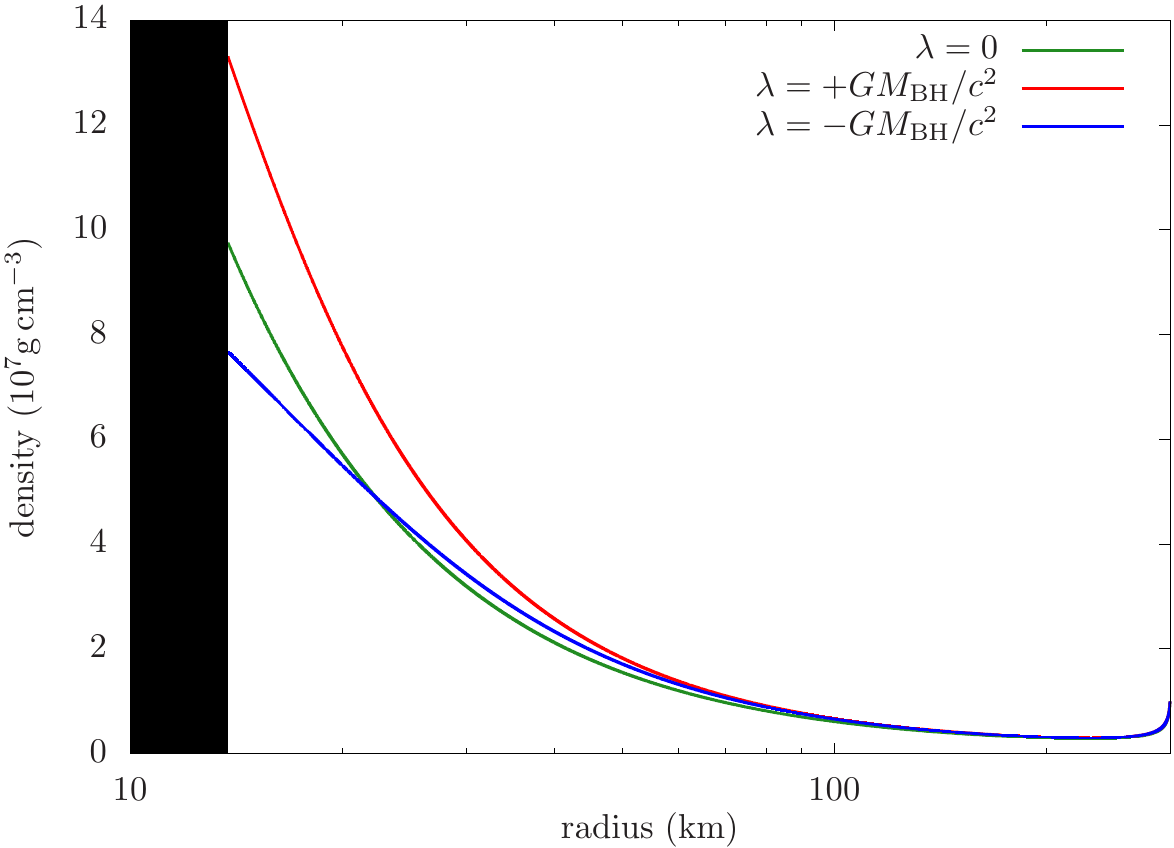}
      \end{minipage} &
      \begin{minipage}[t]{0.31\hsize}
        \centering
        \includegraphics[scale=0.49]{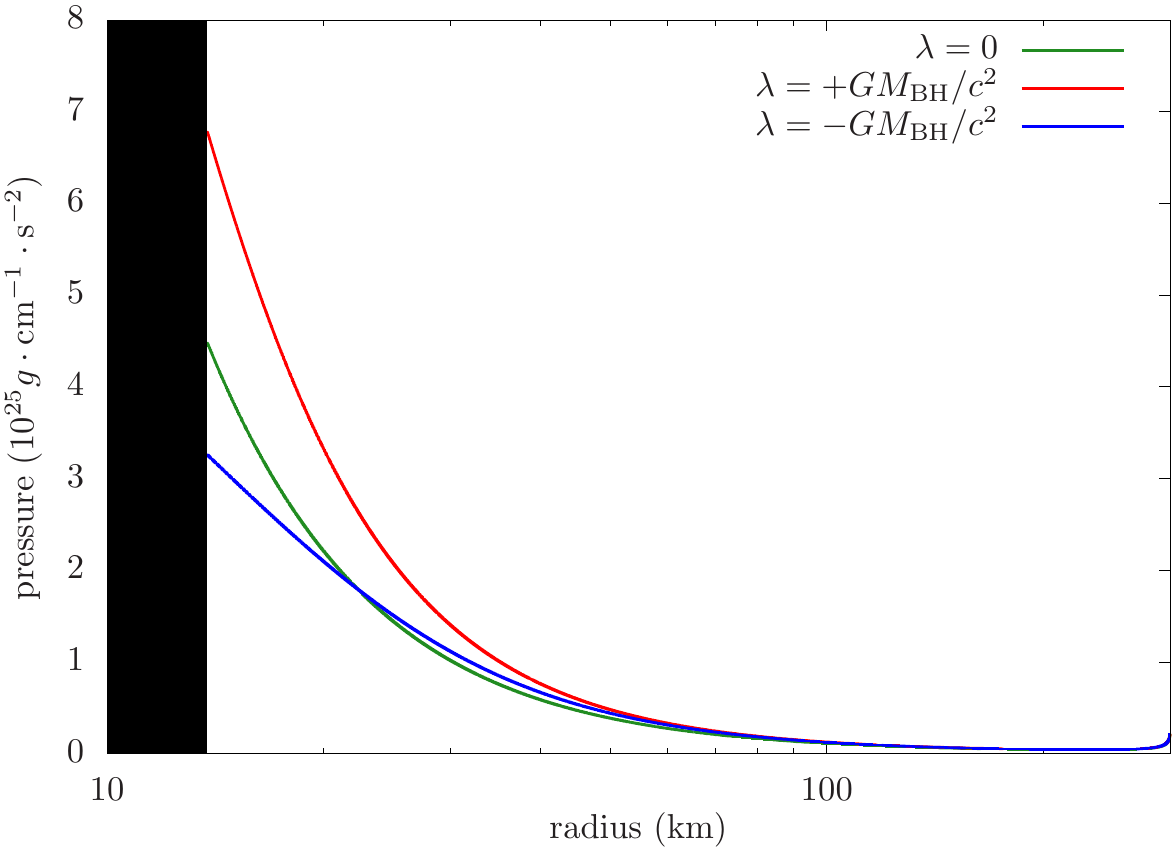}
      \end{minipage} &
         \begin{minipage}[t]{0.31\hsize}
        \centering
        \includegraphics[scale=0.49]{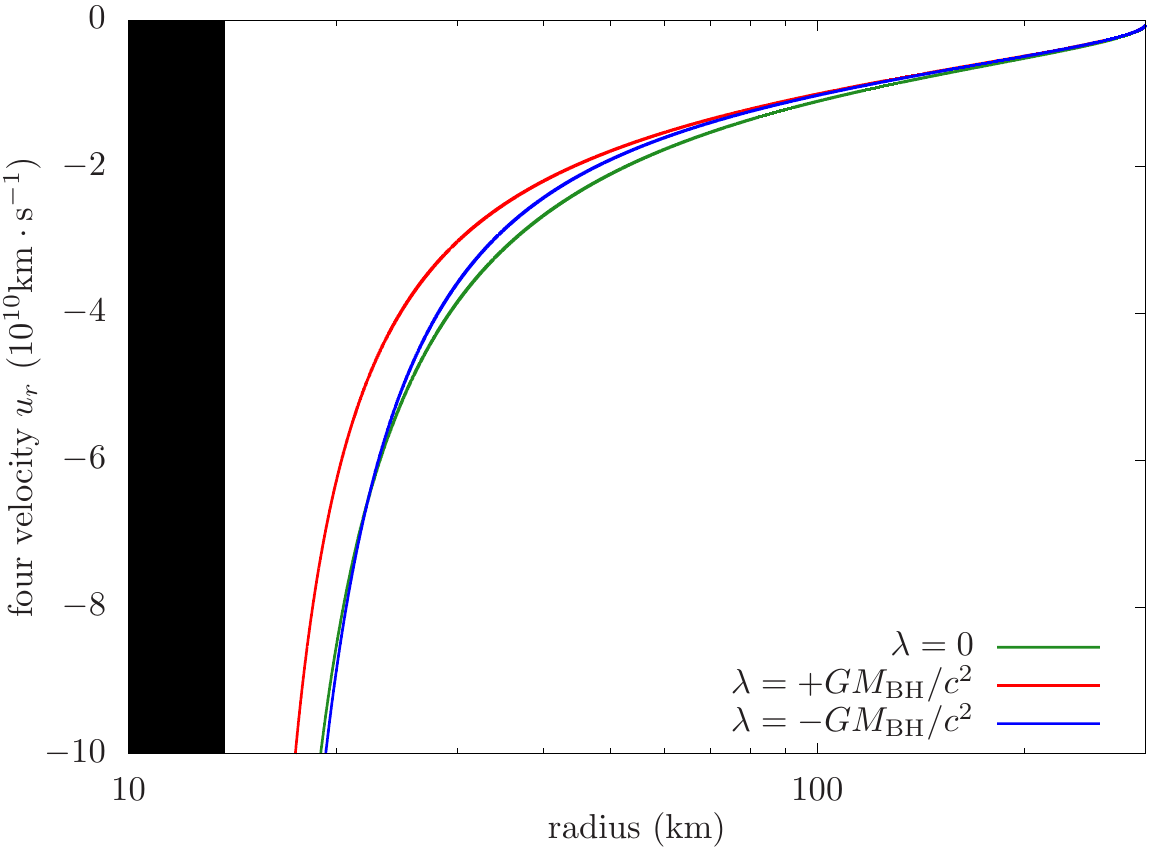}
      \end{minipage}     
    \end{tabular}
    \caption{Radial profiles of density (left), pressure (middle), and the $r$-component of the four velocity (right). The green, red, and blue lines denote the profiles for $\lambda=0$, $\lambda=GM_\mathrm{BH}/c^2$ and $\lambda=-GM_\mathrm{BH}/c^2$, respectively. The black-shaded region shows the horizon of the central BH.}
\end{figure}

In the hydrodynamics calculation, the outermost meshes are fixed to constantly inject the matter. As same as previous tests, we test three resolutions; $N_r=128$, 256, and 384 grid points. The computational region is $r\in[0:300]\,\mathrm{km}$, and the mesh width is varied exponentially so that the resolution gets finer for smaller radius.

Figure B.5 shows the time evolution of the relative error of density. The error at 20km is the order of $o(10^{-3})$, and the error at 50km is the order of $o(10^{-4})$. All calculations show reasonable agreement. In addition, the highest resolution case $N_r=384$ gives the most accurate result, which indicates the resolution convergence.
\begin{figure}[htbp]
    \label{fig_BHacc}
    \begin{tabular}{ccc}
      \begin{minipage}[t]{0.31\hsize}
        \centering
        \includegraphics[scale=0.5]{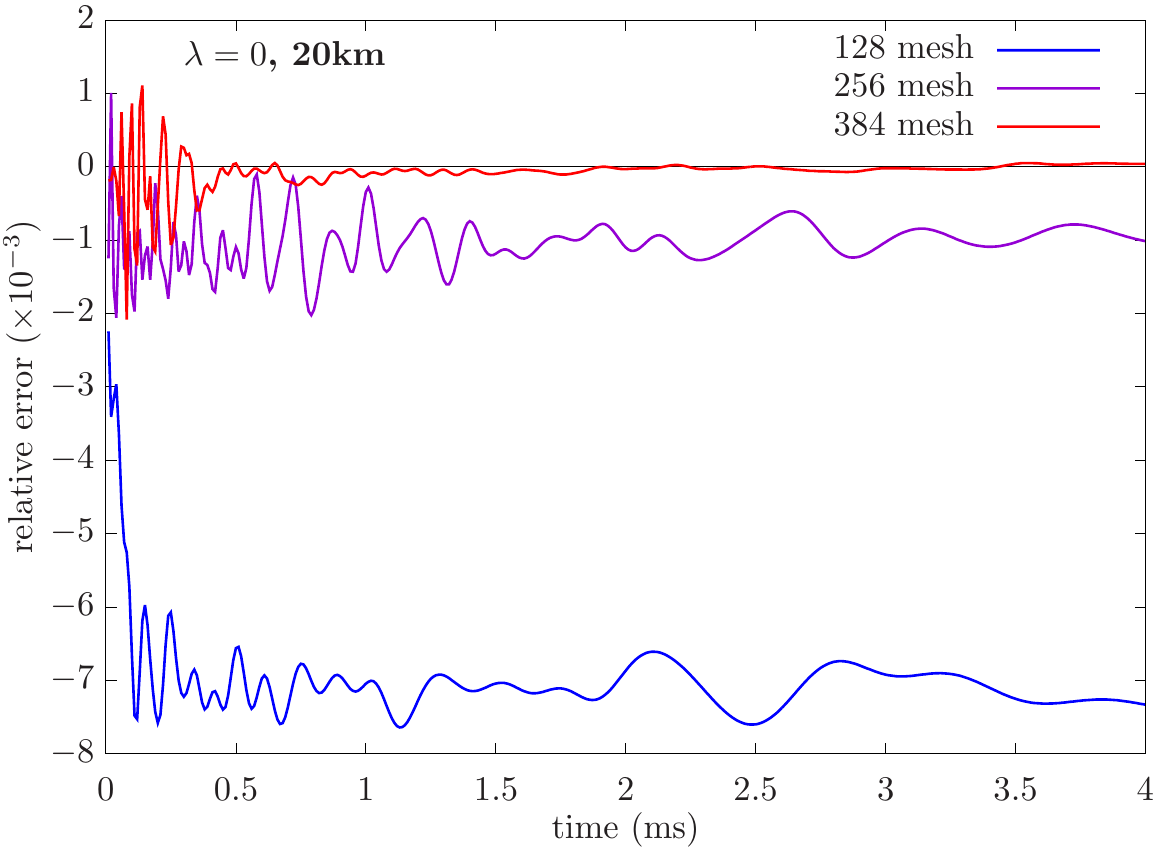}
      \end{minipage} &
      \begin{minipage}[t]{0.31\hsize}
        \centering
        \includegraphics[scale=0.5]{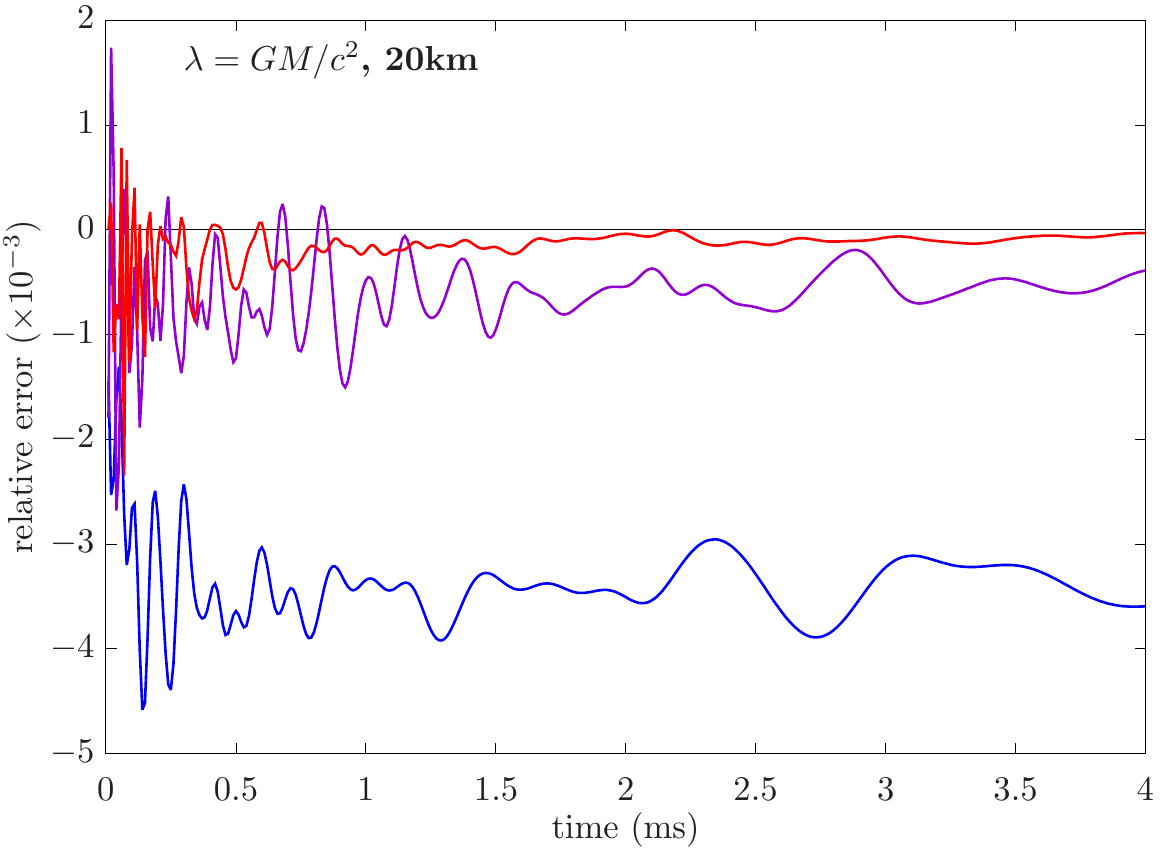}
      \end{minipage} &
         \begin{minipage}[t]{0.31\hsize}
        \centering
        \includegraphics[scale=0.5]{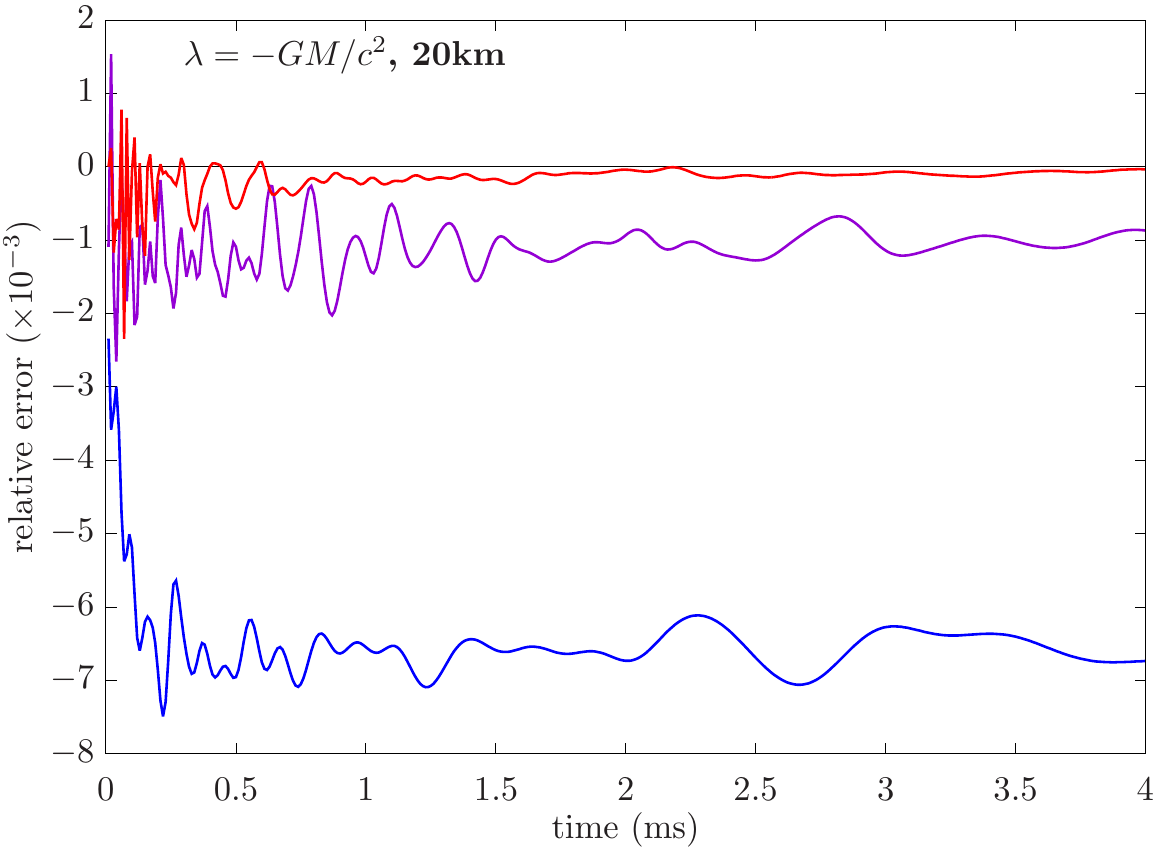}
      \end{minipage} \\ 
      \begin{minipage}[t]{0.31\hsize}
        \centering
        \includegraphics[scale=0.51]{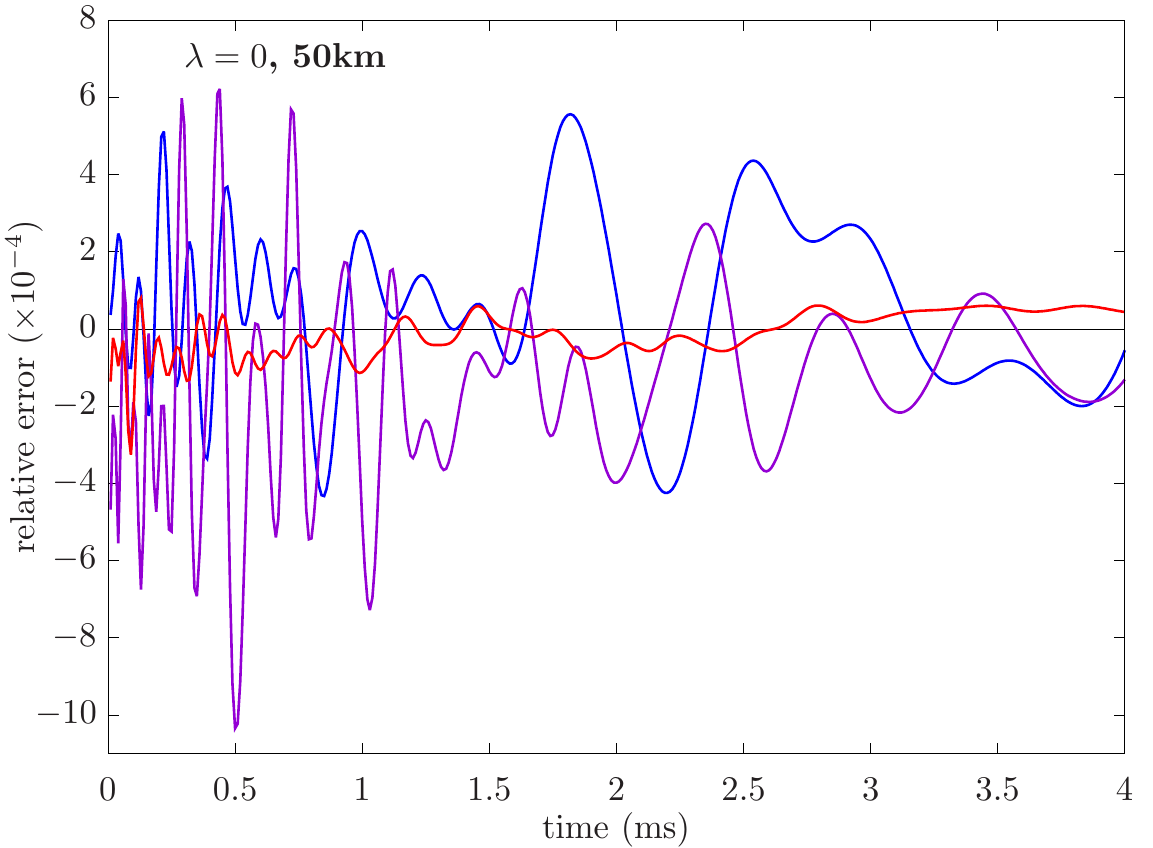}
      \end{minipage} &
      \begin{minipage}[t]{0.31\hsize}
        \centering
        \includegraphics[scale=0.51]{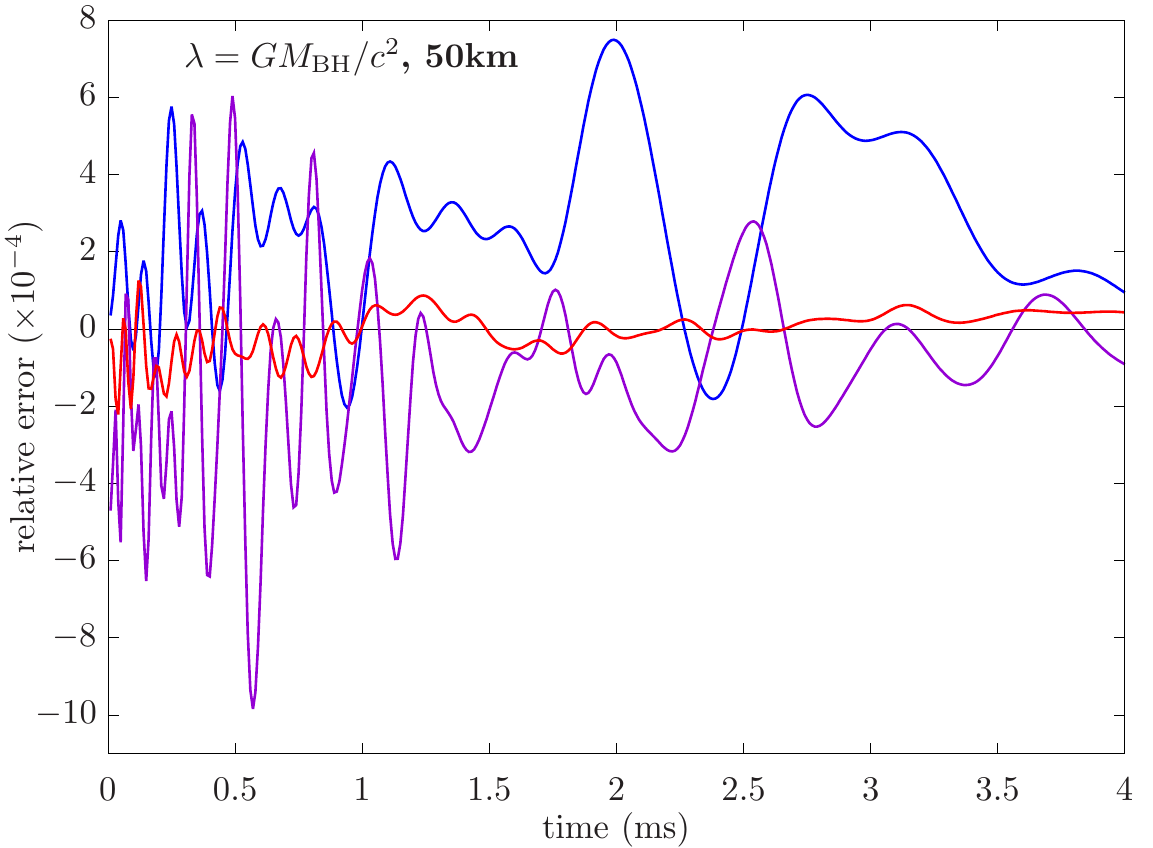}
      \end{minipage} &
         \begin{minipage}[t]{0.31\hsize}
        \centering
        \includegraphics[scale=0.51]{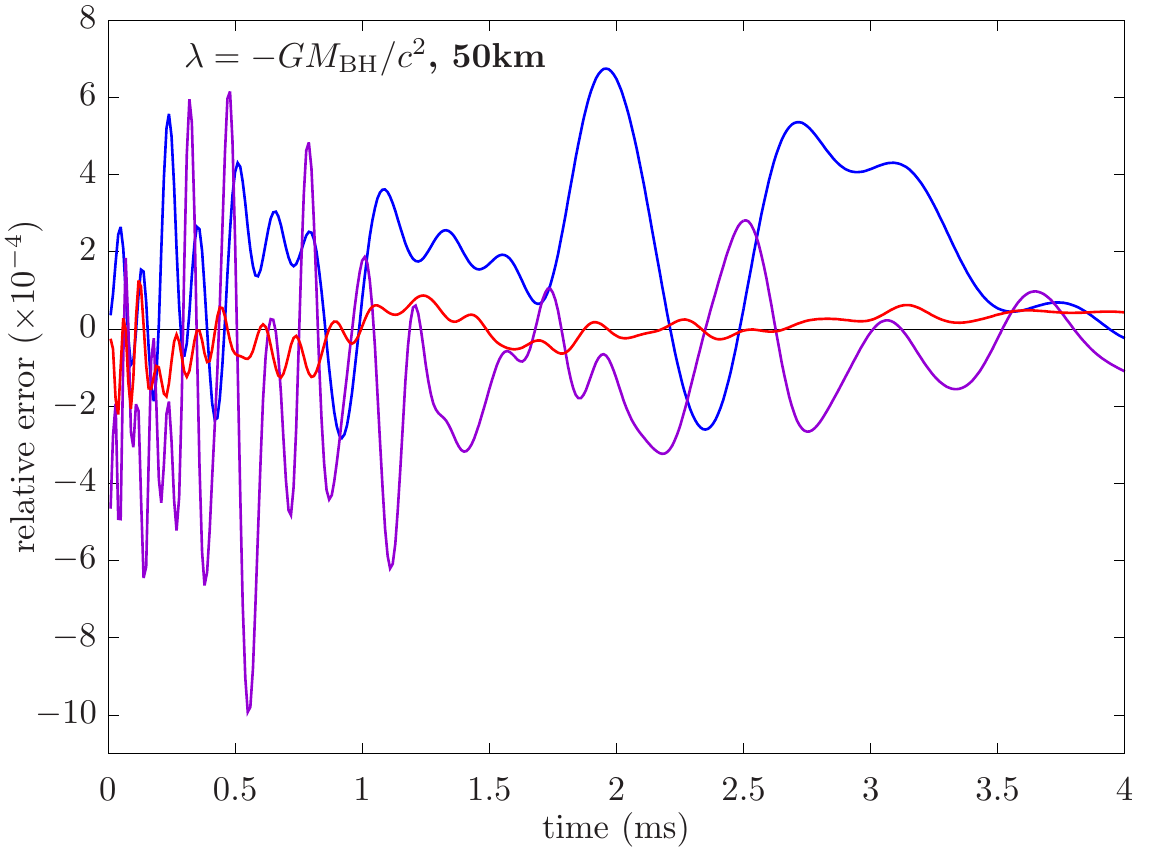}
      \end{minipage}       
    \end{tabular}
    \caption{Relative error of density for different specific angular momentum; non-rotating (left), prograde (middle), and retrograde (right). The spatial position are 20km (top) and 50km (bottom). Red, purple and blue lines denote the results for 384 mesh, 256 mesh, and 128 mesh, respectively.}
\end{figure}


\bibliography{sample631}{}
\bibliographystyle{aasjournal}



\end{document}